# Functional outlier detection for density-valued data with application to robustify distribution to distribution regression

Xinyi Lei[a,b,c,1], Zhicheng Chen[a,b,c,1,*], Hui Li[a,b,c]

[a]Key Lab of Intelligent Disaster Mitigation of the Ministry of Industry and Information Technology, Harbin Institute of Technology, Harbin, 150090, China
[b]Key Lab of Structures Dynamic Behavior and Control of the Ministry of Education, Harbin Institute of Technology, Harbin, 150090, China
[b]School of Civil Engineering, Harbin Institute of Technology, Harbin, 150090, China

## ABSTRACT

Distributional data analysis, concerned with statistical analysis and modeling for data objects consisting of random probability density functions (PDFs) in the framework of functional data analysis (FDA), has received considerable interest in recent years. However, many important aspects remain unexplored, such as outlier detection and robustness. Existing functional outlier detection methods are mainly used for ordinary functional data and usually perform poorly when applied to PDFs. To fill this gap, this study focuses on PDF-valued outlier detection, as well as its application in robust distributional regression. Similar to ordinary functional data, detecting the shape outlier masked by the "curve net" formed by the bulk of the PDFs is the major challenge in PDF-outlier detection. To this end, we propose a tree-structured transformation system for feature extraction as well as converting the shape outliers to easily detectable magnitude outliers, relevant outlier detectors are designed for the specific transformed data. A multiple detection strategy is also proposed to account for detection uncertainties and to combine different detectors to form a more reliable detection tool. Moreover, we propose a distributional-regression-based approach for detecting the abnormal associations of PDF-valued two-tuples. As a specific application, the proposed outlier detection methods are applied to robustify a distribution-to-distribution regression method, and we develop a robust estimator for the regression operator by downweighting the detected outliers. The proposed methods are validated and evaluated by extensive simulation studies or real data applications. Relevant comparative studies demonstrate the superiority of the developed outlier detection method with other competitors in distributional outlier detection.

---

* Corresponding author.
*E-mail address:* zhichengchen@hit.edu.cn (Z. Chen).

[1] Both authors contributed equally to this work.

## 1. Introduction

Functional data analysis (FDA), pertaining to statistical analysis of random objects (e.g., random curves and surfaces) that can be mathematically represented by functions, has made great strides in the past two decades. It has become a particularly important research area in statistics and has gained various applications in other fields (Müller, 2016; Wang et al., 2016; Aneiros et al., 2019). For a detailed introduction to FDA, see the monographs of Ramsay and Silverman (2005), Ferraty and Vieu (2006), Horváth and Kokoszka (2012), Kokoszka and Reimherr (2017). Driven by the rapid development of FDA, statistical methods for random probability density function (PDF)-valued data have also seen a research boom in recent years because such data can be widely encountered in many disciplines such as finance (Horta and Ziegelmann, 2018; Kokoszka et al., 2019), medicine (Petersen et al., 2019; Matabuena et al, 2021), geoscience (Talská et al.,2021) as well as engineering (Chen et al., 2018; Chen et al., 2019a; Chen et al., 2021), see Petersen et al. (2021a) for a recent review. PDF-valued data are special functional data subjected to constraints: (1) taking nonnegative values everywhere and (2) integration over the entire domain must be one. Moreover, the space of the PDFs is not close under ordinary linear operation. These specificities make statistical analysis for PDF-valued data much different from ordinary functional data, and the former is usually more challenging. In some statistics literature, PDF-valued data are also classified as distributional data (Zhu and Müller, 2021; Pegoraro and Beraha, 2021) to distinguish them from ordinary functional data. So far, distributional data analysis mainly concentrates on principal component analysis (Hron et al., 2016; Petersen and Müller, 2016; Cazelles et al., 2018; Pegoraro and Beraha, 2021), regression analysis (Oliva et al., 2013; Arata, 2017; Chen et al., 2018; Talská et al., 2018; Petersen et al., 2019; Chen et al., 2019a; Petersen and Müller, 2019a; Han et al., 2020; Chen et al., 2021a (2020a); Petersen et al., 2021b; Chen et al., 2021; Pegoraro and Beraha, 2021; Ghodrati and Panaretos, 2021), time series analysis (Horta and Ziegelmann, 2018; Guégan and Iacopini, 2018; Kokoszka et al., 2019; Zhang et al., 2021; Zhu and Müller, 2021), correlation analysis (Petersen and Müller, 2019b; Chen et al., 2020; Zhou et al., 2021), spline representations (Machalová et al., 2020; Hron et al., 2020), and so on. Many aspects remain unexplored, such as outlier detection and robustness improvement. The latter is a critical performance requirement in real applications; but unfortunately, it has not received sufficient attention. Outlier detection not only contributes to uncovering abnormal patterns of distributions and providing more insights for the investigated data but also can be exploited to robustify various related statistical analysis methods by downweighting the impacts of detected outliers.

Outlier detection for scalar data has been developed for a long time, but related studies on functional data have only begun in recent years (Kuhnt and Rehage, 2016). For ordinary functional data, outliers can be generally divided into two categories: (1) magnitude outliers and (2) shape outliers (Harris et al., 2020; Dai et al., 2020). The magnitude outlier is much easier to detect. However, detecting shape outliers is much more challenging (Arribas-Gil and Romo, 2014; Nagy et al., 2017; Dai et al., 2020), because the "curve body" of the shape outlier is completely hidden in the majority of the data. So far, many researchers in the statistical community have proposed various approaches for functional outlier detection, most of which are either based on graphical detection tools or functional depth.

The functional boxplot (Sun and Genton, 2011) is the most widely used graphical tool for functional outlier detection. To date, several different modifications of functional boxplots have been developed to accommodate different types of data, such as spatio-temporal data (Sun and Genton, 2012), histogram data (Verde et al.,2014), phase data captured by warping functions (Xie et al.,2017, Xie et al., 2020), multivariate curves (Dai and Genton, 2018a), and trajectory data (Yao et al., 2020). Other representative graphical tools for functional outlier detection include the functional bagplot (Hyndman and Shang, 2010) and the outliergram (Arribas-Gil and Romo, 2014). The standard functional boxplot is more suitable for detecting magnitude outliers rather than shape outliers, while the phase boxplot (Xie et al., 2017) and the outliergram (Arribas-Gil and Romo, 2014) have some potential for shape outlier detection.

Functional depth is a statistical notion that can be used to measure the centrality of curves and has been widely applied in center-outwards rankings as well as outlier detection for functional data (López-Pintado and Romo, 2009; Claeskens et al., 2014; Narisetty and Nair, 2016; Nagy et al., 2017). One popular functional depth is the band depth defined by López-Pintado and Romo (2009), which is also the foundation for constructing some of the aforementioned graphical tools, such as the standard functional boxplot (Sun and Genton, 2011), and the outliergram (Arribas-Gil and Romo, 2014). The weakness of such a band depth includes insensitivity to shape outliers and high computational intensity, despite fast algorithms being available for speeding up calculations (Sun et al., 2012). Researchers have also defined related functional depths specifically for shape outlier detection, such as the functional tangential angle (FUNTA) pseudo-depth (Kuhnt and Rehage, 2016), modified integrated and infimal depths (Nagy et al., 2017), functional directional outlyingness (Dai and Genton, 2018b, 2019), and elastic depth (Harris et al., 2020). The functional directional outlyingness (FDO)-based approach (Dai and Genton, 2019) and the elastic-depth-based approach (Harris et al., 2020) use the strategy of decomposing the total information of curves into magnitude information and shape information, to enhance their effectiveness in shape outlier detection. However, they also have imperfections. For instance, the FDO-based approach is sensitive to the weight function used in directional outlyingness calculation, and improper selection of the detection interval (related to the weight function) may lead to high false positives. The calculation of the elastic depth involves a computationally expensive phase-amplitude separation procedure. Other functional depths that have been employed for functional outlier detection include the functional halfspace depth (Claeskens et al., 2014), extremal depth (Narisetty and Nair, 2016), and others. The shape outliers presented in real datasets are complex and usually exhibit various patterns; hence, detecting them by using a single feature or single depth is generally difficult, and satisfactory results are hard to obtain.

In addition to the graphical tools or depth notations exploited for functional outlier detection, researchers have proposed approaches to extract the geometrical features (e.g., arc-length, curvature, and others) of functions for shape outlier detection (Lejeune et al., 2020). Recently, Dai et al. (2020) proposed a more effective approach for functional outlier detection by applying sequential transformations to the functional data to convert shape outliers to easily detected magnitude outliers. However, the transformations introduced by Dai et al. (2020) are relatively scattered and unsystematic and are not suitable for PDF-valued data. As mentioned above, the PDF-valued data have their specialties, and ordinary functional outlier detectors usually lose their efficacies. So far, a systematic transformation-based detection chain that suits both ordinary functional data and distributional data has not been developed. The literature on PDF-outlier detection is quite rare, and many aspects remain unexplored. A representative existing approach is the modified boxplot proposed by Verde et al. (2014) for histogram data, but it is only able to identify magnitude outliers.

To this end, we propose an alternative tree-structured transformation system for detecting PDF outliers. The properties of the distributional data are taken full account to uncover the shape outliers hidden in the bulk of the curves, and computationally efficient outlier detectors are designed for the related transformed data. We also develop a multiple detection method with variable arguments to cope with the uncertainty induced by model setting and subtly use it to quantify the degree of abnormality of functional outliers based on their frequencies of detection. Such a strategy can also provide an effective solution for fusing different detection methods, wherein their strengths are fostered while their weaknesses are circumvented. Moreover, we propose a nonparametric distributional-regression-based approach for detecting abnormal associations of PDF pairs, which is particularly significant in distributional regression analysis. Finally, we emphasize that outlier detection is usually performed in the data preprocessing stage, thus simplicity, user-friendliness, and high efficiency should take priority for the detection methods. Our methods can meet these requirements.

To illustrate the utility of the proposed functional outlier detection method, we apply it to robustify a distribution-to-distribution regression model (a type of functional regression model where both predictor and response are probability distributions). In recent years, distribution-to-distribution regression is garnering increased research interest in the framework of FDA, existing models can generally be divided into linear and nonlinear models. Representative linear models include the multivariate linear regression model constructed in the Bayes space (Arata, 2017), and the Wasserstein regression first proposed by Chen et al. (2020a, 2021a) and continued by Pegoraro and Beraha (2021). For nonlinear approaches, representative models include the kernel distribution-to-distribution model (Oliva et al., 2013), the distribution-to-warping function regression model (Chen et al., 2018; Chen et al., 2021), and the LQD-RKHS distributional regression model (Chen et al., 2019a). Most recently, Ghodrati and Panaretos (2021) also proposed another distribution-to-distribution regression model based on optimal transport and Wasserstein distances. Other related works include functional regression models with only the response being probability distributions, while the predictors are either scalar data or vector data. Representative models include the compositional regression model (Talská et al., 2018), additive regression model (Han et al., 2020), Fréchet regression model (Petersen and Müller, 2019a; Petersen et al., 2021b), and the spline-based model (Yang, 2020). In addition, regression models for PDF-valued time series have also been developed in recent years (Kokoszka et al., 2019; Zhang et al., 2021; Zhu and Müller, 2021). To date, distributional regression has not been adequately studied. Robustness is a major problem that has great practical importance, but little attention has been paid to this issue. To this end, this study also proposes a robust estimator for the LQD-RKHS distributional regression model (Chen et al., 2019a) in line with the depth-based least squares (DLS) estimator proposed by Martínez-Hernández et al. (2019) for the linear autoregressive model. The DLS estimator is a type of weighted least squares estimator, and its robustness is achieved by downweighting the outlying functional samples detected by appropriate functional outlier detectors. We extend the DLS estimator to nonlinear functional regression using the reproducing kernel Hilbert space (RKHS) theory. In contrast to the work of Martínez-Hernández et al. (2019), the RKHS-based nonlinear functional regression is significantly different from the linear model studied in Martínez-Hernández et al. (2019), which has to be handled quite differently.

The major contributions of our study include: (1) a computationally efficient tree-structured transformation system for feature extraction and facilitating shape outlier detection in distributional anomaly diagnosis; (2) a multiple detection strategy for handling detection uncertainties, which also shows its superiority in abnormality quantification as well as fusing different detectors; (3) a distributional-regression-based diagnostic tool for detecting abnormal associations of density pairs; and (4) a robust estimator for the LQD-RKHS distribution-to-distribution regression model based on our proposed functional outlier detection method.

The remainder of this article is organized as follows. The proposed functional outlier detection methods for distributional data are detailed in Section 2, followed by an application to robustify a distribution-to-distribution regression method in Section 3. Next, using synthetic or real data, the proposed methods are validated and evaluated in Section 4, and the conclusions and discussions are summarized in Section 5.

## 2. Distributional outlier detection methods

### *2.1. Tree-structured transformation system for distributional outlier detection*

Consider a PDF-valued dataset denoted as $\{f_i(x)\}_{i=1}^{n}$ consisting of $n$ different PDFs.

Without loss of generality, all PDFs are assumed to be supported on the compact interval $I = [0,1]$. Generally, such a PDF-valued dataset may contain two types of outliers: (1) the horizontal-shift outlier and (2) the shape outlier (see Fig. 1 for a schematic illustration). The former is much easier to detect, while detection of the latter is much more challenging because they are masked by the "curve net" of the majority of the data.

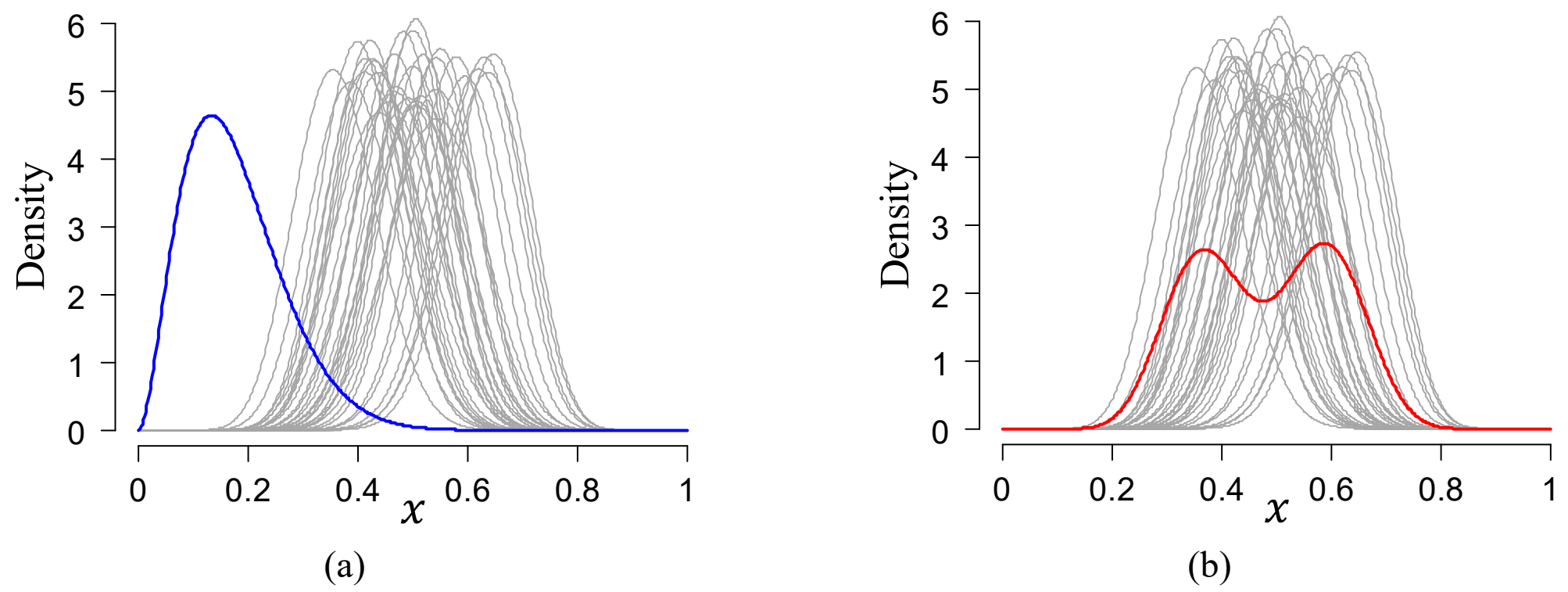

**Fig. 1.** The horizontal-shift outlier (represented by the blue line in the left panel) and shape outlier (represented by the red line in the right panel) contained in a PDF-valued dataset. For comparison, the "good" data are represented by gray lines.

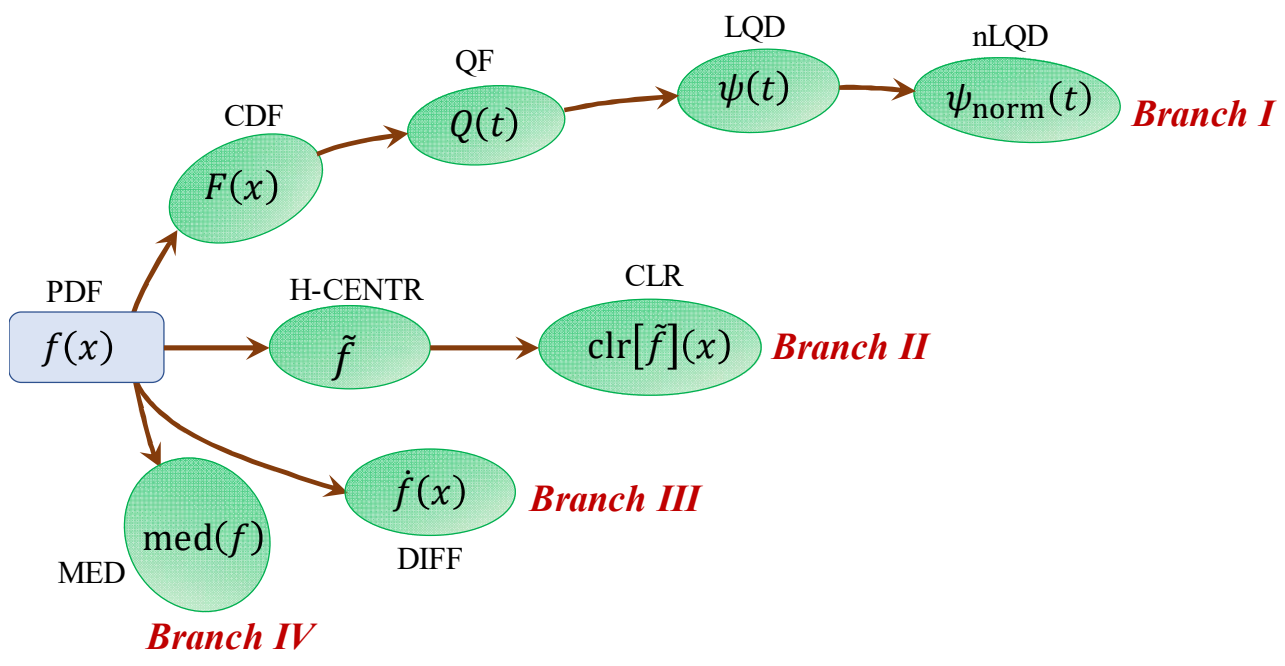

**Fig. 2.** The transformation tree.

In real applications, the shape outliers hidden in functional data would often exhibit various patterns, which are generally difficult to be fully screened out by a single feature or single outlyingness measure. Here, we consider a collection of transformations for extracting different features for outlier detection, as well as for converting the less detectable shape outliers into more detectable magnitude outliers. The considered transformations for the PDF-valued data can be assembled into a tree-structured transformation system, as shown in Fig. 2. With a slight abuse of terminology, such a transformation system is called the transformation tree throughout this study.

At first glance, one can find that the transformation tree is composed of four different branches, whose functions in outlier detection are detailed as follows:

(1) Branch I

The first branch of the transformation tree starts with performing integral on the density functions and consists of four nodes represented by ellipses (see Fig. 2). After we conduct the integral operations on the PDFs $\{f_i(x)\}_{i=1}^{n}$, they can be transformed into cumulative distribution functions (CDFs) denoted as $\{F_i(x)\}_{i=1}^{n}$. Next, we can calculate their corresponding inverse functions to obtain the quantile functions (QFs) denoted as $\{Q_i(t)\}_{i=1}^{n}$ as illustrated in Fig. 3 (b). In the QF-space, the horizontal-shift outliers have been converted to magnitude outliers; however, the shape outliers may still hide in the majority of the data. The main contribution of such a quantile-function-based transformation is that it provides a natural alignment for disordered functional data. To uncover the hidden shape outliers, we further perform the derivative calculation on the quantile functions and take the logarithm of the results (this is actually the log quantile density (LQD) transformation) (Petersen and Müller, 2016), i.e.,

$$\psi_i(t) = \log\left(\frac{dQ_i(t)}{dt}\right) = \log\left(q_i(t)\right) \qquad (1)$$

where $q_i(t)$ represents the quantile density function. The transformed result $\psi_i(t)$ is ordinary functional data that live in the Hilbert space $L^2[0,1]$.

Generally, after the LQD transformation, the shape outliers can significantly stand out in comparison to the majority of the data, as shown in Fig. 3 (c). Moreover, such an LQD transformation has an appealing advantage in exposing the hidden shape outliers attributed to the horizontal variability of the PDF-valued data. It is worth noting that the horizontal translation of a PDF (e.g., $f_1(x) = f(x-c)$, where $c$ stands for any constant) is equivalent to the vertical translation of the corresponding quantile function (e.g., $Q_1(t) = Q(t) + c$). The quantile density function $q(t)$ in the LQD transformation is the derivative of the quantile function $Q(t)$, which is independent of the vertical translation of $Q(t)$ due to $\frac{dQ(t)}{dt} = \frac{d(Q(t)+c)}{dt}$. Consequently, the functional data corresponding to the LQD node in the transformation tree are independent of the horizontal translations of curves in the PDF space (Kokoszka et al., 2019; Petersen et al., 2021a). In other words, the LQD transformation is "blind" to the position shift of the PDF. Such a property makes the LQD transformation to be a powerful tool for revealing the shape outliers masked by the "curve net" formed by the variability in horizontal positions of PDFs.

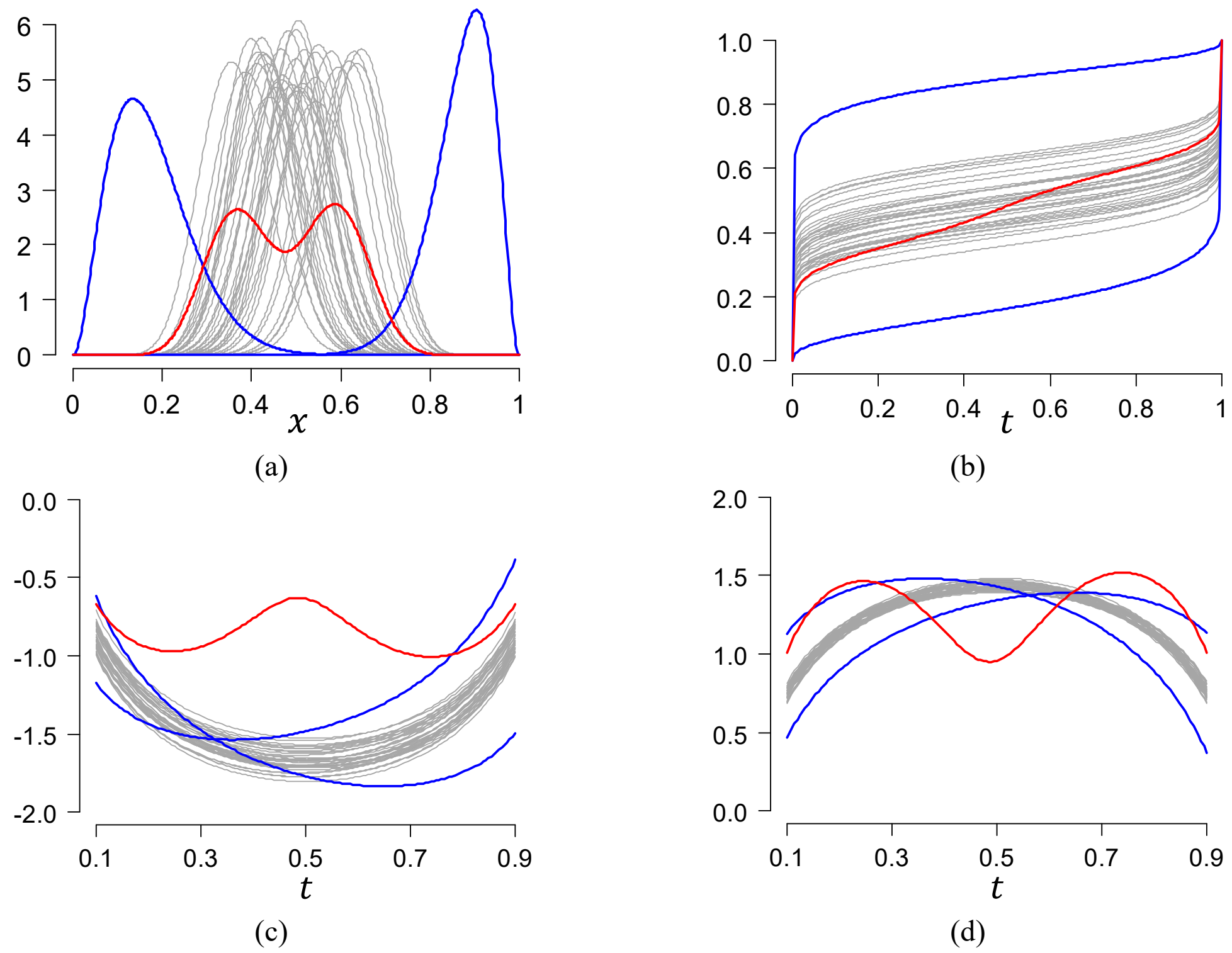

**Fig. 3.** The transformed functional data (of a PDF-valued dataset) associated with nodes QF, LQD, and nLQD in Branch I of the transformation tree. (a) the raw PDF-valued data, (b) the corresponding quantile functions in the QF node, (c) the corresponding LQD transformations in the LQD node, and (d) the corresponding normalized LQD transformations in the nLQD node. The gray lines represent the functional data corresponding to the "good" PDFs, while the colored lines represent the functional data corresponding to the outlying PDFs. For visualization purposes, the lower and upper tails of the curves associated with the LQD and nLQD nodes have been truncated.

Moreover, we recommend normalizing the LQD transformation as follows:

$$\psi_{norm,\,i}(t) = \frac{\psi_i(t)}{\int_0^1 \psi_i(\tau)d\tau} \tag{2}$$

With this normalization, the majority of the data with relatively similar features gather more closely, making the outlying curves more distinguishable, as shown in Fig. 3 (d). According to our experience, this processing usually has favorable effects on shape outlier detection, especially for some real datasets.

It can be concluded that the main contributions of the transformations associated with Branch I are twofold: (1) it aligns the disordered curves into a more ordered form (i.e., align them according to their quantiles); (2) the LQD transformation can eliminate the masking effects (with respect to the shape outliers) induced by horizontal translations of PDFs.

**Remark 1.** In general, when computing the quantile function $Q(t)$ associated with the PDF $f(x)$, it is recommended to perform a preprocessing of the PDF $f(x)$ by adding a small positive constant to it, i.e., $f(x) = (1-\alpha)f(x) + \alpha$, $\alpha > 0$, to avoid the numerical issue in inverse function calculation through interpolation. The magnitude of the constant $\alpha$ depends on the specific application: (1) for outlier detection, $\alpha$ should be smaller (e.g., $\alpha \in [10^{-10}, 10^{-2}]$), otherwise it might increase the variability of the functional data in the LQD node; (2) however, for other applications such as the LQD-RKHS distributional regression described in Chen et al. (2019a), the constant $\alpha$ should be larger (e.g., $\alpha \in [0.2, 0.5]$ as recommended by Chen et al. (2019a)), otherwise the LQD transformation would be "blind" to the horizontal translation of PDFs. Subsection A.2.1 of the online supplement has more details including graphical illustrations.

(2) Branch II

The second branch of the transformation tree consists of two nodes, H-CENTR and CLR, representing the horizontal centralization processing and the centered log-ratio (clr) transformation (Egozcue et al., 2006; Talská et al., 2018) for functional data, respectively.

In the H-CENTR node, all PDFs are translated along the $x$-direction to align them according to pre-specified feature points (e.g., the median, mean, mode, or a combination of them), that is, the PDFs are horizontally translated to make their feature points coincide with each other. Such centralization processing can help expose shape outliers, as shown in the first row of Fig. A-25 in the online supplement for a demonstration. However, it is not the "panacea" for all situations, for instance, performing the same centralization processing on the real dataset shown the second row of Fig. A-25 appears to reveal no outliers, while the latter investigation conducted in Subsection 4.2 finds that this real dataset contains various shape outliers. Generally, after the centralization processing, the resulting PDFs are no longer commonly supported on the interval $[0, 1]$. Let $\mathrm{Htran}[f_i]$ denote the translated PDF of $f_i$ and $[u_i, v_i]$ be the corresponding support of $\mathrm{Htran}[f_i]$; the common support of the translated PDFs can be found as $[u, v] = [\max\{u_1, \cdots, u_n\}, \min\{v_1, \cdots, v_n\}]$. Next, we truncate the translated PDF using the common support to obtain the function $\tilde{f}_i$ in the H-CENTR node, i.e., $\tilde{f}_i(x) = \chi_{\{x\in[u,v]\}}(x) \cdot \mathrm{Htran}[f_i](x), x \in [u,v]$.

The resulting functions $\tilde{f}_i s$ are positive-valued with square-integrable logarithms. If we introduce the linear operations defined in Eqs. (A-1) of the online supplement to the functions $\tilde{f}_i s$, they can be treated as elements of the Bayes space $\mathfrak{B}^2([u,v])$ (Egozcue et al., 2006; Van den Boogaart et al. 2014; Hron et al., 2016). A detailed introduction to the Bayes space is beyond the scope of this article, see Subsection A.1.1 in the online supplement and references therein for more details. It is worth noting that the Bayes space is a quotient space, where the two elements are identical if they are proportional (Hron et al., 2016), i.e., $\tilde{f} =_{\mathfrak{B}} c \cdot \tilde{f}$, $\forall \tilde{f} \in \mathfrak{B}^2([u,v])$ with $c$ standing for an arbitrary positive constant and $=_{\mathfrak{B}}$ standing for the equal sign of the Bayes space. Therefore, although the transformed function $\tilde{f}$ in the H-CENTR node might no longer be a PDF, it is equivalent to the PDF $\tilde{f}/c_0$ with $c_0 = \int_u^v \tilde{f}(x)dx$ if viewed in the Bayes space.

The Hilbert space structure of the Bayes space $\mathfrak{B}^2([u,v])$ naturally induces a distance for measuring the dissimilarity between two PDFs (Talská et al., 2018), i.e.,

$$d_{\mathfrak{B}}(\tilde{f},\tilde{g}) = \|\tilde{f} \ominus \tilde{g}\|_{\mathfrak{B}} = \|\tilde{f} \oplus (-1 \odot \tilde{g})\|_{\mathfrak{B}} \tag{3}$$

where $\oplus$ and $\odot$ denote the perturbation and powering operations defined in Eqs. (A-1a) and (A-1b) of the online supplement, respectively, $\|\cdot\|_{\mathfrak{B}}$ stands for the norm induced by the inner product $\langle\cdot,\cdot\rangle_{\mathfrak{B}}$ (see Eq. (A-2) of the online supplement), i.e., $\|\tilde{f}\|_{\mathfrak{B}} = (\langle\tilde{f},\tilde{f}\rangle_{\mathfrak{B}})^{\frac{1}{2}}$, $\forall \tilde{f} \in \mathfrak{B}^2([u,v])$.

The mathematical expression of the clr transformation is given by Eq. (A-3) of the online supplement, which is the isometric isomorphism between the Bayes space and the $L^2([u,v])$ space (Egozcue et al.,2006; Talská et al., 2018). With this isometric isomorphism, $d_{\mathfrak{B}}(\tilde{f},\tilde{g})$ can be conveniently calculated using:

$$d_{\mathfrak{B}}(\tilde{f},\tilde{g}) = d_{L_2}(\mathrm{clr}[\tilde{f}], \mathrm{clr}[\tilde{g}]) \tag{4}$$

where $\mathrm{clr}[\tilde{f}]$ stands for the clr-representation of $\tilde{f}$, $d_{L_2}$ stands for the $L_2$ distance. This distance with its pseudo versions will be employed to detect outliers in the CLR node.

The second branch of the transformation tree can be regarded as extracting features for PDFs in their own space for outlier detection.

(3) Branch III

The third branch of the transformation tree consists of only one node indexed with DIFF, which directly computes the derivatives to the PDFs. Such a transformation would be helpful to expose the outlying PDFs with significant abnormal slopes with respect to the majority of the data, which is similar to that in Dai et al. (2020) for ordinary functional data.

(4) Branch IV

The fourth branch of the transformation tree also consists of only one node indexed with MED, which uses the median information for outlier detection. Given a PDF $f_i$ defined on the compact interval $[0,1]$, the median denoted as $\mathrm{med}(f_i)$ is defined as

$$\mathrm{med}(f_i) = \inf\left(\left\{x \in [0,1]:\ \int_0^x f_i(t)dt \geq \frac{1}{2}\right\}\right) \tag{5}$$

The median of a horizontal-shift outlier shown in Fig. 1 (a) usually deviates significantly from the medians corresponding to the majority of the PDFs. Hence, performing outlier detection in the median dataset can contribute to the screening of outlying PDFs with abnormal medians (e.g., the horizontal-shift outlier). Note that the median dataset belongs to scalar data, thus the corresponding outliers can be easily detected using the classic boxplot.

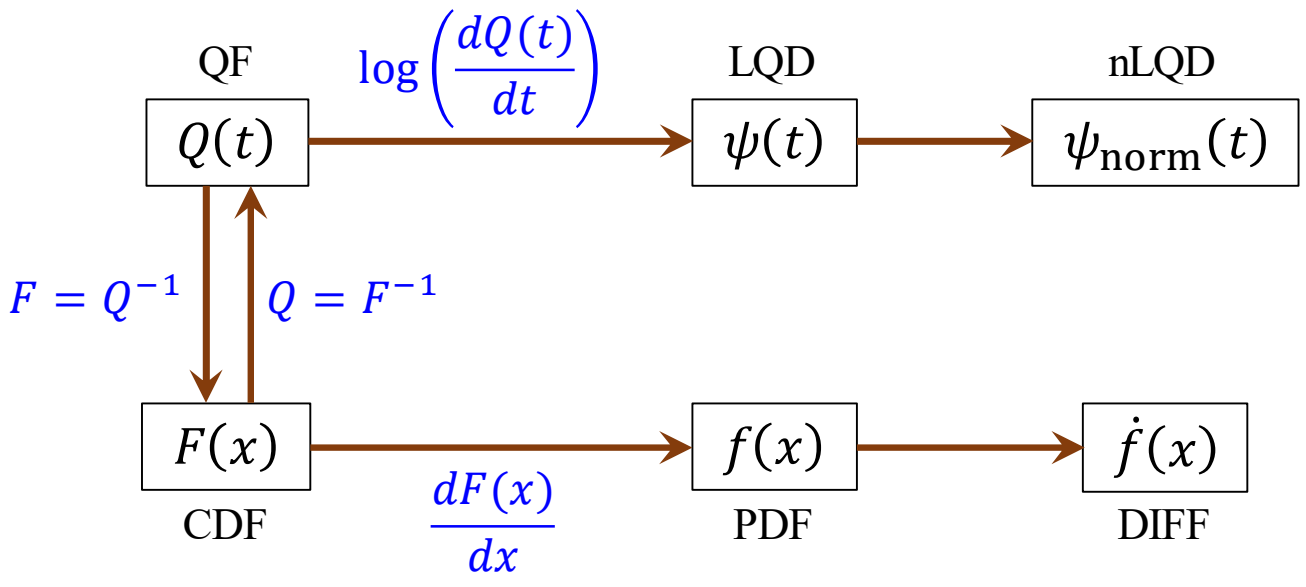

**Fig. 4.** A transformation chain formed by combining Branch I and Branch III of the transformation tree in Fig. 2.

**Remark 2.** The transformations in Branches I and III can be combined to form a transformation chain, as shown in Fig. 4. Note that the quantile function and the CDF are inverse functions with respect to each other, and the information that can be extracted by the chain stems from the distribution function and its reverse function. Another special feature of distributional data is that in contrast to ordinary functional data, one can leverage the information of inverse function for outlier detection.

*2.2. Basic outlier detection principles*

We now present the basic detection principles for functional outliers using the transformed curves, based on which an enhanced detection method will be proposed in the next subsection. Aided by appropriate distance or functional outlyingness, the problem of functional outlier detection can be reduced to the problem of scalar outlier detection. Here, only the detection outlines are provided, and more computational details can be found in the online supplement (Subsections A.3.1~3.3).

We present two efficient approaches for transforming functional outlier to scalar outlier.

(1) Distance-based approach

The first recommended method is the distance-based approach, which is also the preferred approach in this study. For convenience, let $\phi(t)$ be the transformed curves (e.g., $\psi_{norm,i}(t)$ in the nLQD node, $\mathrm{clr}[f_i](x)$ in the CLR node). In addition, let $d(\cdot,\cdot)$ be the user-prescribed distance for outlier detection; the related distances that will be used in this study are listed in Table A-1 in the online supplement. Given the transformed functional dataset denoted as $\mathcal{D} = \{\phi_i(t)\}_{i=1}^{n}$, let $m(t)$ be the center function estimated by the cross-sectional median of the curves, i.e., $m(t) = \underset{1\le i\le n}{\mathrm{median}}\{\phi_i(t)\}$; then, we can calculate the distances of the curves with respect to the center functions

$$\delta_i = d\big(\phi_i(t)\chi_{\{t\in A\}}, m(t)\chi_{\{t\in A\}}\big) \tag{6}$$

where $\chi_{\{\cdot\}}$ denotes the indicator function, and $A = [\omega_1, \omega_2] \subset [0,1]$ is the user-prescribed detection interval. The resulting dataset $\{\delta_i\}_{i=1}^{n}$ is an ordinary scalar dataset, and the elements in $\{\delta_i\}_{i=1}^{n}$ that take anomalously large values correspond to the outlying functions in $\mathcal{D} = \{\phi_i(t)\}_{i=1}^{n}$; thus, the outliers can be detected by the modified one-sided boxplot detector given in Eq. (A-12) of the online supplement. We recommend performing such distance-based detections in nodes nLQD, CLR, and DIFF; more detailed discussions including recommended distances as well as related default argument settings can be found in Subsection A.3.2 of the online supplement. For convenience, this approach is referred to as the Tree-Distance detection method throughout this study.

(2) Functional directional outlyingness (FDO)-based approach

The functional directional outlyingness-based approach was proposed by Dai and Genton

(2019) for ordinary functional outlier detection, which can simultaneously detect magnitude and shape outliers; however, it might suffer an instability issue if the detection interval is improper. Thus, the FDO-based approach is used as a supplementary method for identifying the unexposed shape outliers, and a related strategy to remedy its instability is presented in the next subsection. We recommend performing such FDO-based detections in the quantile function (QF) space (associated with the QF node), and the computational details of the FDO-based approach along with a numerical study to illustrate its instability are provided in Subsection A.3.3 of the online supplement. For convenience, this detection scheme is referred to as the QF-FDO detection method in this study.

*2.3. Multiple detection*

The distributional outlier detection methods discussed above involve various arguments, such as the whisker parameters and the detection region $A = [\omega_1, \omega_2]$. Different argument settings might lead to different detection results. Thus, the detection results contain the uncertainty introduced by the argument settings. This subsection proposes a multiple detection strategy specifically to cope with such uncertainty. It can also provide an effective way to fuse the two detection methods presented in the previous subsection as well as address the instability issue suffered by the FDO-based approach. We emphasize that the distance-based and the FDO-based approaches are highly efficient and capable of implementing multiple detections by setting different arguments.

In multiple detection, the detections for outlying PDFs are duplicated multiple times using different argument combinations. Let $S_O = [l_1, l_2, \cdots, l_{n_O}]$ be the index set of the whole detected outliers, and $\mathcal{J}_O = [J_1, J_2, \cdots, J_{n_O}]$ be their corresponding frequencies of being detected. The outliers that are frequently detected should be those that have higher degrees of outlyingness. Therefore, the frequency of being detected in multiple detection naturally provides a measure for quantifying the degree of outlyingness for the functional outliers (a detailed illustration will be provided later in Subsection 4.2 through application to real data).

Next, we discuss how to remedy the instability issue of the FDO-based approach using the multiple detection strategy. As demonstrated in Subsection A.3.3.2 of the online supplement, the unstable detection mainly leads to an extremely high false-positive rate, which means that the number of detected outliers is considerably higher than the true number of outliers. After independently performing the outlier detections for $m_D$ times in the multiple detection, we can count the number of detected outliers in each duplicated detection and arrange the counts into a vector denoted as $\boldsymbol{N}_{out} = \{n_1^{out}, \cdots, n_k^{out}, \cdots, n_{m_D}^{out}\}$ with $n_k^{out}$ being the number of detected outliers in the $k$th detection. Generally, unstable detections can be identified as abnormally large values in $\boldsymbol{N}_{out}$. Thus, the detection results of the unstable cases can be filtered out; that is, the final results (i.e., the detected outliers and their frequencies of being detected) should exclude the results corresponding to the identified unstable cases.

*2.4. Abnormal association detection for PDF pairs*

We now consider another type of anomaly, that is, the abnormal association of PDF pairs. Consider a PDF-valued dataset consisting of $n$ two-tuples of PDFs denoted as $\mathcal{T} = \{g_i(x), f_i(x)\}_{i=1}^{n}$, and assume that all the PDFs are supported on the compact interval $I = [0,1]$, $f_i(x)$ depends on $g_i(x)$ in some manner. We say that the association of a PDF two-tuple is abnormal if it significantly violates the dependence pattern, followed by the majority of the data. It is worth noting that a PDF pair with an abnormal association can behave as either abnormal or normal in their respective datasets. For instance, if we exchange the positions of $f_k(x)$ and $f_j(x)$ to mismatch them to $g_k(x)$ and $g_j(x)$, then both $\{g_k(x), f_j(x)\}$ and $\{g_j(x), f_k(x)\}$ may become abnormally associated; however, if viewed separately, both $f_k(x)$ and $f_j(x)$ still belong to the normal data in $\{f_i(x)\}_{i=1}^{n}$, unless they are originally outliers. Therefore, the functional outlier detection methods described above have limitations in detecting such abnormal associations because they are designed for a single dataset.

If an appropriate distribution-to-distribution regression (DtDR) model has been used to fit the dependence of the PDF-valued two-tuples, and it is assumed that the DtDR model approximates the dependence pattern of the majority of the data, then the outlying two-tuple might stand out as extraordinary residuals. With a slight abuse of terminology, such outliers are called regression outliers throughout this study, and a DtDR-based approach will be provided for detecting them. Regression-based outlier diagnostics for scalar data started very early, see e.g. the monograph of Rousseeuw and Leroy (1987); however, related research contributions for functional outliers still appear to be quite rare.

In this study, the LQD-RKHS DtDR model proposed by Chen et al. (2019a) is employed for this task. The main reasons for this choice are twofold: (1) the LQD-RKHS DtDR is a nonlinear and nonparametric model, which is more flexible in modeling different dependence patterns of the distributional data; thus, the risk of model misspecification in outlier detection is much lower; (2) it is highly efficient because the regression operator has an explicit expression. At present, only the standard version of the LQD-RKHS DtDR model is available for regression outlier detection. Its robust version will be provided later, but it is not suitable for outlier diagnosis as its robustness relies on downweighting the outliers detected in advance.

The regression operator in the standard LQD-RKHS DtDR is estimated by solving a regularized least squares (LS) problem (Chen et al., 2019a). As noted in Rousseeuw and Leroy (1987), a regression model estimated by LS may mask some outliers as the regression model itself is not robust. To reduce this risk, we can perform a stepwise fitting and deletion procedure for regression outlier detection. Specifically, we first fit the distributional regression model to the raw data and calculate the residuals for all the samples, based on which we perform a residual diagnosis and remove the detected outliers from the training dataset; then, we use the remaining training data to fit the regression model and re-perform residual diagnosis for the PDFs. After several rounds of such stepwise deletion, the regression model at the last step is much more likely to be fitted by using "good" data. Next, we use it as the baseline model to perform the residual diagnosis for all the samples and output the outliers detected by such a final regression model as the regression outliers. Moreover, before performing the regression outlier detection, one can also conduct a first-round outlier filtration processing to the raw PDF-valued datasets by choosing the appropriate single-dataset detection methods discussed above, to filter out the extremely anomalous functional samples that will potentially cause the fitted regression model to misbehave grossly.

The regularization parameter (also called the smoothing parameter) in the LQD-RKHS DtDR is determined by the generalized cross-validation ((GCV), see Wahba (1990) and Lian (2007a) for details) to avoid overfitting, and the estimated regularization parameter is denoted as $\lambda_{GCV}$, see Subsection A.3.4.1 of the online supplement for more details. In contrast to the time-consuming leave-one-out cross-validation (Chen et al., 2019a), GCV is much more efficient. Note that overfitting can significantly reduce the power of residual-diagnosis-based outlier detection. Hence, we also recommend a reinforced strategy to avoid overfitting by setting a threshold $\theta_\lambda$ to prevent the regularization parameter from taking very small values. The final regularization parameter is selected as $\max\{\lambda_{GCV}, \theta_\lambda\}$.

Finally, the algorithm for regression outlier detection is summarized in Algorithm 1, presented in Appendix A, along with additional computational details.

## 3. Application to robustify distributional regression

We apply the proposed distributional outlier detection methods to robustify the LQD-RKHS distribution-to-distribution regression model proposed by Chen et al. (2019a).

Consider the following structured functional dataset consisting of univariate continuous PDFs finitely supported on the compact interval [0, 1]:

$$\begin{Bmatrix} f_1(x) \\ g_1(x) \end{Bmatrix}, \begin{Bmatrix} f_2(x) \\ g_2(x) \end{Bmatrix}, \cdots, \begin{Bmatrix} f_n(x) \\ g_n(x) \end{Bmatrix} \tag{7}$$

where $f_i(x)$ and $g_i(x)$ are PDFs, and support $f_i s$ are dependent on $g_i s$ in some manner. The general form of the distribution-to-distribution regression model can be formulated as:

$$f_i(x) = \mathbf{\Psi}(g_i(x)) + e_i \tag{8}$$

where $\mathbf{\Psi}$ is the regression operator, and $e_i$ is the random error term assumed to be independent of $g_i$.

### 3.1. Functional transformation and dimension reduction

In the LQD-RKHS distributional regression framework (Chen et al. 2019a), the PDFs are transformed into the Hilbert space $L^2[0,1]$ to release their inherent constraints by using the log-quantile-density (LQD) transformation (Petersen and Müller, 2016), i.e.,

$$\begin{aligned} \psi_i^{f^*}(t) &= -\log\left\{f_i^*\left(Q_{f,i}^*(t)\right)\right\} \ with\, f_i^* = (1-\alpha)f_i + \alpha \ \ i = 1,\cdots,n \\ \psi_i^{g^*}(t) &= -\log\left\{g_i^*\left(Q_{g,i}^*(t)\right)\right\} \ with \ \ g_i^* = (1-\alpha)f_i + \alpha, \ \ i = 1,\cdots,n \end{aligned} \tag{9}$$

where $Q_{f,i}^*$ and $Q_{g,i}^*$ are the quantile functions corresponding to the PDFs $f_i^*$ and $g_i^*$, respectively, the parameter $\alpha \in [0.2, 0.5]$ is the ratio of the added uniform density supported on $[0, 1]$ (such processing is aimed at curing the “blindness” of the LQD transformation to horizontal translations of PDFs (see Subsection A.2.1 of the online supplement for more details), and the added effects of the uniform density can be easily cleared by using Eq. (7) in Chen et al. (2019a). Additionally, functional principal component analysis (FPCA) is conducted on the Hilbertian data $\psi_i^{f^*}(t)$ for dimension reduction (Chen et al., 2019a), to avoid the curse of dimensionality in the subsequent nonparametric functional regression. Consequently, the Hilbertian data $\psi_i^{f^*}(t)$ can be approximated by the truncated Karhunen–Loève expansion $\psi_i^{f^*}(t) \approx \mu_{\psi_i^{f^*}}(t) + \sum_{k=1}^{m} \xi_{i.k}^{f^*} \varphi_k(t)$, where $\mu_{\psi_i^{f^*}}(t)$ is the mean function of the Hilbertian data $\left\{\psi_i^{f^*}(t)\right\}_{i=1}^{n}$, $\{\varphi_k(t)\}_{k=1}^{m}$ is the first $m$ orders of the dominant FPCs, and $\left\{\xi_{i.k}^{f^*}\right\}_{k=1}^{m}$ is the corresponding FPCA scores. Thus, the infinite-dimensional Hilbertian data $\psi_i^{f^*}(t)$ can be reduced to a finite-dimensional vector-valued data formed by the FPCA scores denoted by $\boldsymbol{\xi}_i^{f^*} = \left[\xi_{i,1}^{f^*}, \xi_{i,2}^{f^*}, \cdots, \xi_{i,m}^{f^*}\right] \in \mathbb{R}^m$. Consequently, the original distribution-to-distribution regression is converted to function-to-vector regression, which is formulated as follows:

$$\boldsymbol{\xi}^{f^*} = F_{reg}\left(\psi^{g^*}(t)\right) + \boldsymbol{\varepsilon}, \qquad \psi^{g^*} \in L^2[0,1] \ and \ \boldsymbol{\xi}^{f^*}, \boldsymbol{\varepsilon} \in \mathbb{R}^m \tag{10}$$

where $F_{reg}$ is the regression operator and $\boldsymbol{\varepsilon}$ is the random error term. Both $\boldsymbol{\xi}^{f^*}$ and $\boldsymbol{\varepsilon}$ are row vectors of dimension $m$.

*3.2. Robust estimator for the regression operator*

The above processing is essentially the same as the standard LQD-RKHS distributional regression method. Next, we derive the robust estimator for the regression operator using the reproducing kernel Hilbert space (RKHS) theory. A detailed introduction to the reproducing kernels and the corresponding RKHS is beyond the scope of this article; readers can refer to Wahba (1990), Berlinet and Thomas-Agnan (2011), Wu and Lin (2012), and Kadri et al. (2016). The related notations and basic properties of the RKHS that are used in this study are provided in Subsection A.1.2 of the online supplement. The robust LQD-RKHS distributional regression procedure is two-stage: 1) detect outlying PDFs and 2) downweighting the detected outliers in the regression operator estimation, similar to the strategy of Martínez-Hernández et al. (2019).

In the RKHS-based nonlinear functional regression framework (Wahba, 1990; Lian, 2007a), the unknown regression operator $F_{reg}$ is assumed to live in an RKHS denoted as $\mathcal{H}(K_r)$ associated with the operator-valued reproducing kernel $K_r$. To cope with the adverse influence of outlying training samples, the regression operator is estimated by solving the following weighted regularized least squares problem:

$$\min_{F_{reg} \in \mathcal{H}(K_r)} \left\{ \sum_{i=1}^{n} w_i \left\| \boldsymbol{\xi}_i^{f^*} - F_{reg}\left(\psi_i^{g^*}\right) \right\|_2^2 + \lambda_s \left\| F_{reg} \right\|_{\mathcal{H}(K_r)}^2 \right\} = \min_{F_{reg} \in \mathcal{H}(K_r)} J\left(F_{reg}\right) \tag{11}$$

where $\lambda_s$ is the regularization parameter, $\|\cdot\|_2$ and $\|\cdot\|_{\mathcal{H}(K_r)}$ are the norms induced by the inner products of the Hilbert spaces $\mathbb{R}^m$ and $\mathcal{H}(K_r)$, respectively, $w_i$ is the weight corresponding to the $i$th PDF pair $\{g_i, f_i\}$. The value of $w_i$ is 1 if neither $g_i$ nor $f_i$ are outliers; otherwise, it would be a nonnegative real number less than 1. Detailed discussions about how to design the weights based on the detected outliers are provided in Subsection A.4.1 of the online supplement.

The representer theorem (Schölkopf et al., 2001; Lian 2007b) still holds for the above weighted RKHS-based regularized least squares problem; according to which, the optimal solution of the regression operator $F_{reg}$ takes the general form given in the following proposition (proof in Subsection A.4.2 of the online supplementary):

**Proposition 1**. *The solution of the regression operator in Eq. (11) obeys the following general structure:*

$$F_{reg}(\cdot) = \sum_{j=1}^{n} K_r\left(\cdot, \psi_j^{g^*}\right) \boldsymbol{\alpha}_j \quad \Leftrightarrow \quad F_{reg}\left(\psi_i^{g^*}\right) = \sum_{j=1}^{n} K_r\left(\psi_i^{g^*}, \psi_j^{g^*}\right) \boldsymbol{\alpha}_j \tag{12}$$

*where $\boldsymbol{\alpha}_j$ is the undetermined vector-valued coefficient assumed to reside in a Hilbert space, $K_r\left(\psi_i^{g^*}, \psi_j^{g^*}\right)$ is the operator-valued reproducing kernel associated with $\mathcal{H}(K_r)$.*

A commonly used operator kernel is the Gaussian operator kernel defined as follows (Lian, 2007a):

$$K_r\left(\psi_i^{g^*}, \psi_j^{g^*}\right) = \exp\left\{ -\frac{1}{2\sigma^2} \int \left| \psi_i^{g^*}(\tau) - \psi_j^{g^*}(\tau) \right|^2 d\tau \right\} I_{id} = a_{ij} I_{id} \tag{13}$$

where $I_{id}$ denotes the identity operator.

The vector-valued coefficient $\boldsymbol{\alpha}_j$ given in Eq.(12) is also assumed to reside in an RKHS $H(k_r)$ with real reproducing kernel $k_r$. Thus, according to the representer theorem of the real RKHS theory (Schölkopf et al., 2001), $\boldsymbol{\alpha}_j$ has the following representation:

$$\boldsymbol{\alpha}_j(\cdot) = \sum_{l=1}^{m} b_{jl} k_r(\cdot, l) \quad \Longrightarrow \quad \boldsymbol{\alpha}_j(k) = \sum_{l=1}^{m} b_{jl} k_r(k, l), \ \forall k \in \{1,2,\cdots,m\} \tag{14}$$

where $\boldsymbol{\alpha}_j(k)$ denotes the $k$th element of the $m$-dimensional vector $\boldsymbol{\alpha}_j$. Let

$$\mathbf{K} = \{k_r(k,l)\}_{k,l=1}^{k,l=m} \text{ and } \boldsymbol{\beta}_j = [b_{j1}, \cdots, b_{jm}], \ j = 1, \cdots, m \tag{15}$$

where $\mathbf{K}$ is the Gram matrix with elements $\mathbf{K}(k,l) = k_r(k,l)$. According to the RKHS theory (Berlinet and Thomas-Agnan, 2011), $\mathbf{K}$ is a symmetric and positive semidefinite matrix. Hence, Eq.(14) can be expressed in the following matrix form:

$$\boldsymbol{\alpha}_j = \boldsymbol{\beta}_j \boldsymbol{K}^{\mathrm{T}} = \boldsymbol{\beta}_j \boldsymbol{K}, \qquad j = 1, \cdots, m \tag{16}$$

In such settings, the functional-to-vector regression model is a double-kernel model, one operator kernel $K_r$ and one real kernel $k_r$. Note that the vector-valued coefficient $\boldsymbol{\alpha}_j$ lives in the Euclidean space $\mathbb{R}^m$, which is much simpler than the operator space where the regression operator is assumed to live.

The Euclidean space $\mathbb{R}^m$ can be extended to an RKHS by endowing it with a suitable reproducing kernel. It is worth noting that the reproducing kernel depends not only on the vector space $\mathbb{R}^m$ but also on the equipped inner product (Wu and Lin, 2012). The commonly used inner product for $\mathbb{R}^m$ is in the following form:

$$\langle \mathbf{u}, \mathbf{v} \rangle = \sum_{i=1}^{m} u_i v_i, \ \forall \mathbf{u}, \mathbf{v} \in \mathbb{R}^m \tag{17}$$

Then, the corresponding reproducing kernel $k_r$ is

$$k_r(k,l) = \begin{cases} 0, & l \neq k \\ 1, & l = k \end{cases}, \ l, k \in \{1,2,\cdots,m\} \tag{18}$$

Hence, the Gram matrix given in Eq.(15) is the identity matrix $\mathbf{E}$ of size $m \times m$.

**Remark 3.** The inner product for $\mathbb{R}^m$ can also be defined in other forms, and one less popular alternate definition can be found in Wu and Lin (2012) as follows:

$$\langle \mathbf{u}, \mathbf{v} \rangle = u_1 v_1 + \sum_{i=2}^{m} (u_i - u_{i-1})(v_i - v_{i-1}), \ \forall \mathbf{u}, \mathbf{v} \in \mathbb{R}^m \tag{19}$$

In such a case, the Gram matrix corresponding to the reproducing kernel associated with $\mathbb{R}^m$ will no longer be an identity matrix.

In this study, the one in Eq.(17) is selected as the default inner product for $\mathbb{R}^m$ unless otherwise stated; thus, $k_r$ takes the default form of Eq.(18) (i.e., Gram matrix $\mathbf{K}=\mathbf{E}$). However, to ensure that our method can accommodate more general cases, the desired robust estimator for the regression operator is derived for a general form of the Gram matrix $\mathbf{K}$ (i.e., $\mathbf{K}$ is symmetric and positive semidefinite but not necessarily the identity matrix).

For the sake of convenience, the following matrixes are defined:

$$\mathbf{A}=\{a_{ij}\}_{\boldsymbol{i=1}}^{\boldsymbol{n}}{}_{\boldsymbol{j=1}}^{\boldsymbol{n}} \text{with } a_{ij} \text{ given in Eq.(13)} \tag{20a}$$

$$\mathbf{B}=\left[\boldsymbol{\beta}_1^{\mathrm{T}},\cdots,\boldsymbol{\beta}_n^{\mathrm{T}}\right]^{\mathrm{T}} \text{ with } \boldsymbol{\beta}_j^{\mathrm{T}} \text{ given in Eq.(15)} \tag{20b}$$

$$\mathbf{W}=\mathrm{diag}\left(\sqrt{w_1},\cdots,\sqrt{w_n}\right) \tag{20c}$$

$$\mathbf{Y}=\left(\left(\boldsymbol{\xi}_1^{f^*}\right)^{\mathrm{T}},\cdots,\left(\boldsymbol{\xi}_n^{f^*}\right)^{\mathrm{T}}\right)^{\mathrm{T}} \tag{20d}$$

The following proposition presents the condition for the coefficient matrix $\mathbf{B}$ when the objection function in Eq. (11) reaches its minimum value.

**Proposition 2**. *Under conditions (11), (12), (13) and (14), the optimal solution of the coefficient matrix* $\mathbf{B}$ *obeys*

$$\left(\mathbf{C}_1\mathbf{C}_2+\lambda_s(\mathbf{K}\otimes\mathbf{A})\right)\mathrm{vec}(\mathbf{B})=\mathbf{C}_1\mathrm{vec}(\mathbf{WY}) \tag{21}$$

*where* $\mathbf{C}_1=\mathbf{K}\otimes(\mathbf{AW})$ *and* $\mathbf{C}_2=\mathbf{K}\otimes(\mathbf{WA})$.

The proof of this proposition is given in Appendix B.

According to Proposition 2, the coefficient matrix $\mathbf{B}$ corresponding to the regression operator can be estimated as

$$\mathrm{vec}(\widehat{\mathbf{B}})=\left(\left(\mathbf{K}\otimes(\mathbf{AW})\right)\left(\mathbf{K}\otimes(\mathbf{WA})\right)+\lambda_s(\mathbf{K}\otimes\mathbf{A})\right)^{-}\left(\mathbf{K}\otimes(\mathbf{AW})\right)\mathrm{vec}(\mathbf{WY}) \tag{22}$$

where $\mathbf{D}^-$ denotes the Moore-Penrose pseudoinverse of matrix $\mathbf{D}$.

**Remark 4.** When both $\boldsymbol{K}$ and $\boldsymbol{W}$ are identity matrixes, the estimation $\mathrm{vec}(\widehat{\mathbf{B}})$ reduces to its counterpart in the standard LQD-RKHS distributional regression method in Chen et al. (2019a).

**Remark 5.** Given the estimated coefficient matrix $\widehat{\mathbf{B}}=\left[\widehat{\boldsymbol{\beta}}_1^{\mathrm{T}},\cdots,\widehat{\boldsymbol{\beta}}_n^{\mathrm{T}}\right]^{\mathrm{T}}$, the regression result of the new predictor $\psi_0^{g^*}$ is $F_{reg}\left(\psi_0^{g^*}\right) = \sum_{j=1}^{n} K_r\left(\psi_0^{g^*}, \psi_j^{g^*}\right)\widehat{\boldsymbol{\beta}}_j\boldsymbol{K}$, then the corresponding PDF can be obtained by implementing the processing procedure discussed in Subsection 2.7 in Chen et al. (2019a).

## 4. Simulation and real data studies

### *4.1. Distributional outlier detection on simulated data*

This simulation study is conducted to validate the effectiveness of the proposed Tree-Distance outlier detection method, as well as to assess its performance by comparing it to its competitors. We employ a mixture distribution model, formed by taking a convex combination of beta distribution and truncated generalized Pareto distribution (tGPD) to generate the PDF-valued data. Specifically, we consider the following two scenarios:

$$\begin{aligned} &\text{Scenario I:} \quad f_i(x) = (1-\eta)f_{Beta}(x; a_i, b_i) + \eta f_{tGPD}(x; 2.0, 4.0), i = 1,2,\cdots,n \\ &\text{Scenario II:} \quad f_i(x) = (1-\eta)f_{Beta}(x; a_i, b_i) + \eta f_{tGPD}(x; 0.5, 0.5), i = 1,2,\cdots,n \end{aligned} \tag{23}$$

where $\eta \in [0, 1]$ is the combination coefficient, $f_{Beta}$ is the density of the beta distribution, $f_{tGPD}$ is the density of the tGPD obtained as follows:

$$f_{tGPD}(x; \kappa, \sigma) = \frac{f_{GPD}(x; \kappa, \sigma)}{\int_0^1 f_{GPD}(\tau; \kappa, \sigma)d\tau}, \; x \in [0,1] \tag{24}$$

where $f_{GPD}(x; \kappa, \sigma)$ is the density of generalized Pareto distribution (GPD) defined as

$$f_{GPD}(x; \kappa, \sigma) = \left(\frac{1}{\sigma}\right)\left(1 + \kappa\frac{x}{\sigma}\right)^{-1-\frac{1}{\kappa}} \tag{25}$$

The parameters $a_i$ and $b_i$ associated with the beta distribution in both scenarios are i.i.d. realizations of uniform distributions, i.e., $a_i \sim \mathrm{U}[10,35]$ and $b_i \sim \mathrm{U}[14,20]$. In each scenario, we consider four different values for the combination parameter $\eta$, i.e., $\eta = 0, 0.15, 0.30$ and $0.45$, to yield four different models referred to as Models I, II, III, and IV (listed in Table 1) throughout this simulation study.

In each scenario, we independently use the four models to generate four different PDF-valued datasets, with each dataset consisting of $n = 100$ curves. We then employ Algorithm A.4 in the online supplement to introduce 10 outlying PDFs into each simulated dataset (i.e., the contamination ratio is 10%). The parameter $\zeta_{hs}$ in Algorithm A.4 is set to 0, meaning that only the shape outliers are generated while the horizontal-shift outliers are not considered because the latter is much easier to detect. The other parameter $\varpi$ in Algorithm A.4 is set to 0.2. Representative simulated data based on the eight models (listed in Table 1) are visualized in Fig. 5.

**Table 1**
Considered models for distributional data generation.

| | Model I | Model II | Model III | Model IV |
|---|---|---|---|---|
| Scenario I | $\eta = 0$ | $\eta = 0.15$ | $\eta = 0.30$ | $\eta = 0.45$ |
| Scenario II | $\eta = 0$ | $\eta = 0.15$ | $\eta = 0.30$ | $\eta = 0.45$ |

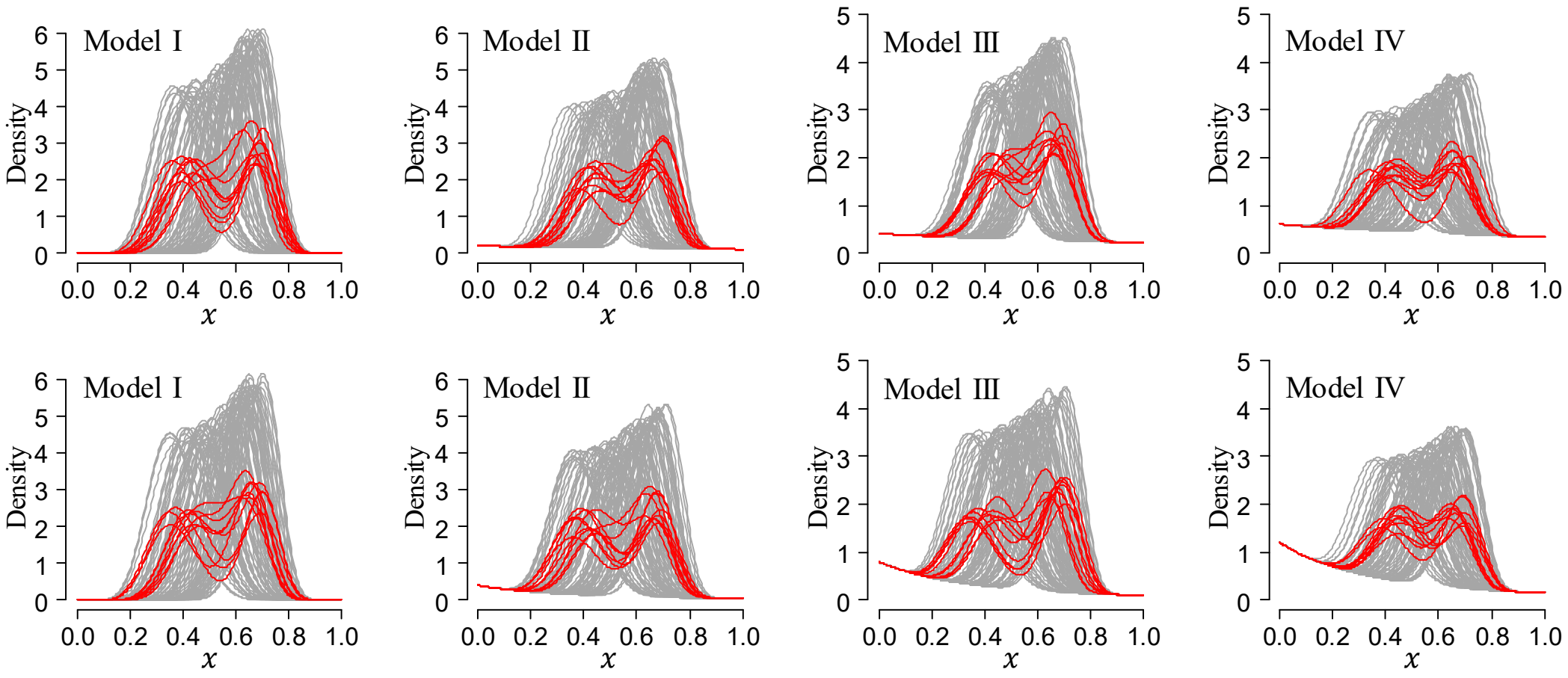

**Fig. 5.** Representative simulated data associated with the eight models, the gray lines and red lines stand for the "good" data and outlying data, respectively. The first row corresponds to Scenario I, while the second row corresponds to Scenario II.

We consider three different detection schemes and compare their performances.

(i) Tree-Distance method

In this detection scheme, we perform outlier detections on the transformed data associated with the nLQD, CLR, DIFF, and MED nodes using the default argument settings listed in Table A-2 in the online supplement. The outliers detected from the four nodes are merged to form the final detection results. Note that before transforming the PDFs to their clr-representations (corresponding to the CLR node), they should be aligned horizontally in the H-CENTR node, and the feature point for such an alignment is calculated by the average of the median and mode.

(ii) QF-FDO method

In the second detection scheme, we consider using the tool of functional directional outlyingness (FDO) defined by Dai and Genton (2019) to perform outlier detection in the QF-space for the simulated distributional data. The computational details are provided in Subsection A.3.3.1 of the online supplement. The detection region is set as [0.2, 0.8]. The outliers in the MO- and VO-directions are detected by the two- and one-sided boxplot-based detectors (see Eq. (A-11) and Eq. (A-12) in the online supplement), respectively. We consider three different whisker parameters, valued at 1.5, 2.0, and 2.5, for VO-outlier detection, while the whisker parameter for MO-outlier detection is fixed at 1.5. Such settings yield three different detection cases, and they are performed and evaluated independently, and the results are not merged.

(iii) Warping-function-based detection using the phase distance

The warping-function-based detection strategy considered here essentially belongs to the elastic-depth-based approach proposed by Harris et al. (2020). The outliers are detected based on the phase information of the CDFs captured by the warping functions. Given two CDFs $F_i(x)$ and $F_j(x)$, the warping function for extracting the phase difference of $F_i(x)$ with respect to $F_j(x)$ can be directly calculated as $\gamma_{ij}(x) = \left(F_i^{-1} \circ F_j\right)(x)$ rather than using the time-consuming DP algorithm (Srivastava et al., 2011; Srivastava and Klassen, 2016). Section A.5 of the online supplement provides additional computational details regarding the same. Next, the phase distance (defined by Harris et al. (2020)) for the two CDFs can be calculated as $d_p\left(F_i, F_j\right) = \arccos\langle q_e, q_{ij}\rangle$ with $q_e(x) = 1$ and $q_{ij}(x) = \sqrt{d\gamma_{ij}(x)/dx}$ being the square-root slope functions (SRSFs) of the warping functions $\gamma_e(x) = x$ (identity warping function) and $\gamma_{ij}(x)$, respectively. With such phase distance at hand, we can calculate the phase depth defined by Harris et al. (2020). Then, the shape outliers hidden in the CDFs can be identified as phase anomalies by using Algorithm 1 described by Harris et al. (2020), only replacing the amplitude depth by the phase depth. The parameter $k$ (in Algorithm 1 of Harris et al. (2020)) for controlling the whisker is set to 2.0. As pointed out in Section A.5 of the online supplement, if the PDFs take near-zero values, the corresponding warping functions obtained using the direct computation method may be problematic. This defect can be remedied by adding an appropriate ratio of uniform distribution, i.e., $f_i^* = (1 - \alpha_i)f_i + \alpha_i$ where $\alpha_i$ is the mixing proportion. According to our experience, choosing $\alpha_i = 0.1$ can effectively alleviate the numerical issue; therefore, if the minimum value of the PDFs in a dataset is less than 0.1, all the PDFs in the same dataset will be processed using the above principle with $\alpha_i = 0.1$.

The above three different detection schemes are applied to detect the outlying PDFs independently. In the simulated PDF dataset consisting of 100 curves, let $n_{out}$ and $n_{norm}$ denote the number of outlying PDFs and non-outlying PDFs, respectively, and $n_c^{DT}$ and $n_f^{DT}$ denote the number of correctly and falsely detected outlying PDFs, respectively. Similar to Dai and Genton (2019), the performance of the detection methods can be assessed based on the correct and false detection rates, defined as follows:

Correct detection rate: $p_c = \frac{n_c^{DT}}{n_{out}} \times 100\%$

False detection rate: $p_f = \frac{n_f^{DT}}{n_{norm}} \times 100\%$

We repeat the above outlier detection experiment 1000 times, and the functional data in each repeated experiment are re-simulated to calculate the average correct and false detection rates over the 1000 repeated experiments for the above three detection schemes, and the results are reported in Tables 2~4. For comparison, we also calculate the average correct (false) detection rates associated with the outliers detected in the four considered nodes (i.e., MED, nLQD, CLR, and DIFF) on the tree of the first detection scheme, and the corresponding results are listed in columns 3~6 of Table 2. It is worth noting that the data-generating processes of Model IV are the same in Scenarios I and II; therefore, the detection results of Model I are only reported for Scenario I.

For the first detection scheme, it can be seen from Table 2 that the detected outliers are mainly from the nodes of nLQD and CLR, and rarely from the nodes of MED and DIFF. In Scenario I, the CLR node slightly outperforms the nLQD node, whereas, in Scenario II, the opposite is the case. Recall that the outliers detected by the tree are obtained by merging the outliers identified in the four considered nodes, the calculated average correct detection rates (false detection rates) associated with the tree (listed in the last column of Table 2) are the final results of the Tree-Distance detection scheme. It is evident from the results that the average correct detection rates of the tree are higher than those associated with the nLQD and CLR nodes, especially for Model IV of Scenario II, indicating that the outliers detected in the nLQD and CLR nodes are different. Thus, the two transformations have complementarity in outlier detection. Note that the role of the MED node is mainly to detect the horizontal-shift outliers; thus, the outliers detected from this node are mainly false positives because the simulated data only contain shape outliers with no significant horizontal deviations with respect to the bulk of the data. The DIFF node shows almost no contribution to the outlier detection in this experiment; however, we cannot conclude that the DIFF node is useless and should be removed from the transformation tree considering that it may have its merits in other situations. To demonstrate this, we provide another simulated PDF-valued dataset in Subsection A.7.1 of the online supplement (i.e., Additional Simulation Study I), where the DIFF node plays the dominant role in uncovering the outliers. From the results of this additional simulation study listed in Table A-5 in the online supplement, it can be seen that the calculated average correct detection rates associated with MED, nLQD, CLR, DIFF, and TREE are 41.71%, 45.37%, 28.59%, 62.22%, and 99.28%, respectively, which provides strong evidence that the considered transformations have good complementarity in outlier detection.

**Table 2**
The calculated average correct detection rates and false detection rates (in brackets) associated with the Tree-Distance detection scheme. The data in the third, fourth, fifth, and sixth columns correspond to the results detected in the nodes of MED, nLQD, CLR, DIFF, respectively, while the last column corresponds to the merged results detected in the four considered nodes on the tree. Both $p_c$ and $p_f$ (in brackets) are presented in percentage terms.

| Scenario | Model | MED $p_c$(%) ($p_f$)(%) | nLQD $p_c$(%) ($p_f$)(%) | CLR $p_c$(%) ($p_f$)(%) | DIFF $p_c$(%) ($p_f$)(%) | TREE $p_c$(%) ($p_f$)(%) |
|---|---|---|---|---|---|---|
| Scenario I | Model I | 0.00 (0.20) | 98.20 (0.00) | 98.76 (0.00) | 0.00 (0.00) | 99.29 (0.20) |
| | Model II | 0.00 (0.12) | 96.11 (0.00) | 98.17 (0.00) | 0.00 (0.00) | 98.88 (0.12) |
| | Model III | 0.00 (0.06) | 93.07 (0.00) | 97.95 (0.00) | 0.00 (0.00) | 98.61 (0.06) |
| | Model IV | 0.00 (0.05) | 87.53 (0.00) | 97.21 (0.00) | 0.00 (0.00) | 97.86 (0.05) |
| Scenario II | Model I | — | — | — | — | — |
| | Model II | 0.00 (0.11) | 97.48 (0.03) | 94.13 (0.00) | 0.00 (0.00) | 98.57 (0.12) |
| | Model III | 0.00 (0.05) | 97.42 (0.25) | 84.74 (0.00) | 0.00 (0.00) | 97.88 (0.28) |
| | Model IV | 0.00 (0.01) | 83.48 (0.01) | 72.56 (0.00) | 0.00 (0.00) | 88.92 (0.02) |

For the second detection scheme, the calculated average correct detection rates are significantly lower than those of the first detection scheme. Comparing the average false detection rates between Table 2 and Table 3, it appears that the second detection scheme also has a higher risk of false detection, especially for Model IV in Scenario II. The striking false detection rate associated with Model IV in Scenario II is attributable to the unstable detection of the FDO-based method mentioned in Subsection 2.2. In general, the second detection scheme performs relatively poorly in this simulation study. Reducing the whisker parameter of the boxplot associated with the VO-outliers can increase its outlier detection power; however, only a slight improvement is observed in this experiment, which also leads to a higher risk of false detection. In Subsection A.7.2 of the online supplement, we provide an additional comparative study using another data-generating process. The QF-FDO method performs much better with average detection rates higher than 85% in most cases (see Tables A-6~8 in the online supplement); however, it is still inferior to the Tree-Distance detection scheme, and a similar unstable detection phenomenon has also been observed.

**Table 3**
The calculated average correct detection rates and false detection rates (in brackets) associated with the QF-FDO detection scheme. The data in the third, fourth, and fifth columns correspond to the results detected by setting the whisker parameters associated with the VO-outliers to be 1.5, 2.0, and 2.5, respectively, while the whisker parameter associated with the MO-outliers is fixed at 1.5. Both $p_c$ and $p_f$ (in brackets) are presented in percentage terms.

| Scenario | Model | 1.5IQR (VO) $p_c$(%) ($p_f$)(%) | 2.0IQR (VO) $p_c$(%) ($p_f$)(%) | 2.5IQR (VO) $p_c$(%) ($p_f$)(%) |
|---|---|---|---|---|
| Scenario I | Model I | 61.12 (2.85) | 54.39 (1.46) | 48.98 (0.73) |
| | Model II | 65.38 (2.76) | 58.82 (1.34) | 52.65 (0.68) |
| | Model III | 68.04 (2.13) | 61.55 (0.97) | 55.26 (0.44) |
| | Model IV | 71.00 (6.21) | 63.06 (4.56) | 55.67 (3.40) |
| Scenario II | Model I | — | — | — |
| | Model II | 63.99 (2.90) | 57.06 (1.46) | 51.03 (0.74) |
| | Model III | 71.59 (2.41) | 65.10 (1.13) | 58.66 (0.52) |
| | Model IV | 59.25 (16.02) | 55.99 (15.32) | 53.47 (14.78) |

**Table 4**
The calculated average correct detection rates and false detection rates (in brackets) associated with the warping-function-based detection scheme. Both $p_c$ and $p_f$ (in brackets) are presented in percentage terms.

| Scenario | Model I $p_c$(%) ($p_f$)(%) | Model II $p_c$(%) ($p_f$)(%) | Model III $p_c$(%) ($p_f$)(%) | Model IV $p_c$(%) ($p_f$)(%) |
|---|---|---|---|---|
| Scenario I | 43.08 (8.11) | 43.15 (8.35) | 38.09 (8.48) | 32.33 (8.79) |
| Scenario II | — | 42.18 (8.73) | 38.74 (9.03) | 35.78 (9.04) |

The results associated with the third detection scheme, listed in Table 4, show that the average correct detection rates are lower than 50% for the considered cases, meaning that fewer than half (on average) of the outliers have been successfully identified by the warping-function-based method. Moreover, the calculated average false detection rates associated with the third detection scheme are significantly higher than those of the other two detection schemes, except for the unstable detection cases that emerged in the second detection scheme.

Comparing Tables 2~4, we see that the first detection scheme (i.e., the proposed Tree-Distance method) performs excellently with high accuracy and low risk of false detection, while the second detection scheme (i.e., the QF-FDO method) significantly underperforms in comparison, and the third detection scheme (i.e., the warping-function-based approach conducted in the CDF space) is the worst performer. We reiterate that the FDO-based approach can identify the unexposed outlying curves and is potentially complementary to the distance-based method if the unstable detection risk can be effectively controlled. The multiple detection strategy described in Subsection 2.3 can provide an effective solution for this purpose, to which we now turn.

### *4.2. Multiple detection on real data*

This subsection aims to illustrate and validate the multiple detection method described in Subsection 2.3, by applying it to a real dataset. The real dataset used in this study consists of PDFs estimated from the monitoring data collected by a strain sensor installed on the bottom plate of a steel box girder of a long-span bridge. The PDFs have been estimated in Chen et al. (2019a, 2020), where they used such PDF-valued data in distributional regression and distributional dependence analysis. In this study, we select 150 PDFs associated with one sensor (i.e., Sensor A in Chen et al. (2019a)) for investigation. We consider using the multiple detection method to fuse the Tree-Distance and QF-FDO detection schemes to implement a combined detection.

Before conducting the multiple detection, we perform an initial detection on the data by independently applying the Tree-Distance and QF-FDO detection schemes to them, aiming to explore their complementarity by checking whether they can uncover different types of outliers. Detailed descriptions of the initial detection, as well as the related results, are provided in Subsection A.6.2 of the online supplement. As expected, the results show that the two detection schemes have good complementarity in outlying PDF detection.

Next, we perform the multiple detection on the data. The parameter settings considered for the multiple detection are listed in Table 5. The other arguments that are not listed in Table 5, such as the feature point for the horizontal centralization processing conducted in the H-CENTR node (before the CLR node), are the same as their counterparts in the previous subsection. In such settings, the Tree-Distance and QF-FDO schemes comprise a total of $3^3 \cdot (1 \cdot 3) \cdot (1 \cdot 3) \cdot 1 = 243$ and $3 \cdot 3 \cdot 4 \cdot 1 = 36$ parameter combinations, respectively. Therefore, 279 unique independent detections can be conducted on the dataset. Let $n_i^{out}$ stand for the count of the detected outliers in the $i$th detection, the count series $\{n_1^{out}, n_2^{out}, \cdots, n_{279}^{out}\}$ associated with the 279 detections are summarized as the histogram shown in Fig. 6. As previously mentioned, the FDO-based approach may suffer from the risk of instability, that is, when the parameter setting is improper, the false detection rate might be drastically increased. One advantage of the multiple detection strategy is that it has the potential to screen unstable detections, as discussed in Subsection 2.3. It is evident from Fig. 6 that there are cases where the detected outliers count is significantly larger than the majority cases, which is most likely caused by unstable detections. Hence, the related detection results should be filtered out. In practice, the outlying counts shown in Fig. 6 can also be easily detected using the one-sided boxplot-based detector given in Eq. (A-12) of the online supplement. Here, the whisker of the boxplot for this purpose is set to be 2.5IQR.

**Table 5**
Summary of considered parameters in the multiple detection.

| Detection scheme | | Arguments |
|---|---|---|
| Tree-Distance | nLQD node | PDF preprocessing parameter $\alpha$: $10^{-10}$, $10^{-6}$, $10^{-2}$<br>Detection region: [0.001,0.999], [0.1, 0.9], [0.2, 0.8]<br>Whisker ($L_1$, $L_\infty$): (2.0IQR, 3.0IQR), (2.5IQR, 3.5IQR), (3.5IQD, 4.5IQR) |
| | CLR node | PDF preprocessing parameter $\alpha$: 0.1<br>Whisker ($L_2$): 2.0IQR, 2.5IQR, 3.5 IQR |
| | DIFF node | Detection region: [0, 1]<br>Whisker ($L_1$, $L_\infty$): (2.0IQR, 3.0IQR), (2.5IQR, 3.5IQR), (3.5IQD, 4.5IQR) |
| | MED node | Whisker: 1.5IQR |
| QF-FDO | | PDF preprocessing parameter $\alpha$: $10^{-10}$, $10^{-6}$, $10^{-2}$<br>Detection region: [0.001,0.999], [0.1, 0.9], [0.2, 0.8]<br>Whisker (VO direction): 1.5IQR, 2.0IQR, 2.5IQR, 3.0 IQR<br>Whisker (MO direction): 1.5IQR |

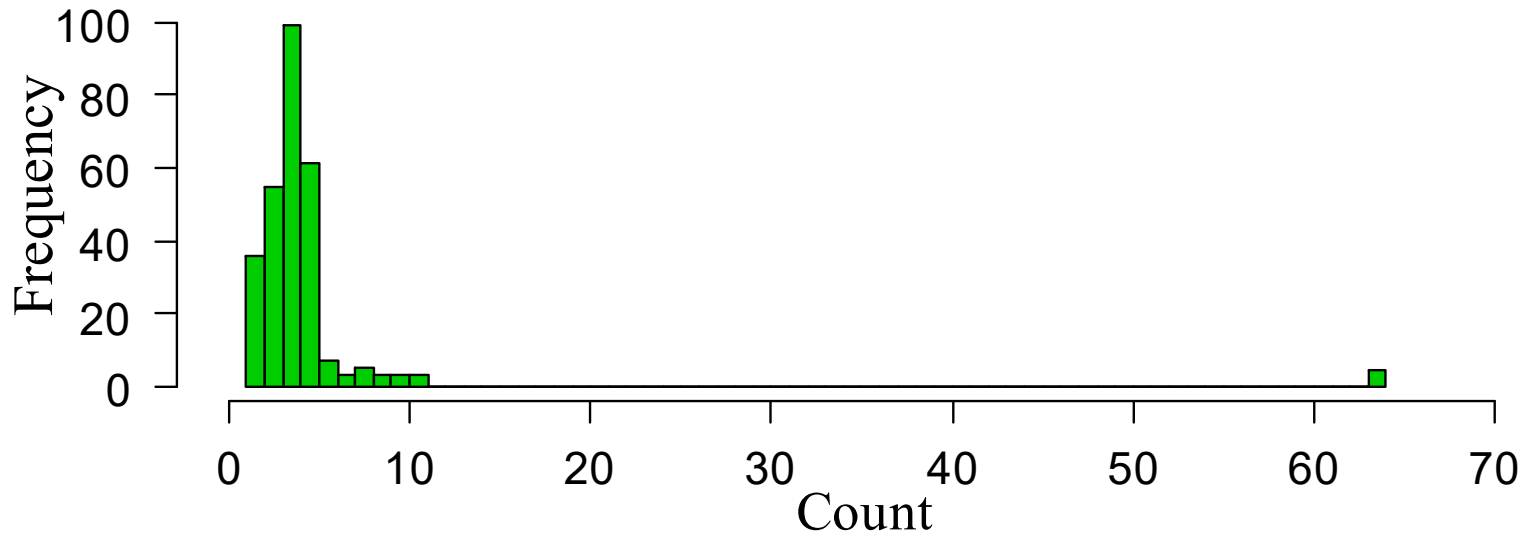

**Fig. 6.** Histogram of the counts $\{n_1^{out}, n_2^{out}, \cdots, n_{****}^{out}\}$ associated with the multiple detection.

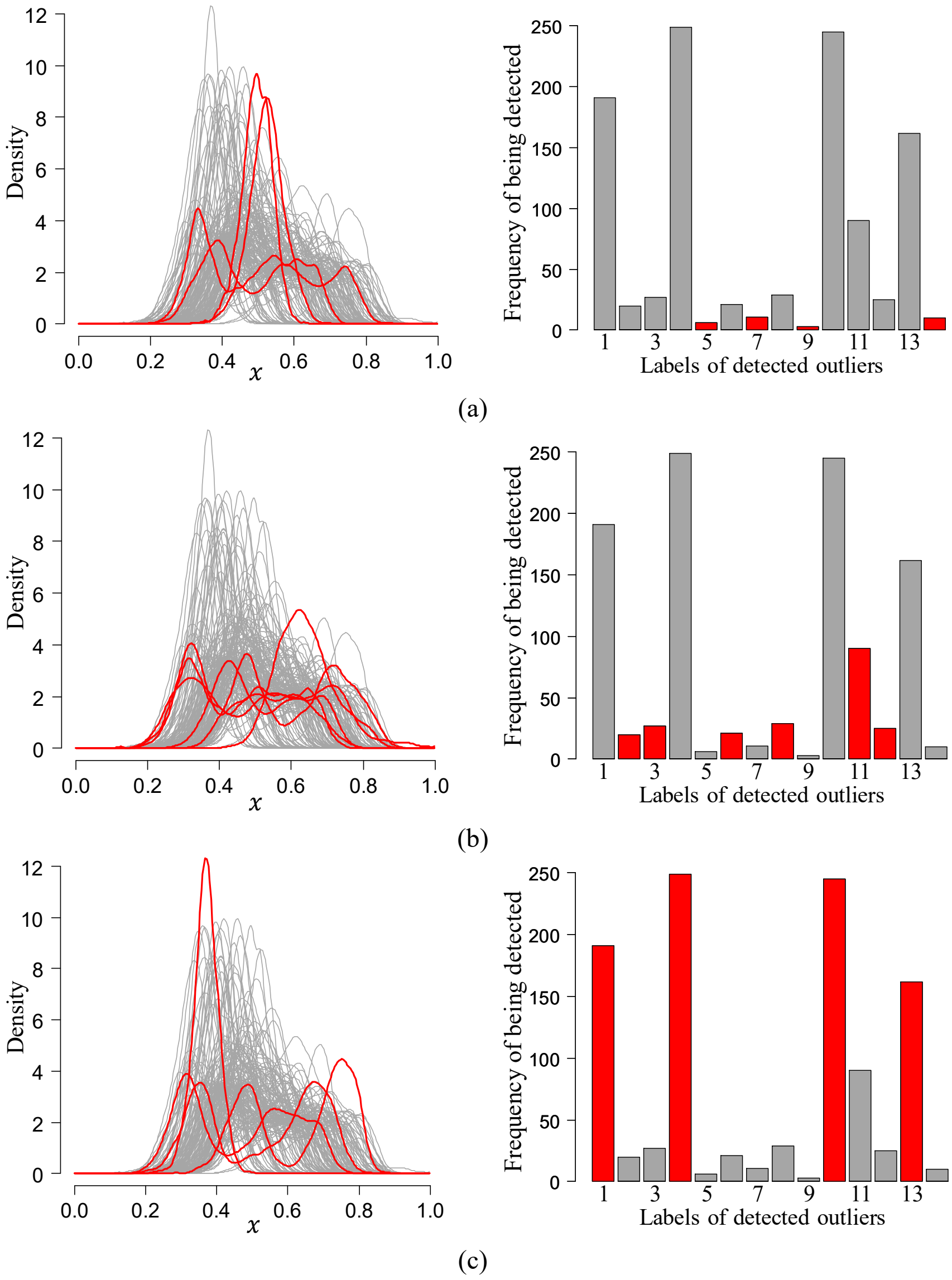

**Fig. 7.** Visualizations for the three categories of detected outliers in the multiple detection according to their frequencies of being detected. (a) Mild abnormal, (b) moderate abnormal, and (c) heavy abnormal. The left column corresponds to the PDFs (the outlying PDFs of a given category are highlighted by red curves), and the right column corresponds to the bar plots of the frequencies being detected (the results of a given category are highlighted by red bars).

After removing the potentially unreliable detection results, we use the remaining results to compute the detection frequencies for the identified outlying PDFs, which results in a vector denoted as $\mathcal{J}_O = [J_1, J_2, \cdots, J_{n_O}]$ with $J_i$ being the frequency (of being detected) corresponding to the $i$th identified outlying PDF and $n_O$ being the number of identified outlying PDFs. A high value of $J_i$ means that the $i$th outlying PDF is frequently being detected, intuitively, this PDF will have a high degree of outlyingness. Based on the frequencies (of being detected) stored in $\mathcal{J}_O$, the detected outlying PDFs are classified into three categories: mild abnormal, moderate abnormal, and heavy abnormal. Graphical visualizations of the three classes of detected outliers are shown in Fig. 7.

The above distributional outlier detection experiment using real data provides evidence to show that the multiple detection method is an effective strategy for coping with parameter uncertainties, unstable detections, and fusing different detectors, which can significantly improve the reliability of distributional outlier detection and provide a measure for quantifying the degree of outlyingness.

*4.3. Regression outlier detection on simulated data*

This subsection presents simulation studies to validate the effectiveness of the distributional-regression-based approach in abnormal association detection (i.e., regression outlier detection).

First, we use Algorithm A.5 in the online supplement to simulate $n$ groups of correlated PDF-valued two-tuples denoted as $\{g_i, f_i\}_{i=1}^{n}$. Each PDF element is generated from a mixture beta distribution model with the mixture coefficient $q_i$ being randomly sampled from the uniform distribution $\mathrm{U}(0,0.5)$. The parameters of the beta distributions corresponding to $f_i, i = 1,2,\cdots,n$ are nonlinearly dependent on those corresponding to $g_i, i = 1,2,\cdots,n$. Representative functional samples of the simulated PDF-valued data $\{g_i\}_{i=1}^{100}$ and $\{f_i\}_{i=1}^{100}$ are visualized in the left column of Fig. 8, and the right column presents five typical curves selected from each PDF dataset. Both PDF datasets contain unimodal and bimodal curves, and such distributional data are relatively complex from the angle of the curve shapes.

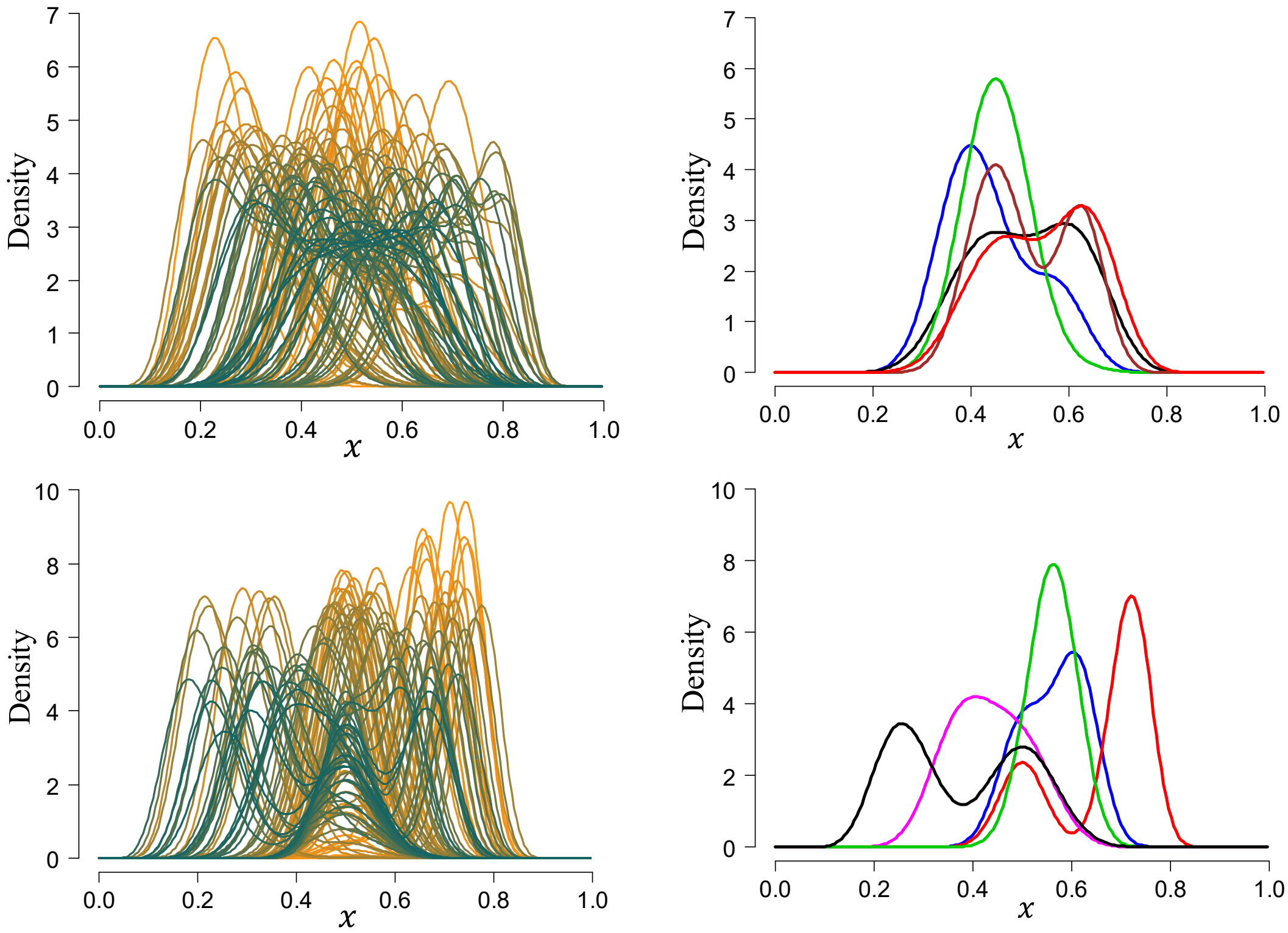

**Fig. 8.** Representative simulated functional samples of the PDF-valued datasets $\{g_i\}_{i=1}^{100}$ (the upper row) and $\{f_i\}_{i=1}^{100}$ (the lower row). The left column corresponds to the whole curves contained in the functional datasets, while the right column corresponds to five selected representative curves from the PDFs shown in the left panel of the same row.

We adopt an intra-set element exchange strategy to generate the abnormal associations of PDFs. For example, while specifically taking the intra-set element exchange in the dataset $\{g_i\}_{i=1}^{n}$ we first construct two PDF-valued subsets using the PDFs from $\{g_i\}_{i=1}^{n}$ based on their peak values defined as $\pi_i^g = \sup_{x\in[0,1]} g_i(x), i = 1, \cdots, n$: (1) the first subset, denoted as $S_1$, consists of PDFs with peak values lower than the 20th percentile of $\Pi^g = \{\pi_1^g \cdots, \pi_n^g\}$; (2) the second subset, denoted as $S_2$, consists of PDFs with peak values higher than the 80th percentile of $\Pi^g$. If we randomly select two PDFs from $S_1$ and $S_2$ to exchange their places, it can produce two abnormal associations in the PDF-valued two-tuples $\{g_i, f_i\}_{i=1}^{n}$. Performing similar element exchanges in the other PDF dataset, i.e., $\{f_i\}_{i=1}^{n}$, can also produce abnormal associations in $\{g_i, f_i\}_{i=1}^{n}$. We use the bivariate parameter $(M_g, M_f)$ to control for the number of element exchanges in $\{g_i\}_{i=1}^{n}$ and $\{f_i\}_{i=1}^{n}$. For instance, $(M_g, M_f) = (2, 3)$ means that two pairs of PDFs in $\{g_i\}_{i=1}^{n}$ will be randomly selected to perform pairwise element exchange using the above principles, and three pairs of PDFs in $\{f_i\}_{i=1}^{n}$ will be processed similarly. Based on this strategy, the implementation details for generating abnormal associations in $\{g_i, f_i\}_{i=1}^n$ are summarized in Algorithm A.7 in the online supplement.

Note that the intra-element exchange can only rearrange the order of the elements. If viewed independently, the curve plots of $\{g_i\}_{i=1}^n$ and $\{f_i\}_{i=1}^n$ after processing by Algorithm A.7 (in the online supplement) are the same as those in the upper and lower left panels of Fig. 8, respectively. To visualize the simulated abnormal associations, we plot the curves of the abnormal PDF-valued two-tuple (denoted as $\{g_j, f_j\}$) along with the histogram of random samples generated from the original distribution (the one before performing the element exchange operation), as shown in Fig. 9 (a), where the abnormal PDF can be distinguished as it no longer fits the histogram well. For comparison, Fig. 9 (b) also illustrates another representative PDF-valued two-tuple (denoted as $\{g_k, f_k\}$) with a normal association.

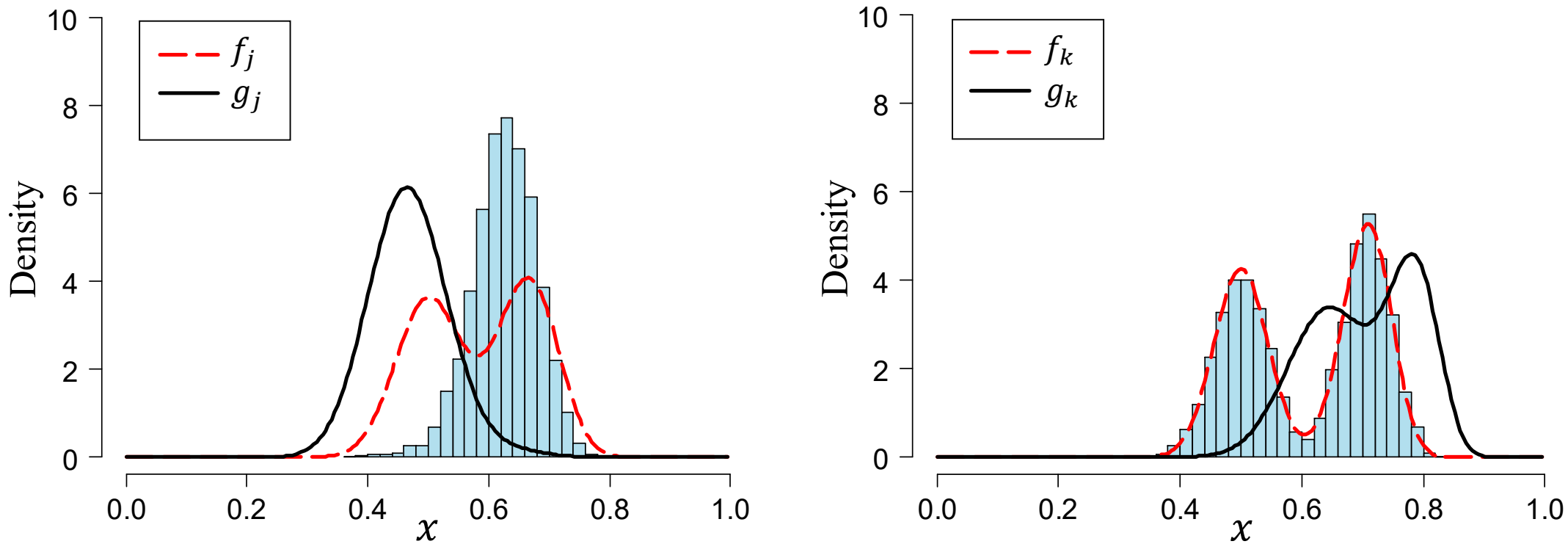

(a) (b)

**Fig. 9.** (a) Visualization for a representative simulated abnormal PDF-valued two-tuple and (b) visualization for a representative simulated "good" PDF-valued two-tuple. The histogram is drawn based on 5000 random samples drawn from the original distribution associated with $f_j$ (or $f_k$) before the PDF exchange.

In the following, the distributional-regression-based approach described in Subsection 2.4 with implementation details summarized in Algorithm 1 of Appendix A, is employed to detect the regression outliers. The associated argument settings are listed in Table 6. Similar to Subsection 4.1, the performance is assessed based on the correct and false detection rates averaged over a series of repeated detection experiments. In each repetition, a total of 100 groups of PDF-valued two-tuples are simulated using Algorithm A.5 (in the online supplement), and 10 abnormal associations are generated by Algorithm A.7 (in the online supplement) to contaminate the original distributional data. We set $(M_g, M_f) = (0, 5)$, (2, 3), (4, 1), and (5, 0), considering four different contamination scenarios. Using the results of 500 repeated detection experiments, Table 7 lists the calculated average correct and false detection rates for the four considered contamination scenarios. The correct detection rates are all greater than 80%, with 92.24% being the best, indicating that the distributional-regression-based approach can effectively detect regression outliers. The false detection rates are approximately 6%, implying that out of the 90 “good” curves, approximately six curves will be falsely identified as outliers, on average. At first glance, the false detection rate is relatively high, which might be attributed to the regression error. Recall that a function-to-function regression model can be generally written as $f(x) = \Gamma(g(x)) + \varepsilon(x)$ with $\Gamma$ and $\varepsilon(x)$ being the regression operator and functional error term (assumed to be zero mean), respectively, and the functional response $f(x)$ is independent of the error term $\varepsilon(x)$. Thus, given a specified predictor $g_k$, only the conditional mean $\mathrm{E}(f_k | g_k) = \Gamma(g_k)$ can be predicted by the fitted regression model, whereas the quantity of the error is unpredictable. If the simulated error $\varepsilon_k$ in generating the experimental data is considerably large, the corresponding PDF-valued two-tuple $\{g_k, f_k\}$ may also be an outlier. Although we have leveraged a horizontal threshold $\theta_h$ in Algorithm 1 to reduce the risk of false detection caused by horizontal shift errors of PDFs (see Subsection A.3.4.2 of the online supplement for details), the remaining shape errors may also lead to false detections.

**Table 6**
Argument settings for the regression outlier detection.

| $(\alpha_{mix}^{\mathrm{LQD}}, \alpha_{mix}^{\mathfrak{B}})$ | $(r_2^{\mathrm{LQD}}, r_2^{\mathfrak{B}})$ | $(\theta_\lambda, \theta_h)$ | $m_{\mathrm{FPCA}}$ | $N_{iters}^{reg}$ |
|---|---|---|---|---|
| (0.3, 0.1) | (1.5, 1.5) | (0.01, 0.15) | 5 | 4 |

**Table 7**
The calculated correct and false detection rates associated with the regression outlier detection for four different contamination scenarios.

| $(M_g, M_f)$ | (0,5) | (2,3) | (4,1) | (5,0) |
|---|---|---|---|---|
| $p_c$ (%) | 92.24 | 81.76 | 81.20 | 84.06 |
| $p_f$ (%) | 5.57 | 7.01 | 6.81 | 5.95 |

In general, with average correct detection rates greater than 80%, the proposed distributional-regression-based outlier detection method performs well, as the simulated data are quite complex (see Fig. 8), and the false detection rates are also acceptable. If we reduce the complexity of the simulated data, the detection method can be expected to perform better. For comparison, we also conduct an additional simulation study with less complicated PDF-valued data generated by Algorithms A.9~10 in the online supplement. The argument settings for the regression outlier detector are the same as those in Table 6. As expected, the detection performance is much better, and the correct detection rates (averaged over 500 replicated detection experiments) are all greater than 98% (see Table A-9 in the online supplement).

### *4.4. Robust distributional regression on real data*

We conduct a real data study to evaluate the proposed robust LQD-RKHS distributional regression method. The investigated real data are strain measurements collected by two strain gauges installed on the bottom plate of the steel box of a long-span bridge (Chen et al., 2019b), see Subsection A.6.4 of the online supplement for a more detailed description. For convenience, these two sensors are called Sensor A and Sensor B throughout the rest of this study, and the regression model is built for regressing PDFs from Sensor A to Sensor B. Using the data processing procedures described in Subsection A.6.4 of the online supplement, we obtain 120 pairs of PDFs for regression analysis. The estimated PDFs are denoted as $\{\hat{g}_i\}_{i=1}^{120}$ and $\{\hat{f}_i\}_{i=1}^{120}$ for Sensor A and Sensor B and are visualized in Fig. 10 (a) and (b), respectively. One can see from the figure that the resulting PDF-valued datasets contain various outlying curves, especially those associated with Sensor B.

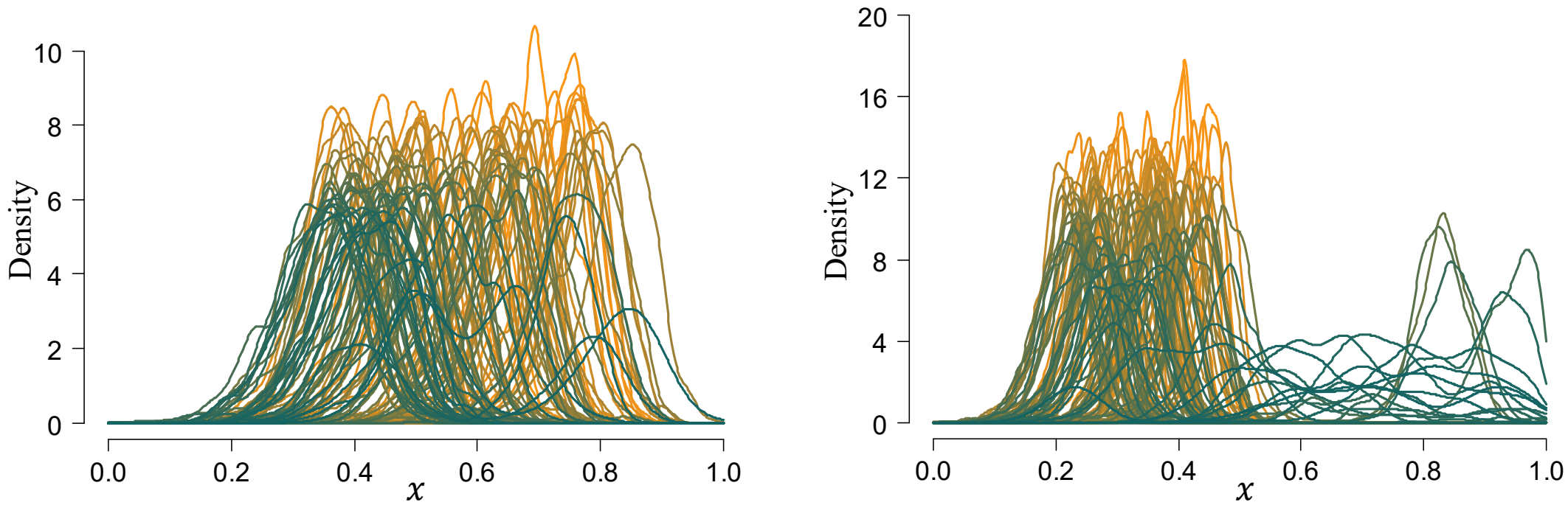

**Fig. 10.** The estimated PDFs of post-processed strain measurements: (a) $\{\hat{g}_i\}_{i=1}^{120}$ associated with Sensor A and (b) $\{\hat{f}_i\}_{i=1}^{120}$ associated with Sensor B.

Before conducting the distributional regression, we perform a two-stage initial outlier detection on the datasets $\{\hat{g}_i\}_{i=1}^{120}$ and $\{\hat{f}_i\}_{i=1}^{120}$:

(i) Single dataset outlier detection: outlier detections for the datasets $\{\hat{g}_i\}_{i=1}^{120}$ and $\{\hat{f}_i\}_{i=1}^{120}$ are conducted independently by using the Tree-Distance method, and the relevant argument settings are the same as their counterparts in Subsection 4.1.

(ii) Regression outlier detection: the outlier detection for the datasets $\{\hat{g}_i\}_{i=1}^{120}$ and $\{\hat{f}_i\}_{i=1}^{120}$ is conducted jointly by using the distributional-regression-based approach described in Subsection 2.4 after the outliers detected in the first stage have been removed. The relevant argument settings are the same as those listed in Table 6 except that the whisker parameters $\left(r_2^{\mathrm{LQD}}, r_2^{\mathfrak{B}}\right)$ are set to $(3.0,\ 3.0)$.

The outliers detected in the first and second stages are called Type I and Type II outliers, and the relevant detection results are presented in Figs. A-21 and A-22 in the online supplement. One can see that the investigated PDFs contain all types of outliers described in this study, namely, shape outliers, horizontal-shift outliers, and regression outliers.

According to these detection results, 23 pairs of PDFs with either one or two PDFs detected as Type I or Type II outliers are classified into the abnormal dataset, and the remaining 97 pairs of PDFs are regarded as “good” data. For method validation and performance evaluation, 35 pairs of PDFs are randomly selected from the “good” data to serve as the test functional samples, while the remaining 85 pairs of PDFs (including the abnormal functional samples) are used as the training functional samples. Both the standard and robust LQD-RKHS methods are employed to construct the distributional regression models from sensor A to sensor B and subsequently employed to predict the 35 test PDFs associated with Sensor B.

**Table 8**
Common argument settings for the standard and robust LQD-RKHS distributional regression models

| Argument | Setting |
|---|---|
| operator kernel $K_r$ in Eq. (12) | the Gaussian kernel given in Eq. (13) |
| $\sigma$ in Eq. (13) | determined according to Eq.(26) |
| $\alpha$ in Eq. (9) | $\alpha = 0.3$ |
| truncation order of FPCA | $m = 5$ |

The common argument settings for the standard and robust LQD-RKHS distributional regression models are presented in Table 8. For the robust version, the weights $w_i s$ in Eq.(11) (to dampen the impacts of the detected outlying PDFs) are determined according to the principles described in Subsection A.4.1 in the online supplement, and the relevant parameters $\rho_1$ and $\rho_2$ in the weight functions given in Eqs. (A-18) and (A-21) are set to 1. The argument settings for the Type I and Type II outlier detections are the same as the aforementioned initial detection. Recall that the regularization parameter $\lambda_s$ has a significant impact on the RKHS-based regression. To avoid the strength of the standard version being suppressed by improper parameter setting, we consider nine different regularization parameters, valued at 0.001, 0.01, 0.05, 0.1, 0.5, 1.0, 2.0, 4.0, and 8.0, for the standard LQD-RKHS method. The regularization parameter for the robust version is fixed at 0.1. Consequently, 10 different regression models can be constructed, where the first nine models are standard LQD-RKHS models with different regularization parameters, and the last one is the proposed robust model. Given a pair of test PDFs denoted as $\{\hat{g}_k, \hat{f}_k\}$, the prediction error is quantified by the integrated absolute error (IAE), defined as $\varepsilon_k^{\mathrm{IAE}} = \int \left| \hat{f}_k^{\mathrm{pred}}(\tau) - \hat{f}_k(\tau) \right| d\tau$, where $\hat{f}_k^{\mathrm{pred}}$ stands for the prediction of $\hat{f}_k$ obtained by the regression model using $\hat{g}_k$ as the predictor. The boxplots for the prediction errors associated with the 35 test PDFs are plotted in Fig. 11 for the 10 considered models. As expected, the robust approach is significantly more accurate than the standard approach.

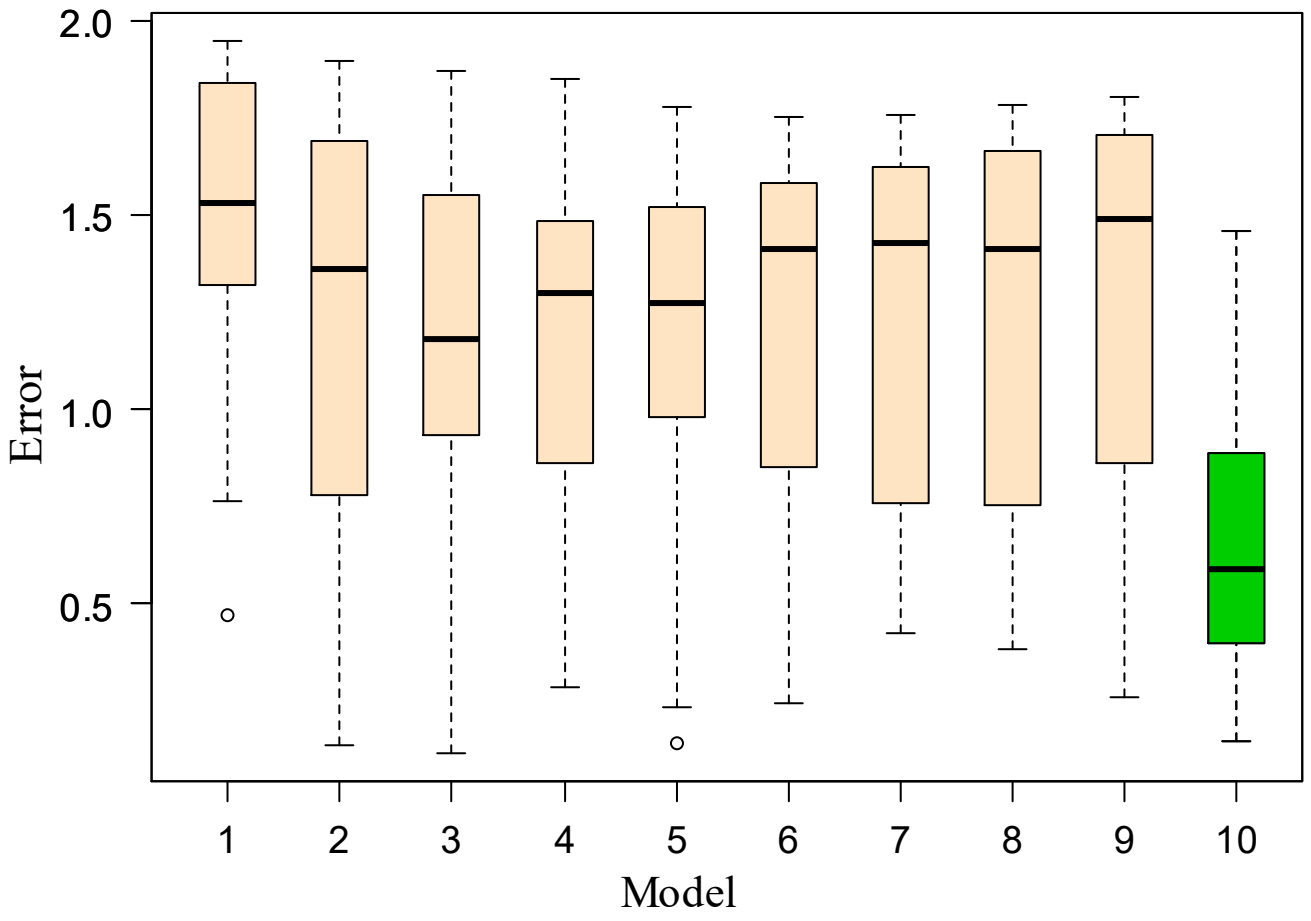

**Fig. 11.** Boxplots for the prediction errors of the test PDFs, Models 1~9 correspond to the standard LQD-RKHS method with regularization parameters $\lambda_s = 0.001, 0.01, 0.05, 0.1, 0.5, 1.0, 2.0, 4.0$ and $8.0$, respectively, and Model 10 corresponds to the robust LQD-RKHS method with regularization parameter $\lambda_s = 0.1$.

For comparison, 12 representative predicted PDFs obtained by Model 3 (standard version) and Model 10 (robust version) are shown in Fig. 12 along with the true PDFs represented as gray shadows. Most of the predicted PDFs obtained by the standard LQD-RKHS method exhibit significant deviations from the true PDFs to the right side, indicating that the estimated regression operator has been seriously distorted by the outlying PDFs, which can be attributed to abnormally large measurements. However, the proposed robust estimator can be effectively immune to such sample contaminations; hence, the corresponding predicted PDFs agree with the true densities much better.

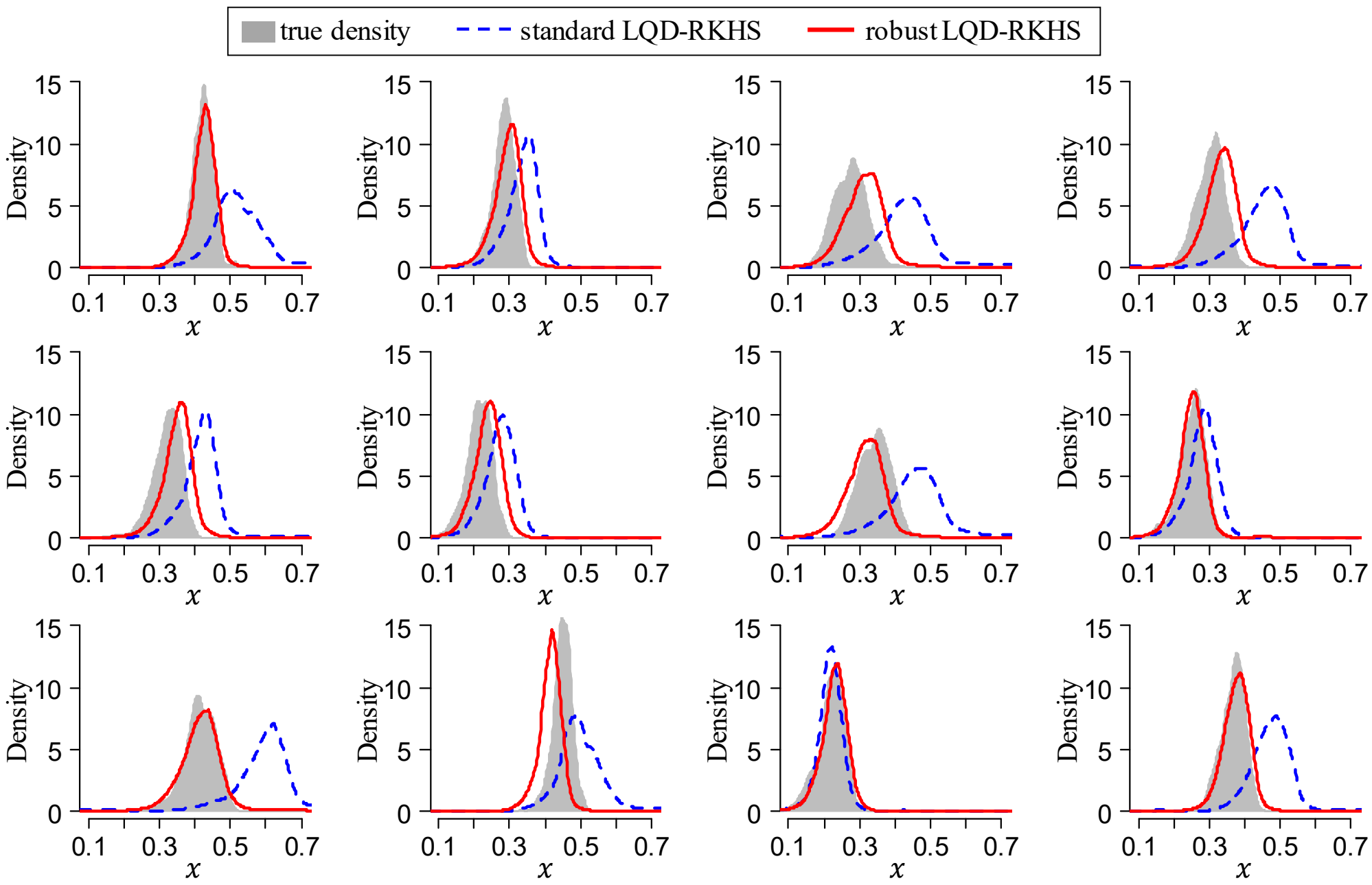

**Fig. 12.** Comparisons for 12 representative predicted PDFs obtained by the standard LQD-RKHS method with $\lambda_s = 0.05$ (i.e., Model 3) and robust LQD-RKHS method with $\lambda_s = 0.1$ (i.e., Model 10).

## 5. Conclusions and discussions

This study develops various functional outlier detection methods for detecting different types of outlying curves hidden in PDF-valued datasets by considering the specificities of distributional data. Based on the proposed functional outlier detection methods, we develop a robust estimator for the LQD-RKHS distribution-to-distribution regression model by dampening the impacts of detected outlying PDFs. The feasibility and usefulness of the proposed method are validated by simulation or real data studies.

The Tree-Distance approach leverages various distributional transformations to convert the less detectable shape outliers to more detectable magnitude outliers, and the “exposed” outliers can be easily detected using appropriate distances. Such a detection method is simple and easy to use, and it enables the extraction of different features for outlier detection. Related simulation studies demonstrate that the Tree-Distance method can effectively detect various patterns of outlying PDFs, and it also significantly outperforms the competitors. Moreover, the simulation studies provide strong evidence that the nLQD, CLR, DIFF, and MED nodes associated with the four branches of the transformation tree have good complementarity in distributional outlier detection.

The QF-FDO based approach can detect the shape and magnitude outliers simultaneously in the quantile function space but can lead to issues such as unstable detection. Related simulation studies have shown that it has a higher risk of false detection. Such an approach has the potential to cooperate with the Tree-Distance method in uncovering the "unexposed" shape outliers; however, effective measures (e.g., the multiple detection) are required to deal with the unstable detection issue.

Both the Tree-Distance and QF-FDO methods are highly efficient, which makes it possible to implement the multiple detection by selecting different parameter combinations. The multiple detection can not only account for the uncertainties attributed to parameter settings but also fuse different detection tools (e.g., the Tree-Distance and QF-FDO methods) to perform a combined detection. Moreover, it enables the identification and filtering of unstable detection results. The frequencies of detection associated with the outliers uncovered in multiple detection can provide a natural measure for quantifying their outlyingness. Applying it to real data shows that such a multiple detection method is more reliable in uncovering the various outliers that exhibit complex patterns.

The distributional-regression-based detection method provides an effective tool for detecting abnormal associations of PDF-valued two-tuples. The latter is another anomaly that may hide in the distributional data when PDFs from the two datasets are of concern. The outlier detection methods designed for single dataset inspection (e.g., the Tree-Distance method and QF-FDO method) are generally ineffective in detecting such abnormal associations, whereas the proposed distributional-regression-based approach fills this gap based on residual diagnosis, and the detected outliers are referred to as regression outliers in this study. Related simulation and real data studies have demonstrated the effectiveness of the proposed distributional-regression-based detection method in abnormal association detection for PDF-valued two-tuples.

In the robust distribution-to-distribution regression application, robustness is achieved by downweighting the impacts of the detected outliers in disturbing the regression operator estimation. Compared to the "crude" hard rejection in terms of directly removing the detected outliers, such a downweighting approach belongs to a more flexible soft rejection strategy. The latter enables us to treat the outliers separately by designing different weights for different types of outliers. Related real data applications show that the standard LQD-RKHS distributional regression method is highly sensitive to outliers, while the proposed robust version exhibits good performance even when the raw data are seriously contaminated by outliers.

The proposed transformation tree for outlying PDF detection can also be extended to outlier detection for ordinary functional data. Given a contaminated ordinary functional dataset, one can first use the classic functional outlier detection method to filter out the easily detectable magnitude outliers from the raw data and then convert the remaining functions to PDFs that are to be nested in the distributional outlier detection framework. As mentioned previously, the presented distributional transformations have their advantages in (1) naturally aligning curves according to quantiles and (2) eliminating the masking effects attributed to horizontal translations of PDFs. Thus, transforming ordinary functional data into distributional data for outlier detection may be more effective for some functional datasets. This can be investigated in future studies.

At present, both the Tree-Distance and QF-FDO detection methods are only applicable to univariate PDFs because the quantile function is unavailable for multivariate distribution (Petersen et al., 2021a). Developing effective functional outlier detection methods for PDFs associated with multivariate distributions can also be investigated in future studies.

## Acknowledgments

This work was financially supported by the National Natural Science Foundation of China (Grant Nos. 51908166 and 51638007), China Postdoctoral Science Foundation (Grant No. 2019M661287), and Postdoctoral Science Foundation of Heilong Jiang province.

## Appendix A. Computational details for the regression outlier detection

Let $\{(g_i, f_i): i = 1, \cdots, n\}$ be the PDF dataset for regression outlier detection. Based on the LQD-RKHS distributional regression method detailed in Chen et al. (2019a), we can construct two DtDR models for regression outlier detection. The first one is the forward regression model for regressing PDFs from $\{g_i\}$ to $\{f_i\}$. For convenience, we denote this regression model as $\Gamma^{gf}_{\mathrm{LQD-RKHS}}$. The other is the reverse regression model for regressing PDFs from $\{f_i\}$ to $\{g_i\}$, denoted as $\Gamma^{fg}_{\mathrm{LQD-RKHS}}$. For a more effective regression outlier detection, the forward and reverse regression models should be used jointly, and the process is summarized in Algorithm 1 with $\mathfrak{D}^{gf}_{\mathrm{LQD-RKHS}}$ and $\mathfrak{D}^{fg}_{\mathrm{LQD-RKHS}}$ denoting the regression-outlier detectors corresponding to $\Gamma^{gf}_{\mathrm{LQD-RKHS}}$ and $\Gamma^{fg}_{\mathrm{LQD-RKHS}}$, respectively. Algorithm 2 provides the implementation details for $\mathfrak{D}^{gf}_{\mathrm{LQD-RKHS}}$ using the forward regression model. The implementation of $\mathfrak{D}^{fg}_{\mathrm{LQD-RKHS}}$ is similar to $\mathfrak{D}^{gf}_{\mathrm{LQD-RKHS}}$ by switching $g_i$ and $f_i$ in Algorithm 2, and the regularization parameters $\lambda^{gf}_s$ and $\lambda^{fg}_s$ (corresponding to $\lambda_s$ in Eq. (18) in Chen et al. (2019a)) for the two regression models. $\alpha^{\mathrm{LQD}}_{mix}$ is the PDF preprocessing parameter, $\alpha$ in Eq. (8), from Chen et al. (2019a). For the LQD-RKHS DtDR, the Gaussian kernel given in Eq. (20) of Chen et al. (2019a) is selected as the reproducing operator kernel, and the related parameter $\sigma$ is determined by:

$$\sigma = \frac{1}{|I_{tr}|^2} \sum_{i \in I_{tr}} \sum_{j \in I_{tr}} \left( \int \left| \psi^{g^*}_i(\tau) - \psi^{g^*}_j(\tau) \right|^2 d\tau \right)^{\frac{1}{2}} \tag{26}$$

where $I_{tr}$ stands for the index set of the training data, $|I_{tr}|$ represents the number of elements contained in $I_{tr}$.

---

**Algorithm 1** Regression Outlier Detector

---

**Input:** PDFs pairs $\{(g_i, f_i): i = 1, \cdots, n\}$, PDF preprocessing parameters $\alpha^{\mathrm{LQD}}_{mix}$ and $\alpha^{\mathfrak{B}}_{mix}$, FPCA order $m_{\mathrm{FPCA}}$, whisker parameters $r^{\mathrm{LQD}}_2$ and $r^{\mathfrak{B}}_2$, thresholds $\theta_\lambda$ and $\theta_h$, number of interactions $N^{reg}_{iters}$

**Output:** Index set of the detected outliers $\mathcal{O}_{\mathrm{index}}$

1: Determine the regularization parameter using the generalized cross-validation for the LQD-RKHS regression models $\Gamma^{gf}_{\mathrm{LQD-RKHS}}$ and $\Gamma^{fg}_{\mathrm{LQD-RKHS}}$, denote the corresponding results as $\lambda^{gf}_{GCV}$ and $\lambda^{fg}_{GCV}$, respectively

2: Set $\lambda^{gf}_s = max\{\lambda^{gf}_{GCV}, \theta_\lambda\}$, $\lambda^{fg}_s = max\{\lambda^{fg}_{GCV}, \theta_\lambda\}$ and $TR_{\mathrm{index}} = \{1, \cdots, n\}$

3: **for** $k = 1$ **to** $N^{reg}_{iters}$ **do**

a: Regression outlier detection using the forward regression model (implemented in Algorithm 2)

$\mathcal{O}^{gf}_{\mathrm{index}} = \mathfrak{D}^{gf}_{\mathrm{LQD-RKHS}}(g_i, f_i \,|\, TR_{\mathrm{index}}, \alpha^{\mathrm{LQD}}_{mix}, \alpha^{\mathfrak{B}}_{mix}, \lambda^{gf}_s, m_{\mathrm{FPCA}}, r^{\mathrm{LQD}}_2, r^{\mathfrak{B}}_2, \theta_h)$

b: Regression outlier detection using the reverse regression model

$\mathcal{O}^{fg}_{\mathrm{index}} = \mathfrak{D}^{fg}_{\mathrm{LQD-RKHS}}(f_i, g_i \,|\, TR_{\mathrm{index}}, \alpha^{\mathrm{LQD}}_{mix}, \alpha^{\mathfrak{B}}_{mix}, \lambda^{fg}_s, m_{\mathrm{FPCA}}, r^{\mathrm{LQD}}_2, r^{\mathfrak{B}}_2, \theta_h)$

c: Set $\mathcal{O}_{\mathrm{index}} = \mathcal{O}^{gf}_{\mathrm{index}} \cup \mathcal{O}^{fg}_{\mathrm{index}}$

d: Set $TR_{\mathrm{index}} = TR_{\mathrm{index}} \setminus \mathcal{O}_{\mathrm{index}}$

**end for**

4. Output $\mathcal{O}_{\mathrm{index}}$

---

**Algorithm 2** LQD-RKHS Detector using the forward regression model

$$\mathcal{O}_{\text{index}}^{gf} = \mathfrak{D}_{\text{LQD-RKHS}}^{gf}\big(g_i, f_i \mid TR_{\text{index}}, \alpha_{mix}^{\text{LQD}}, \alpha_{mix}^{\mathfrak{B}}, \lambda_s^{gf}, m_{\text{FPCA}}, r_2^{\text{LQD}}, r_2^{\mathfrak{B}}, \theta_h\big)$$

**Input:** PDFs pairs $\{(g_i, f_i): i = 1, \cdots, n\}$, index set of training PDFs $TR_{\text{index}}$, PDF preprocessing parameters $\alpha_{mix}^{\text{LQD}}$ and $\alpha_{mix}^{\mathfrak{B}}$, regularization parameter $\lambda_s^{gf}$, FPCA order $m_{\text{FPCA}}$, whisker parameters $r_2^{\text{LQD}}$ and $r_2^{\mathfrak{B}}$, threshold $\theta_h$

**Output:** Index set of the detected outliers $\mathcal{O}_{\text{index}}$

1: Fit the LQD-RKHS DtDR model $\Gamma_{\text{LQD-RKHS}}^{gf}\big(g_i, f_i \mid TR_{\text{index}}, \alpha_{mix}^{\text{LQD}}, \lambda_s^{gf}, m_{\text{FPCA}}\big)$ using the training dataset $\{(g_j, f_j): j \in TR_{\text{index}}\}$

2: Use the fitted model to obtain predictions for the PDFs $\{f_1, \cdots, f_n\}$, and the predicted PDFs are denoted as $\{\hat{f}_1, \cdots, \hat{f}_n\}$

3: Compute the residual using Algorithms A.2 and A.3 in the online supplement, respectively, to obtain the following quantities

$$\varepsilon_i^{\mathfrak{B}} = d_{\mathfrak{B}}\big(\hat{f}_i, f_i \big| \theta_h, \alpha_{mix}^{\mathfrak{B}}\big), i = 1, \cdots, n$$

$$\varepsilon_i^{\text{LQD}} = d_{L_1}\big(\text{LQD}[\hat{f}_i | \alpha_{mix}^{\text{LQD}}], \text{LQD}[f_i | \alpha_{mix}^{\text{LQD}}]\big), i = 1, \cdots, n$$

4: Outlier detection for the residual datasets $\boldsymbol{\mathcal{E}}^{\mathfrak{B}} = \{\varepsilon_1^{\mathfrak{B}}, \cdots, \varepsilon_n^{\mathfrak{B}}\}$ and $\boldsymbol{\mathcal{E}}^{\text{LQD}} = \{\varepsilon_1^{\text{LQD}}, \cdots, \varepsilon_n^{\text{LQD}}\}$ by using the one-sided boxplot-based detector given in Eq.(A-12) in the online supplement

$$\mathcal{O}_{\text{index}}^{\mathfrak{B}} = \text{OneSided_BoxPlotDetector}\big(\boldsymbol{\mathcal{E}}^{\mathfrak{B}} | r_2^{\mathfrak{B}}\big)$$

$$\mathcal{O}_{\text{index}}^{\text{LQD}} = \text{OneSided_BoxPlotDetector}\big(\boldsymbol{\mathcal{E}}^{\text{LQD}} | r_2^{\text{LQD}}\big)$$

5: Output $\mathcal{O}_{\text{index}}^{gf} = \mathcal{O}_{\text{index}}^{\mathfrak{B}} \cup \mathcal{O}_{\text{index}}^{\text{LQD}}$

More computation details consisting of regularization parameter determination using the generalized cross-validation (GCV), and residual calculations for $\varepsilon_i^{\text{LQD}}$ and $\varepsilon_i^{\mathfrak{B}}$ are provided in Subsection A.3.4 of the online supplement. Detailed discussions on how to use the threshold $\theta_h$ (in calculating $\varepsilon_i^{\mathfrak{B}}$) to reduce the false detection risk caused by regression errors are also provided.

## Appendix B. Proof of Proposition 2

**Proof.** According to the reproducing properties given in Eqs. (A-5) and (A-7) in the online supplement, it follows that

$$\langle b_1 k_r(s,\cdot), b_2 k_r(l,\cdot)\rangle_{H(k_r)} = b_1 b_2 k_r(s,l),\ \forall b_1, b_2 \in \mathbb{R}, s, l \in \{1,2,\cdots,m\} \tag{27}$$

and

$$\langle K_r(\psi_1,\cdot)\boldsymbol{\alpha}_1, K_r(\psi_2,\cdot)\boldsymbol{\alpha}_2\rangle_{\mathcal{H}(K_r)} = \langle K_r(\psi_1,\psi_2)\boldsymbol{\alpha}_1, \boldsymbol{\alpha}_2\rangle_{H(k_r)},$$
$$\forall \psi_1, \psi_2 \in L^2[0,1],\ \boldsymbol{\alpha}_1, \boldsymbol{\alpha}_2 \in H(k_r) \tag{28}$$

It is worth noting that $\mathcal{H}(K_r)$ and $H(k_r)$ are two different Hilbert spaces, $L^2[0,1]$ and $H(k_r)$ in Eq. (28) corresponds to $G$ and $\tilde{G}$ in Eq. (A-7) in the online supplement, respectively.

Using the properties given in Eqs. (27) and (28), the objective function given in Eq.(11) can be transformed into the following matrix form by substituting Eqs. (12) and (14) into Eq. (11).

$$J(\mathbf{B}) = \|\mathbf{W}(\mathbf{Y} - \mathbf{ABK})\|_F^2 + \lambda_s \text{trace}(\mathbf{ABKB}^T) \tag{29}$$

where $\mathbf{A}$, $\mathbf{B}$, $\mathbf{W}$, and $\mathbf{Y}$ are the matrixes defined in Eq. (20), $\mathbf{K}$ is the Gram matrix of the real kernel (see Eq. (15)), and $\|\cdot\|_F$ is the Frobenius norm defined as $\|\mathbf{U}\|_F = \left(\sum_{i=1}^{n}\sum_{j=1}^{m} u_{ij}^2\right)^{\frac{1}{2}}$. Let $\frac{\partial J(\mathbf{B})}{\partial \mathbf{B}} = 0$, which yields the following matrix equation:

$$\mathbf{AW}^2\mathbf{ABK}^2 + \lambda_s \mathbf{ABK} = \mathbf{AW}^2\mathbf{YK} \tag{30}$$

This matrix equation can be solved using vectorization. Let $\text{vec}(\mathbf{U})$ denote the vectorization of the matrix $\mathbf{U} = \{u_{ij}\}_{i=1}^{n}{}_{j=1}^{m}$ defined as

$$\text{vec}(\mathbf{U}) = (u_{11}\, u_{21}\, \cdots\, u_{n1}\, u_{12}\, u_{22}\, \cdots\, u_{n2} \cdots u_{1m}\, u_{2m}\, \cdots\, u_{nm})^T \tag{31}$$

Noting further the properties of $\text{vec}(\mathbf{AXC}) = (\boldsymbol{C}^T \otimes \boldsymbol{A})\text{vec}(\mathbf{X})$ and $(\mathbf{AC}) \otimes (\mathbf{BD}) = (\mathbf{A} \otimes \mathbf{B})(\mathbf{C} \otimes \mathbf{D})$, then Eq.(30) can be converted to

$$\Big(\big(\mathbf{K} \otimes (\mathbf{AW})\big)\big(\mathbf{K} \otimes (\mathbf{WA})\big) + \lambda_s(\mathbf{K} \otimes \mathbf{A})\Big)\text{vec}(\mathbf{B}) = \big(\mathbf{K} \otimes (\mathbf{AW})\big)\text{vec}(\mathbf{WY}) \tag{32}$$

Let $\mathbf{C}_1 = \mathbf{K} \otimes (\mathbf{AW})$ and $\mathbf{C}_2 = \mathbf{K} \otimes (\mathbf{WA})$ complete the proof.

### Appendix C. Supplementary material

Online Supplement: Supplementary materials for “Functional outlier detection for density-valued data with application to robustify distribution to distribution regression”.

# ONLINE SUPPLEMENT

Supplementary materials for "Functional outlier detection for density-valued data with application to robustify distribution to distribution regression"

# Contents

## A.1. Theoretical background

### *A.1.1. Bayes space*

The Bayes space, denoted as $\mathfrak{B}^2(I)$, consists of positive functions defined on the common compact interval $I = [a, b]$ with square-integrable logarithms. The Bayes space $\mathfrak{B}^2(I)$ is a separable Hilbert space under the following linear operations and inner product (Egozcue et al.,2006; Van den Boogaart et al. 2014; Hron et al., 2016):

(1) Linear operation

Perturbation: $(f \oplus g)(x) = \frac{f(x)g(x)}{\int_I f(\tau)g(\tau)d\tau}$, x $\in I$, and $f, g \in \mathfrak{B}^2(I)$ (A-1a)

Powering: $(\beta \odot f)(x) = \frac{f(x)^\beta}{\int_I f(\tau)^\beta d\tau}$, x $\in I$ and $\beta \in \mathbb{R}$, $f \in \mathfrak{B}^2(I)$ (A-1b)

where $\mathbb{R}$ stands for the set of real numbers. The perturbation and powering are analogous to the point-wise addition and scalar multiplication of $L^2(I)$ space.

(2) Inner product

$$\langle f, g \rangle_{\mathfrak{B}} = \frac{1}{2\eta} \int_I \int_I \log \frac{f(t)}{f(s)} \log \frac{g(t)}{g(s)} dtds, \quad f, g \in \mathfrak{B}^2(I) \tag{A-2}$$

where $\eta = b - a$ is the Lebesgue measure of the compact interval $I = [a, b]$.

The Bayes space $\mathfrak{B}^2(I)$ is isometrically isomorphic to the $L^2(I)$ space with the following centered log-ratio (clr) transformation as the isomorphic mapping (Egozcue et al.,2006; Talská et al., 2018):

$$\text{clr}[f](x) = \log f(x) - \frac{1}{\eta} \int_I \log f(\tau) d\tau, x \in I, \ f \in \mathfrak{B}^2(I) \tag{A-3}$$

Obviously, the univariate continuous PDF supported on $I = [a, b]$ is an element of the Bayes space $\mathfrak{B}^2(I)$.

### *A.1.2. Notations and basic properties of the reproducing kernel Hilbert space (RKHS)*

*(1) Real reproducing kernel*

Let $D$ be an arbitrary set, and let $H(k_r)$ be the real RKHS associated with the real reproducing kernel $k_r$. According to the real RKHS theory (Berlinet and Thomas-Agnan, 2011; Lian, 2007a), $H(k_r)$ is a subspace of $\{f: D \to \mathbb{R}\}$ consisting of functions defined on $D$. The reproducing kernel $k_r$ is a symmetric semi-definite function defined on $D \times D$, i.e., $k_r: D \times D \to \mathbb{R}$, satisfying the following properties (Berlinet and Thomas-Agnan, 2011; Lian, 2007a)

$$k_r(x,\cdot) \in H(k_r),\ \forall x \in D \tag{A-4}$$

and

$$f(x) = \langle k_r(x,\cdot), f \rangle_{H(k_r)},\ \forall x \in D, f \in H(k_r) \tag{A-5}$$

where, $\langle \cdot,\cdot \rangle_{H(k_r)}$ is the inner-product endowed to the RKHS $H(k_r)$.

*(2) Operator-valued reproducing kernel*

Let $G$ and $\tilde{G}$ be two arbitrary Hilbert spaces, let $\mathcal{H}(K_r)$ denote the RKHS associated with operator-valued reproducing kernel $K_r$. According to the theory of operator-valued kernels (Kadri et al.,2016; Lian 2007a), $\mathcal{H}(K_r)$ is a subspace of the operator space $\{F: G \to \tilde{G}\}$ consisting of mappings from $G$ to $\tilde{G}$. The operator-valued kernel $K_r$ is a symmetric semi-definite mapping from the product Hilbert space $G \times G$ to the other Hilbert space $\tilde{G}$, i.e., $K_r: G \times G \to \tilde{G}$, satisfying (Kadri et al.,2016; Lian 2007a)

$$K_r(\psi,\cdot) \in \mathcal{H}(K_r), \forall\, \psi \in G \tag{A-6}$$

and

$$\langle F(\psi), \alpha \rangle_{\tilde{G}} = \langle K_r(\psi,\cdot)\alpha, F \rangle_{\mathcal{H}(K_r)},\ \forall \psi \in G,\ \alpha \in \tilde{G},\ F \in \mathcal{H}(K_r) \tag{A-7}$$

where $\langle \cdot\,,\,\cdot \rangle_{\tilde{G}}$ and $\langle \cdot\,,\,\cdot \rangle_{\mathcal{H}(K_r)}$ are the inner-products endowed to the Hilbert space $\tilde{G}$ and the RKHS $\mathcal{H}_{rep}(K)$, respectively.

## A.2. PDF preprocessings for LQD and clr transformations

### *A.2.1. PDF preprocessing for LQD transformation*

Given a PDF $f(x)$ finitely supported on the compact interval [0,1], its log quantile density (LQD) transformation (Petersen and Müller, 2016) is defined as

$$\psi(t) = \log\left(\frac{dQ(t)}{dt}\right) = -\log\{f(Q(t))\} \tag{A-8}$$

where $Q(t)$ is the quantile function associated with the PDF $f(x)$. As pointed in Subsection 2.1 of the manuscript, the LQD transformation is "blind" to the horizontal translation of PDFs. Such a "blindness" property is leveraged to "expose" the shape outliers in this study. However, for other applications such as the LQD-RKHS distributional regression (Chen et al., 2019a) involved in the regression outlier detection (Subsection 2.4) and the robust distributional regression (Section 3), effective measures should be taken to cure such a "blindness" issue.

On the other hand, if the PDF $f(x)$ takes near-zero values, computing the inverse function of the CDF $F(x)$ to obtain the quantile function (involved in the LQD transformation) might also suffer from a numerical issue in the interpolation process.

Fortunately, both the issues raised above can be easily addressed by performing the following preprocessing to the PDF $f(x)$ (Chen et al., 2019a)

$$f^*(x) = (1-\alpha)f(x) + \alpha,\ x \in [0,1] \tag{A-9}$$

where $\alpha$ is a prescribed small positive constant referred to as the PDF preprocessing parameter throughout this study. Note that the PDF $f(x)$ is finitely supported on the compact interval [0,1], thus the above processing is equivalent to mixing the original distribution by a proportion of uniform distribution $\mathrm{U}(0,1)$. The magnitude of the constant $\alpha$ depends on the specific applications: (1) for outlier detection, $\alpha$ should be smaller (e.g., $\alpha \in [10^{-10}, 10^{-2}]$), otherwise it might increase the variability of the functional data in the LQD node; (2) however, for the LQD-RKHS distributional regression (Chen et al., 2019a), the constant $\alpha$ should be larger (e.g., $\alpha \in [0.2, 0.5]$ as recommended by Chen et al. (2019a)), otherwise the LQD transformation might be "blind" to the horizontal translation of PDFs.

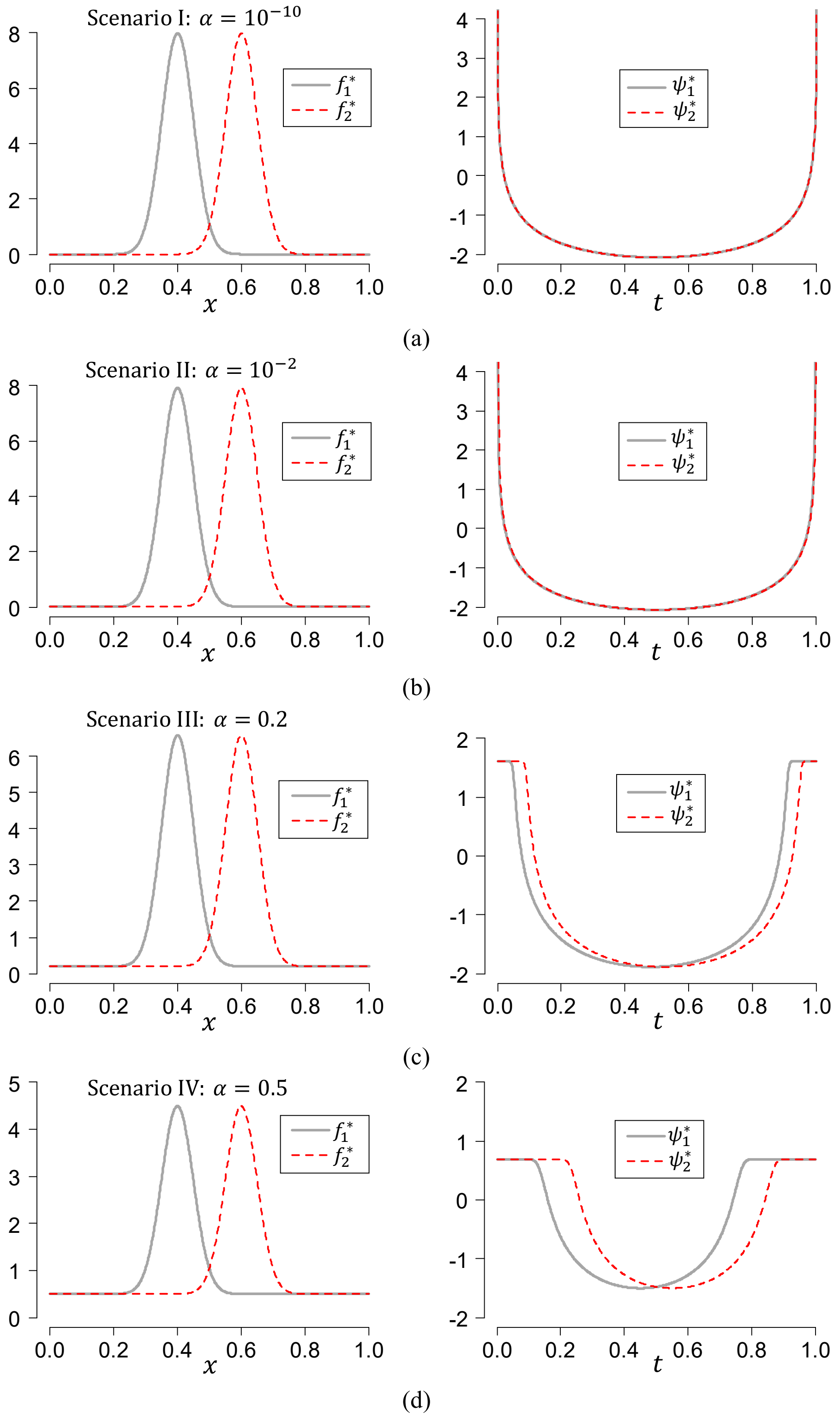

**Fig. A-1.** Comparisons of the LQD transformations associated with two truncated normal densities after they are preprocessed by using Eq.(A-9) with $\alpha$ taking four different values. (a) Scenario I: $\alpha = 10^{-10}$, (b) Scenario II: $\alpha = 10^{-2}$, (c) Scenario III: $\alpha = 0.2$, and (d) Scenario IV: $\alpha = 0.5$. The left column corresponds to the PDFs, while the right column corresponds to the LQD transformations.

For illustration purposes, we consider two PDFs (denoted as $f_1(x)$ and $f_2(x)$) respectively obtained by truncating the densities of norm distributions $N(0.4, 0.05^2)$ and $N(0.6, 0.05^2)$ within the domain of [0,1], the calculated LQD transformations associated with four different values of $\alpha$ (i.e., $\alpha = 10^{-10}, 10^{-2}$, 0.2 and 0.5) are displayed in Fig. A-1. Comparing the LQD transformations shown in Fig. A-1, one can see that the "blindness" phenomenon happens to the scenarios of $\alpha = 10^{-10}$ and $\alpha = 10^{-2}$, but disappears in the scenarios of $\alpha = 0.2$ and $\alpha = 0.5$ as the corresponding LQD transformations of the two PDFs can be clearly distinguished from each other.

For the distributional regression application, the regression is performed to the preprocessed PDF. Therefore, to recover the desired regression prediction associated with the target PDF $f(x)$, one should remember to clear the added uniform distribution through performing the following post-processing (Chen et al., 2019a):

$$\hat{f}(x) = \frac{1}{W}\left|\frac{\hat{f}^*(x) - \alpha}{1-\alpha}\right| \text{ with } W = \int_0^1 \left|\frac{\hat{f}^*(\tau) - \alpha}{1-\alpha}\right| d\tau \qquad \text{(A-10)}$$

where $\hat{f}^*(x)$ stands for the prediction of $f^*(x)$ obtained by the regression model. For more detailed discussion, readers are referred to Chen et al. (2019a).

### *A.2.2. PDF preprocessing for clr transformation*

The clr transformation given in Eq.(A-3) for a PDF taking near-zero values might suffer from a numerical issue in the logarithmic computation or have significant boundary effects, the latter is mainly attributed to the sharp change of the logarithmic function $\log(u)$ near $u = 0$. Therefore, when calculating the clr transformations associate with the investigated PDF-valued datasets $\{f_i(x)\}_{i=1}^n$, all the functional samples contained in $\{f_i(x)\}_{i=1}^n$ will be preprocessed in a unified manner as $f_i(x) = (1-\alpha)f_i(x) + \alpha, i = 1,2,\cdots,n$, if the minimum value of the PDFs (i.e., $\min_{0\le i\le 1} \inf_{x\in[0,1]} \{f_i(x)\}$) is less than 0.1. Unless otherwise stated, such a PDF-prepressing will be performed by default for the clr transformations involved in this study, and the default value of the PDF preprocessing parameter $\alpha$ is 0.1.

## A.3. Supplemental materials for distributional outlier detection

### *A.3.1. Modified boxplot-based detectors for scalar outlier detection*

Let $\{\theta_i\}_{i=1}^{n}$ be a scalar dataset, the following two detectors modified from standard boxplot will be used to identify the potential outliers contained in $\{\theta_i\}_{i=1}^{n}$ according to specific situations:

(i) Detector I (two-sided detector):

$$\mathrm{OUT_{ID}} = \{i \in \{1,2,\cdots,n\} | \theta_i \leq q_{0.25}(\theta) - r_1 \cdot \mathrm{IQR} \ \text{ or } \ \theta_i \geq q_{0.75}(\theta) + r_1 \cdot \mathrm{IQR}\} \qquad \text{(A-11)}$$

where $\mathrm{OUT_{ID}}$ stands for the index set of the detected outliers, $q_{0.25}(\theta)$ and $q_{0.75}(\theta)$ denote the 25th and 75th percentiles of the dataset $\{\theta_i\}_{i=1}^{n}$, respectively, IQR is the interquartile range defined as $\mathrm{IQR}=q_{0.75}(\theta) - q_{0.25}(\theta)$, and $r_1$ is a user prescribed parameter.

(ii) Detector II (one-sided detector):

$$\mathrm{OUT_{ID}} = \{i \in \{1,2,\cdots,n\} | \theta_i \geq q_{0.75}(\theta) + r_2 \cdot \mathrm{IQR}\} \qquad \text{(A-12)}$$

Such a one-sided detector is designed specifically for the scenario that only the one taking abnormally large value can be regarded as the outlier, such as the case with the distance-based detection approach discussed in this study.

### *A.3.2. Supplemental materials for the distance-based detection method*

The distance-based detection method described in Subsection 2.2 of the manuscript is mainly applicable for detecting the functional outliers that have been “exposed” as magnitude outliers after performing the data transformations. We recommend to perform such distance-based detections for the functional data associated with the nLQD, CLR and DIFF nodes on the transformation tree shown in Fig. 2 of the manuscript. Moreover, the outlier detection is also required to be conducted to the medians residing in the MED node, which plays the key role in identifying the horizontal shift outliers. In contrast to the functional data in nodes nLQD, CLR and DIFF, the data in the MED node are scalar data; thus, the corresponding outliers can be directly detected by using the two-sided boxplot detector given in Eq. (A-11). However, for the functional data associated with the nLQD, CLR and DIFF nodes, appropriate distances are required to be selected for converting the functional outlier detection problem into the scalar outlier detection problem by using Eq. (6) in the manuscript. It is worth noting that the functional data in nodes nLQD and DIFF are both ordinary functional data, for which we recommend the $L_1$ and $L_\infty$ distances (see Table A-1 for their corresponding definitions). The $L_1$ distance is mainly for quantifying the global dissimilarities of functional data, while the $L_\infty$ distance is mainly for quantifying the local dissimilarities of functional data. For the same functional dataset, the outlier detections using different distances are conducted independently, then we merge the detected outliers to form the final result. For the functional data associated with the CLR node, we recommend the $L_2$ distance (see Table A-1 for the definition) in consideration of the isometric isomorphism, which is consistent with Eq. (4) in the manuscript.

**Table A-1**
Distances used in this study

| Distance | Definition | |
|---|---|---|
| $L_1$ | $d_{L_1}(f,g) = \int_I \lvert f(\tau) - g(\tau)\rvert d\tau,$ | $f,g \in L^1(I)$ |
| $L_2$ | $d_{L_2}(f,g) = \left(\int_I \left(f(\tau) - g(\tau)\right)^2 d\tau\right)^{1/2},$ | $f,g \in L^2(I)$ |
| $L_\infty$ | $d_{L_\infty}(f,g) = \sup_{\tau \in I} \lvert f(\tau) - g(\tau)\rvert,$ | $f,g \in L^\infty(I)$ |

With the above recommended distances, the default argument settings of outlier detections for the resulting transformed data associated with nodes nLQD, CLR, DIFF and MED are summarized in Table A-2. The functional data in the nLQD node are computed from the quantile functions, all the functional data after the QF node (in the Branch I of the transformation tree) have also been naturally aligned according to the quantiles; thus, one can choose appropriate detection intervals to conduct the curve truncation as shown in Fig. A-2(c). The main reason for performing such a truncation is twofold: (1) reduce the boundary effects in disturbing the outlier detection; (2) restrict the outlier detection within the region of interested. According to our experience, we recommend to independently perform two rounds outlier detections for the functional data in the nLQD node respectively using the detection regions [0.2, 0.8] and [0.4, 0.6] for both the $L_1$ and $L_\infty$ distances, and then the detected outliers are merged to form the final detection result of the nLQD node.

**Table A-2**

Default settings of outlier detections using the distance-based method for the transformed data associated with nodes nLQD, CLR, DIFF, MED on the transformation tree. The parameter $\alpha$ in the second column is the PDF preprocessing parameter described in Section A.2. The detection region $[u, v]$ associated with the CLR node is the common support of the translated PDFs in the horizontal centralization processing (performed in the H-CENTR node) described in Subsection 2.1 of the manuscript. Detector I and Detector II stand for the two- and one-sided boxplot-based detectors given in Eq. (A-11) and Eq. (A-12), respectively.

| Node | $\alpha$ | Distance | Detector | Whisker | Detection region |
|---|---|---|---|---|---|
| nLQD | $10^{-10}$ | $L_1$ and $L_\infty$ | Detector II | $L_1$: 2.5IQR<br>$L_\infty$: 3.5IQR | $[0.2, 0.8]$ and $[0.4, 0.6]$ |
| CLR | 0.1 | $L_2$ | Detector II | $L_2$: 2.5IQR | $[u, v]$ |
| DIFF | — | $L_1$ and $L_\infty$ | Detector II | $L_1$: 2.5IQR<br>$L_\infty$: 3.5IQR | $[0, 1]$ |
| MED | — | — | Detector I | 1.5IQR | — |

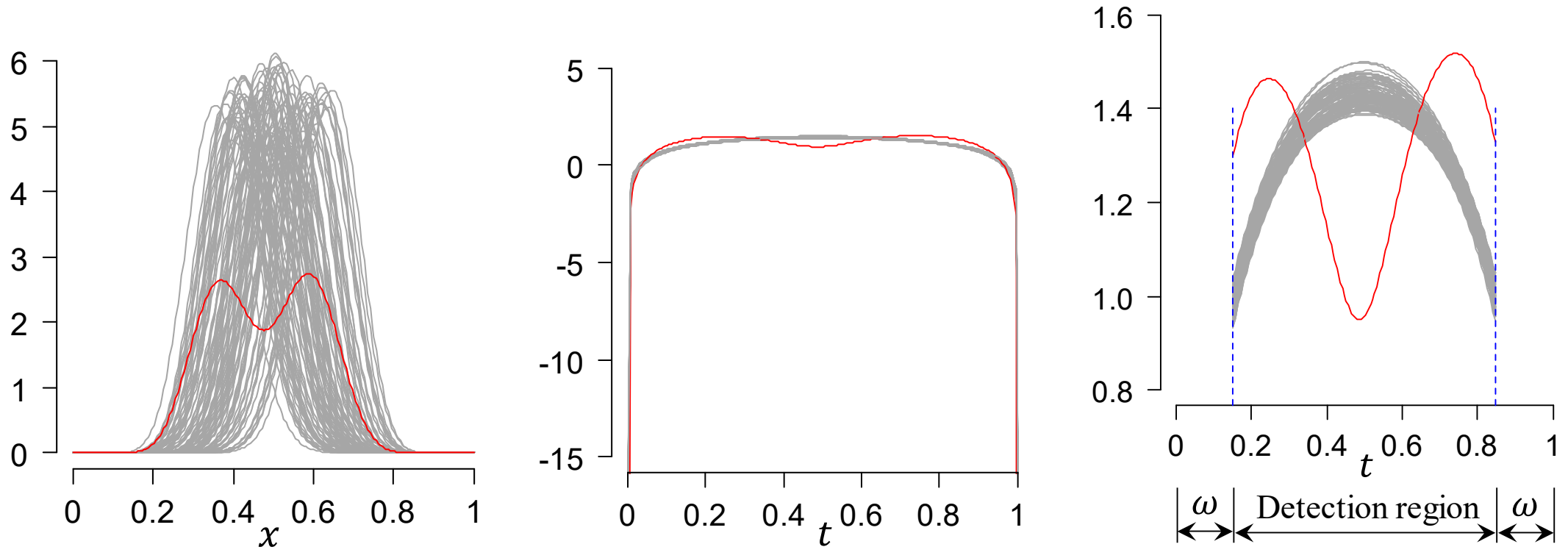

(a) (b) (c)

**Fig. A-2.** Illustration of the curve truncation using detection region $A = [\omega, 1 - \omega]$ for the normalized LQD transformations associated with a PDF-valued dataset. (a) The raw PDF-valued dataset, (b) the corresponding normalized LQD transformations before curve truncation, and (c) the corresponding normalized LQD transformations after curve truncation. The gray lines and the red line represent the "good" data and the outlying curve, respectively.

If $\phi_i(t)$ in Eq. (6) (in the manuscript) is the clr-representation, the center function $m(t)$ calculated by the cross-sectional median (i.e., $m(t) = \underset{1 \le i \le n}{\text{median}}\{\phi_i(t)\}$) may fail to satisfy the constraint of the clr-representation (i.e., integral to zero). Consequently, the distance calculated using Eq. (6) with the $L_2$ distance can be regarded as a pseudo version of the $d_{\mathfrak{B}}$ given in Eq. (4) (in the manuscript). In outlier detection applications, the mathematically rigid restrictions can be loosened properly.

### *A.3.3. Supplemental materials for the FDO-based detection method*

*A.3.3.1 Computational details*

The functional directional outlyingness (FDO) was defined by Dai and Genton (2019), then it has been successfully applied in outlier detection for ordinary functional data with potential for uncovering magnitude outliers and shape outliers simultaneously. However, according to our test, directly performing outlier detection in the PDF space using the FDO tool usually results in poor performance (related illustrations will be given latter). To remedy this embarrassing, we take a strategy of applying the FDO tool in the quantile function (QF) space (i.e., the QF node in the transformation tree) for outlier detection, which is referred to as the QF-FDO method throughout this study. As illustrated in Figs. 3 (a) and (b) of the manuscript, after converting the PDFs to quantile functions, the horizontal-shift outlying PDFs have become the magnitude outliers, while the shape outliers are still hidden in the bulk of the data. In contract to the disordered PDFs, the quantile functions are in a much more organized pattern as they have been registered according to quantiles. In the following, we first present the computational details for the QF-FDO method, followed by related detection results using simulated data.

Consider a functional dataset denoted as $\{f_i(x)\}_{i=1}^{n}$ with each element being a PDF defined on the compact interval $[0,1]$, let $Q_i(t)$ be the quantile function corresponding to $f_i(x)$. Also, let $\underset{1\le k\le n}{\text{median}}\{a_k\}$ denote the sample median of the real-valued dataset $\{a_i\}_{i=1}^{n}, a_i \in \mathbb{R}$. Note that when fixing $t$ at $t = t_0$, the set of $\{Q_i(t_0)\}_{i=1}^{n}$ is a real-valued dataset. Then, the pointwise outlyingness of $Q_i(t)$ at $t = t_0$ can be calculated by (Dai and Genton, 2019)

$$\text{dSDO}\big(Q_i(t_0)\big) = \frac{Q_i(t_0) - \underset{1\le k\le n}{\text{median}}\{Q_k(t_0)\}}{\text{MAD}(Q(t_0))} \tag{A-13}$$

where $\text{MAD}(Q(t_0))$ is the median absolute deviation (MAD) calculated by $\text{MAD}\big(Q(t_0)\big) = c \cdot \underset{1\le k\le n}{\text{median}}\left\{\left|Q_k(t_0) - \underset{1\le j\le n}{\text{median}}\{Q_j(t_0)\}\right|\right\}$ with $c$ being a constant (we set $c$ to be its default value 1.4826 throughout this study).

Based on the FDO theory (Dai and Genton, 2019), the magnitude and shape anomalies of a given quantile function $Q_i(t)$ can be respectively measured by the mean outlyingness (MO) and the variation of outlyingness (VO) calculated as follows:

$$\mathrm{MO}(Q_i) = \int_0^1 \mathrm{dSDO}\big(Q_i(t)\big)\omega(t)dt$$
$$\mathrm{VO}(Q_i) = \int_0^1 \left|\mathrm{dSDO}\big(Q_i(t)\big) - \mathrm{MO}(Q_i)\right|^2 \omega(t)dt \tag{A-14}$$

where $\omega(t)$ is the weight function, which is commonly chosen as $\omega(t) = 1/\lambda(I)$ with $\lambda(I)$ being the Lebesgue measure of the quantile function's domain of definition denoted as $I$.

Generally, the magnitude outliers and shape outliers in the collection of quantile functions would stand out as MO-outliers (abnormal in the MO direction) and VO-outliers (abnormal in the VO direction), respectively. The MO- and VO-outliers can be efficiently identified by the two- and one-sided boxplot-based detectors given in Eqs. (A-11) and (A-12), respectively.

*A.3.3.2 Illustration of the instability issue and related mitigation measure*

We conduct a simulation study to examine the detection performance of the FDO method. To begin with, we generate a PDF-valued dataset composed of 100 functions by using Algorithm A. 1 with the input arguments setting to $n = 100$, $\delta_1 = 36$ and $\delta_2 = 63$. Then, we use Algorithm A. 4 to generate and insert 5 outlying PDFs into the simulated functional dataset by setting $N_O = 5$, $\zeta_{hs} = 0$, and $\varpi = 0.2$. Let $S_{\text{contam}} = \{f_i\}_{i=1}^n$ denote the contaminated PDF dataset, then we use it to produce the following four different PDF datasets:

Model I: $\mathrm{S}_{\text{contam}}^{\mathrm{I}} = \{f_i^{\mathrm{I}}\}_{i=1}^n$ with $f_i^{\mathrm{I}} = f_i,\ f_i \in S_{\text{contam}}$

Model II: $\mathrm{S}_{\text{contam}}^{\mathrm{II}} = \{f_i^{\mathrm{II}}\}_{i=1}^n$ with $f_i^{\mathrm{II}} = 0.9 * f_i + 0.1,\ f_i \in S_{\text{contam}}$

Model III: $\mathrm{S}_{\text{contam}}^{\mathrm{III}} = \{f_i^{\mathrm{III}}\}_{i=1}^n$ with $f_i^{\mathrm{III}} = 0.7 * f_i + 0.3,\ f_i \in S_{\text{contam}}$

Model IV: $\mathrm{S}_{\text{contam}}^{\mathrm{IV}} = \{f_i^{\mathrm{IV}}\}_{i=1}^n$ with $f_i^{\mathrm{IV}} = 0.5 * f_i + 0.5,\ f_i \in S_{\text{contam}}$

Fig. A-3 displays the four simulated datasets.

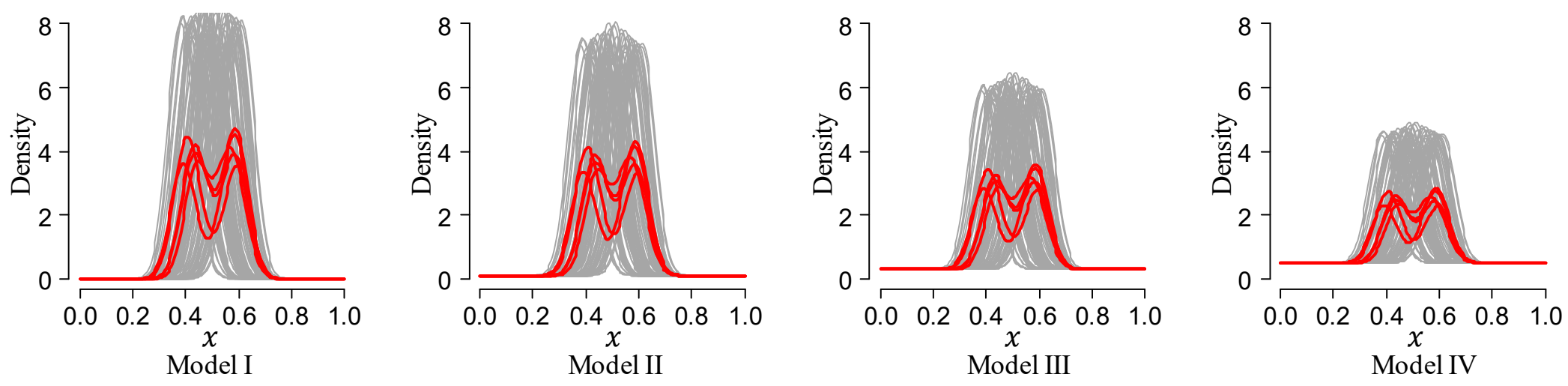

**Fig. A-3.** Simulated PDF-valued datasets, each dataset consists of 100 curves with bold red curves standing for synthetic outlying PDFs.

| **Algorithm A. 1**: PDF generation procedure |
|---|
| **Input**: Number of PDFs $n$, parameters $\delta_1$ and $\delta_2$ |
| **Output**: PDF-valued dataset $\{f_i\}_{i=1}^n$ |
| 1: Independently generate $a_i \sim \mathrm{U}(\delta_1, \delta_2)$, $i = 1,2,\cdots,n$ |
| 2: Sort $\{a_1, a_2, \cdots, a_n\}$ in ascending order and denote the resulting series as $\{b_1, b_2, \cdots, b_n\}$ |
| 3: Generate PDF $f_i(x) = \mathrm{BetaPdf}(x; a_i, b_i), i = 1,2,\cdots,n$, where BetaPdf stands for the PDF of the beta distribution with parameters $a_i$ and $b_i$. |
| 4: Output $\{f_i\}_{i=1}^n$ |

The whisker parameters of the boxplot-based detectors for the MO- and VO-outliers are set to $r_1 = 1.5$ and $r_2 = 2.5$, respectively. We first directly apply the FDO method to the PDFs (just replace the quantile function $Q_i(t)$ in Eqs. (A-13) and (A-14) by the corresponding PDF $f_i(x)$), the detection results are shown in Fig. A-4. Then, we perform the FDO-based outlier detection in the QF-space with the same parameter settings, the detection results are shown in Fig. A-5. Unfortunately, excepting the QF-space detection for Model I exhibits a satisfactory detection result, grossly high false detection phenomena occur to the remaining scenarios. Comparing the MS-plots (the scatter plot of VO versus MO) as shown in the second row of Fig. A-4 and Fig. A-5 for the eight cases, it is evident that only the QF-space detection for Model I exhibits normal pattern. Further investigation found that such a poor performance is mainly attributed to the curve overlapping occurred at the lower and upper tails of the collection of the functions (PDFs or QFs).

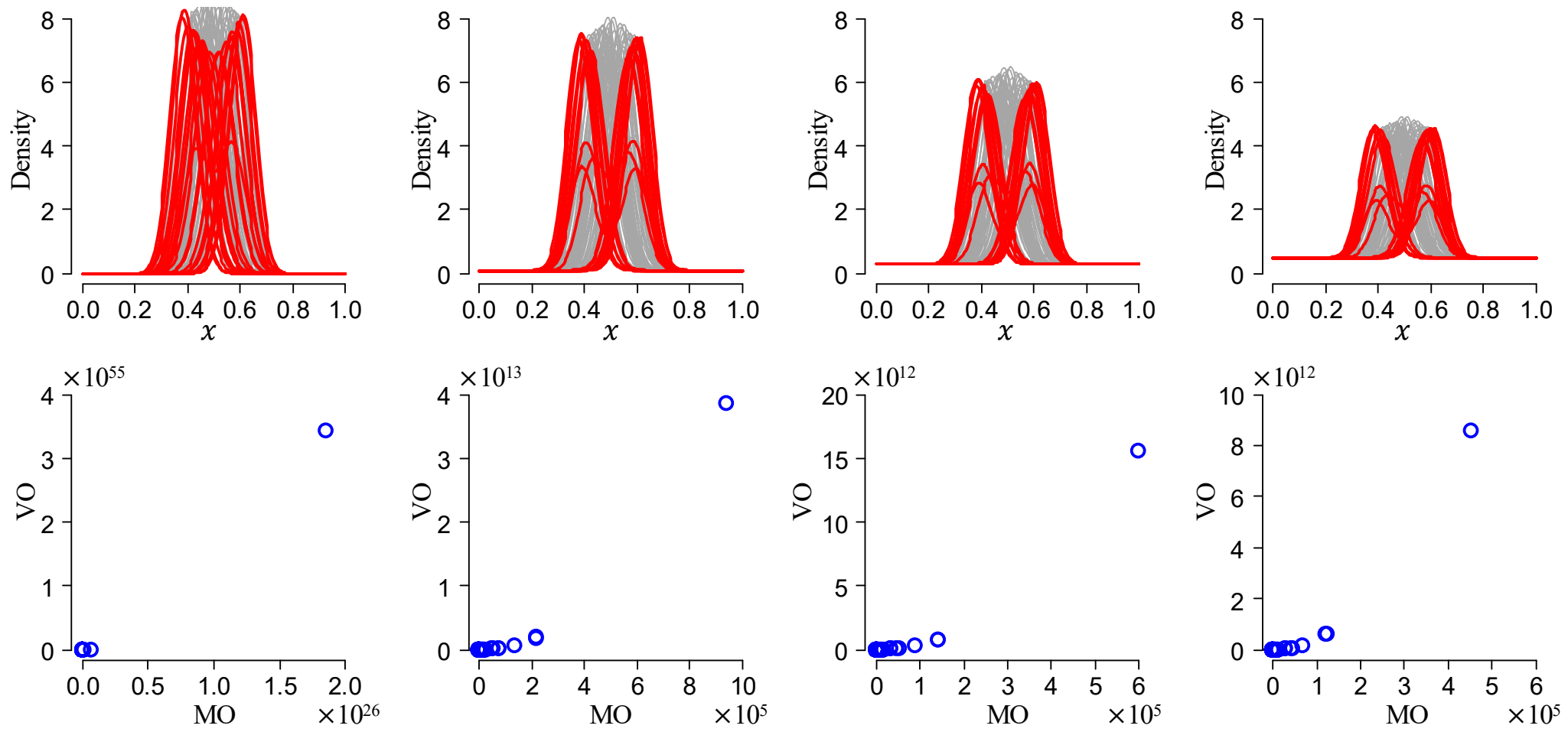

**Fig. A-4.** Outlier detection results for the four datasets presented in Fig. A-3 by directly applying the FDO-based method to PDFs. First row corresponds to the PDFs with the detected outliers represented by red curves, second row corresponds to the associated MS-plots (scatter plots of the MO- and VO-values).

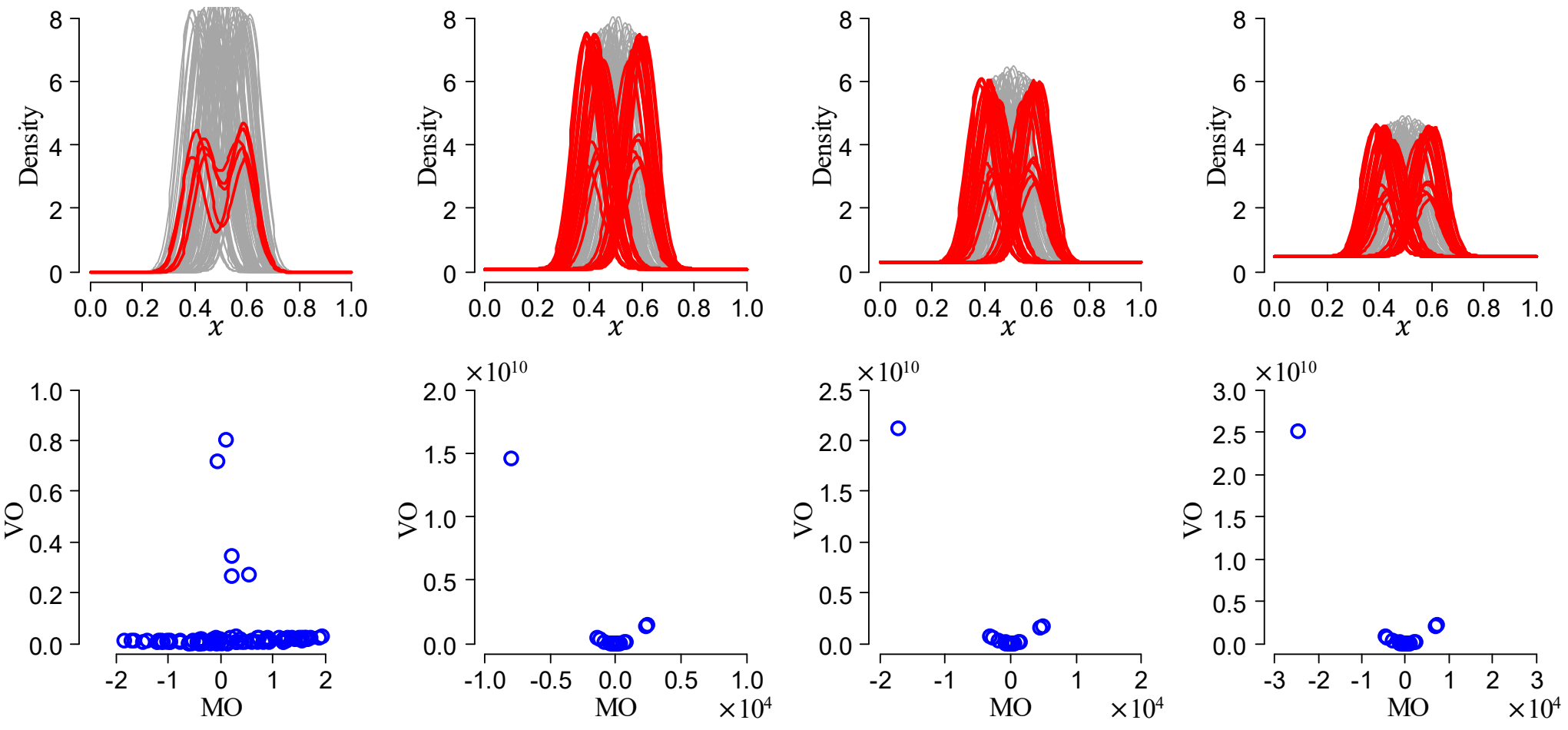

**Fig. A-5.** Same as Fig. A-4 except that the outliers are detected by performing the FDO-based method in the QF-space.

Fortunately, the curves of quantile functions have been naturally aligned according to their quantiles, facilitating us to choose a unified detection interval by truncating the lower and upper parts of $I$ (i.e., the domain of definition of the quantile function). Denote the selected detection interval as $A = [\omega, 1-\omega]$; the curves outside this interval are truncated as illustrated in the left panel of Fig. A-6, which is equivalent to cut off the lower and upper tails of the PDFs as shown in the right panel of Fig. A-6. To apply the FDO-based approach to the resulted quantile functions, we only need to replace the $Q_i(t)$ and $\omega(t)$ in Eq. (A-14) by $Q_i(t)\chi_A(t)$ and $\omega(t) = 1/\lambda(A)$, respectively. We consider three different detection intervals with $\omega$ valued at 0.05, 0.1, 0.2, the corresponding detection results are shown in Fig. A-7 ~ Fig. A-10. Noticeable performance improvements have been observed in several scenarios, but not all cases have obtained the satisfactory results. This simulation study indicates selecting appropriate detection interval is an effective strategy to make the FDO-based detector work well; however, the feasible detection intervals for different scenarios are different, it is hard for the user to select a common detection interval universally suited for all functional data.

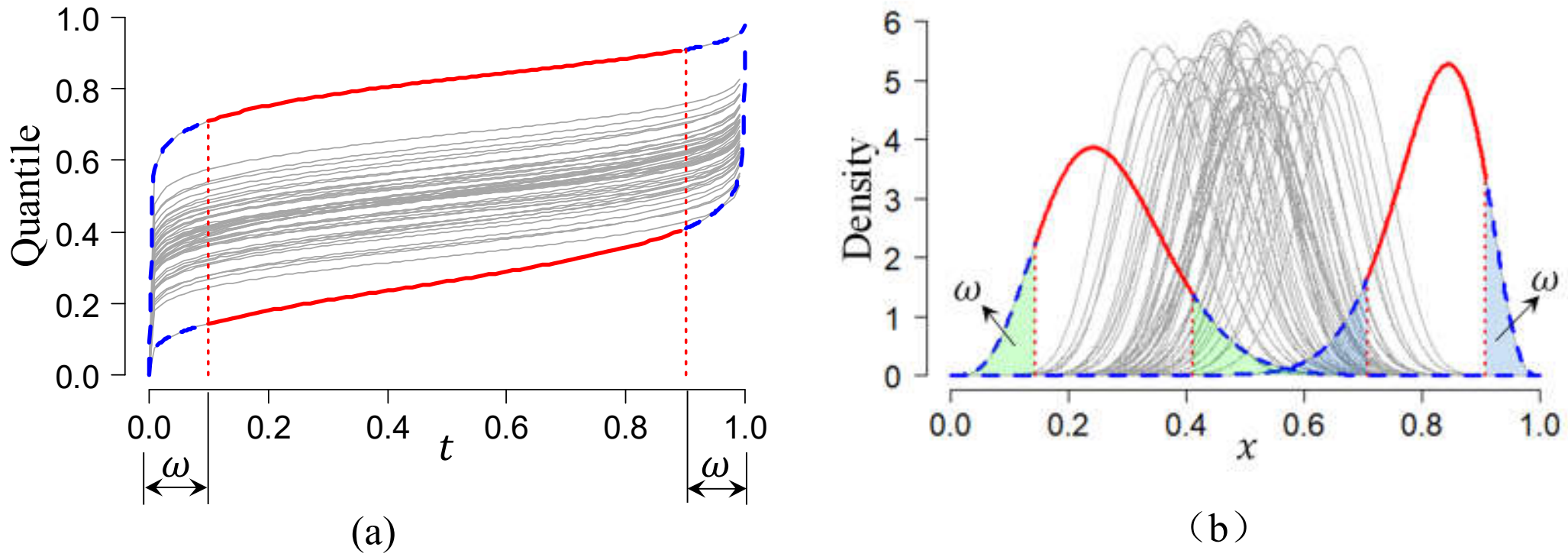

**Fig. A-6.** Illustration of curve truncation using the selected detection region. (a) in quantile function space and (b) in PDF space.

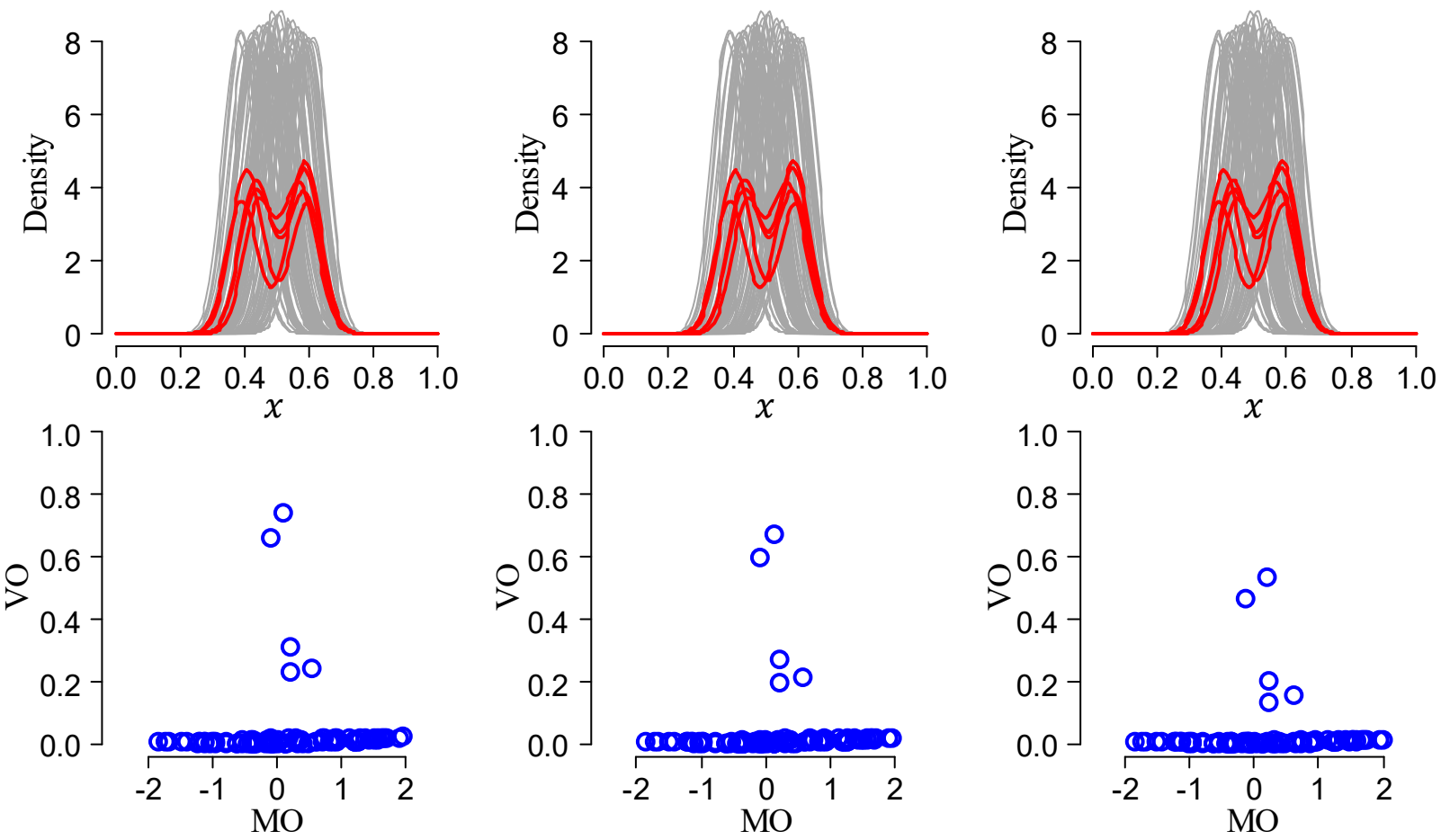

**Fig. A-7.** Outlier detection results for the datasets of Model I shown in Fig. A-3 by performing the FDO-based method in the QF-space using three different detection regions. First column corresponds to detection region [0.05, 0.95], second column corresponds to detection region [0.1, 0.9], third column corresponds to detection region [0.2, 0.8].

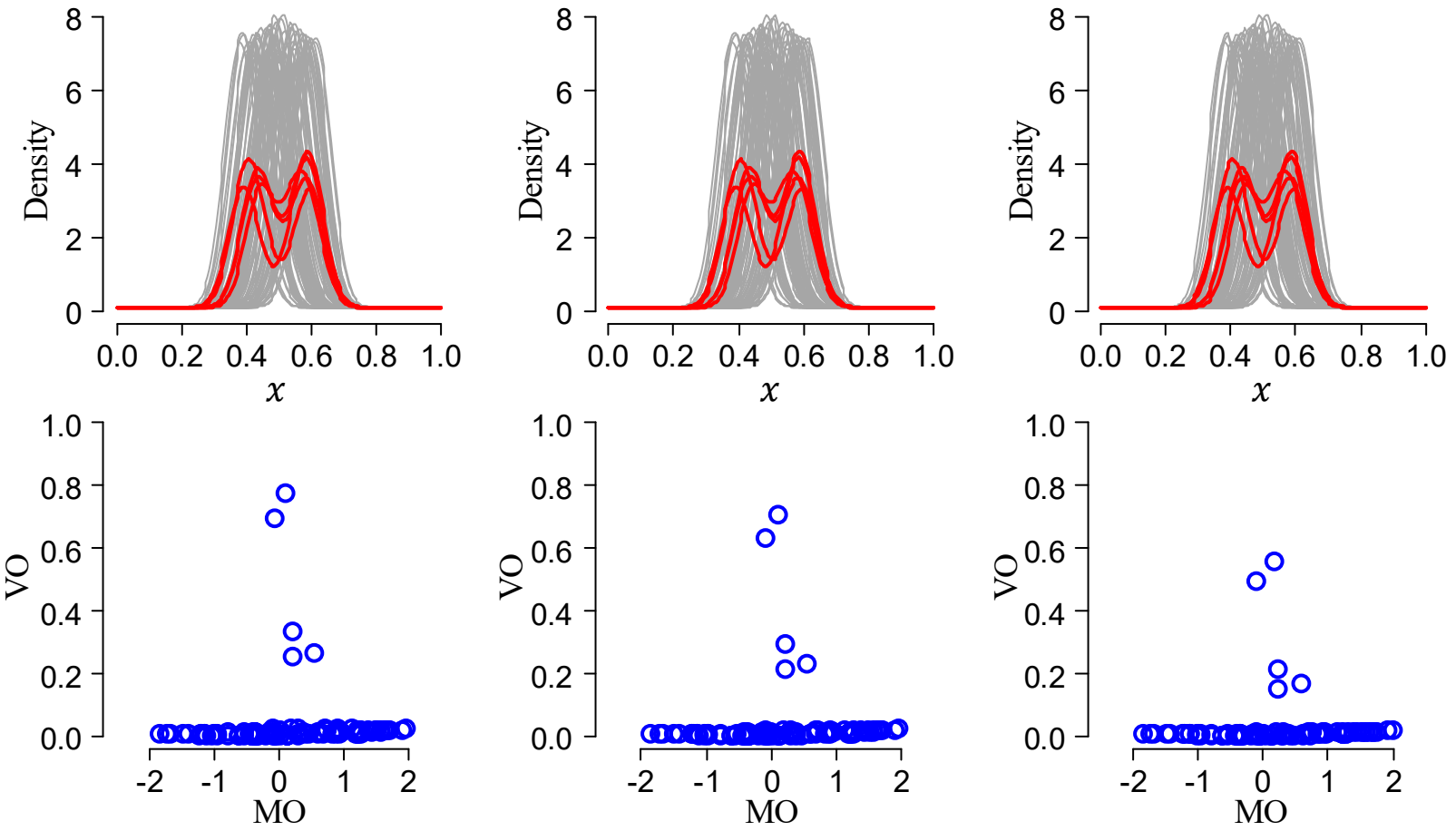

**Fig. A-8.** Same as Fig. A-7 except that the outlier detection is performed to the PDF dataset of Model II.

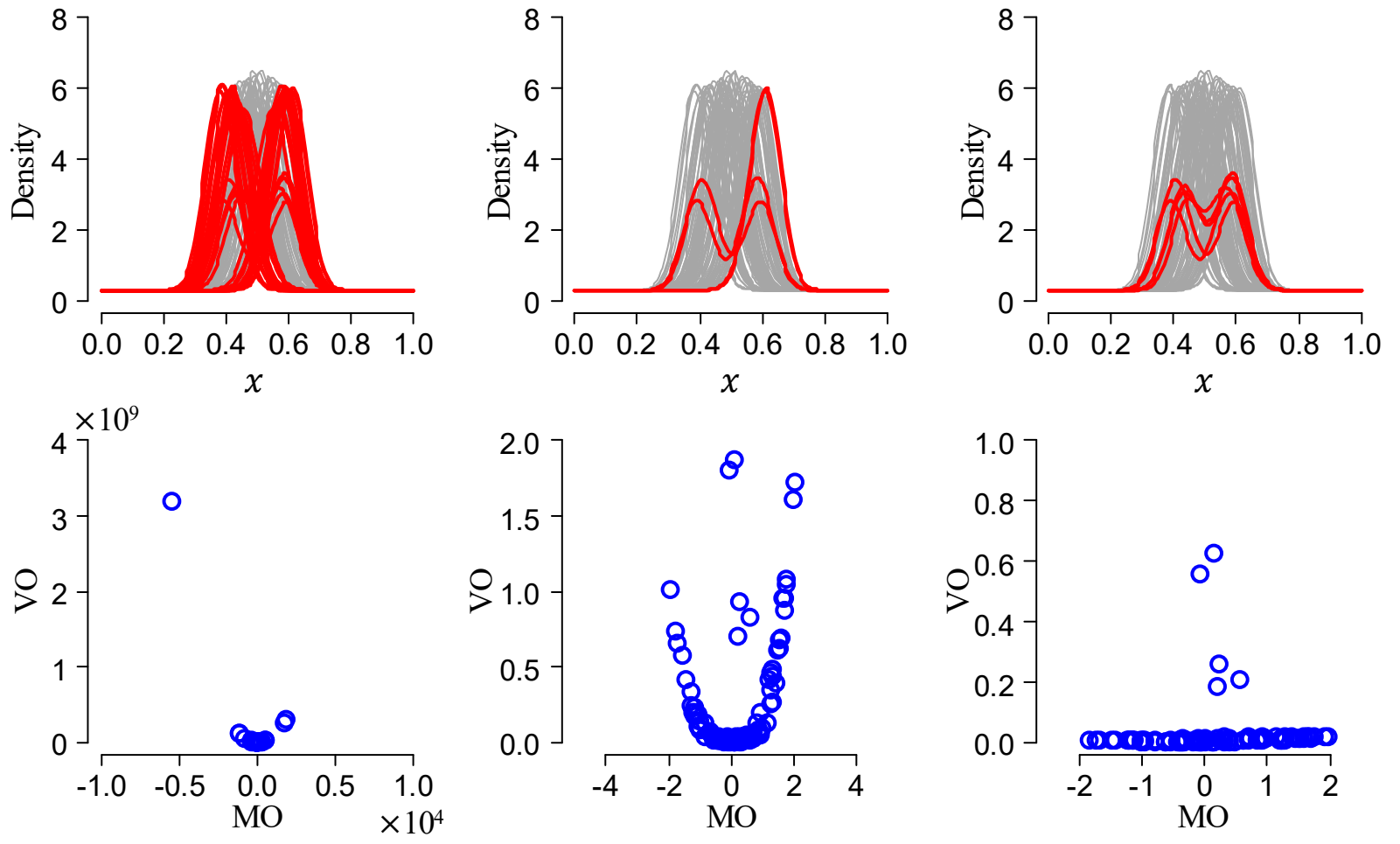

**Fig. A-9.** Same as Fig. A-7 except that the outlier detection is performed to the PDF dataset of Model III.

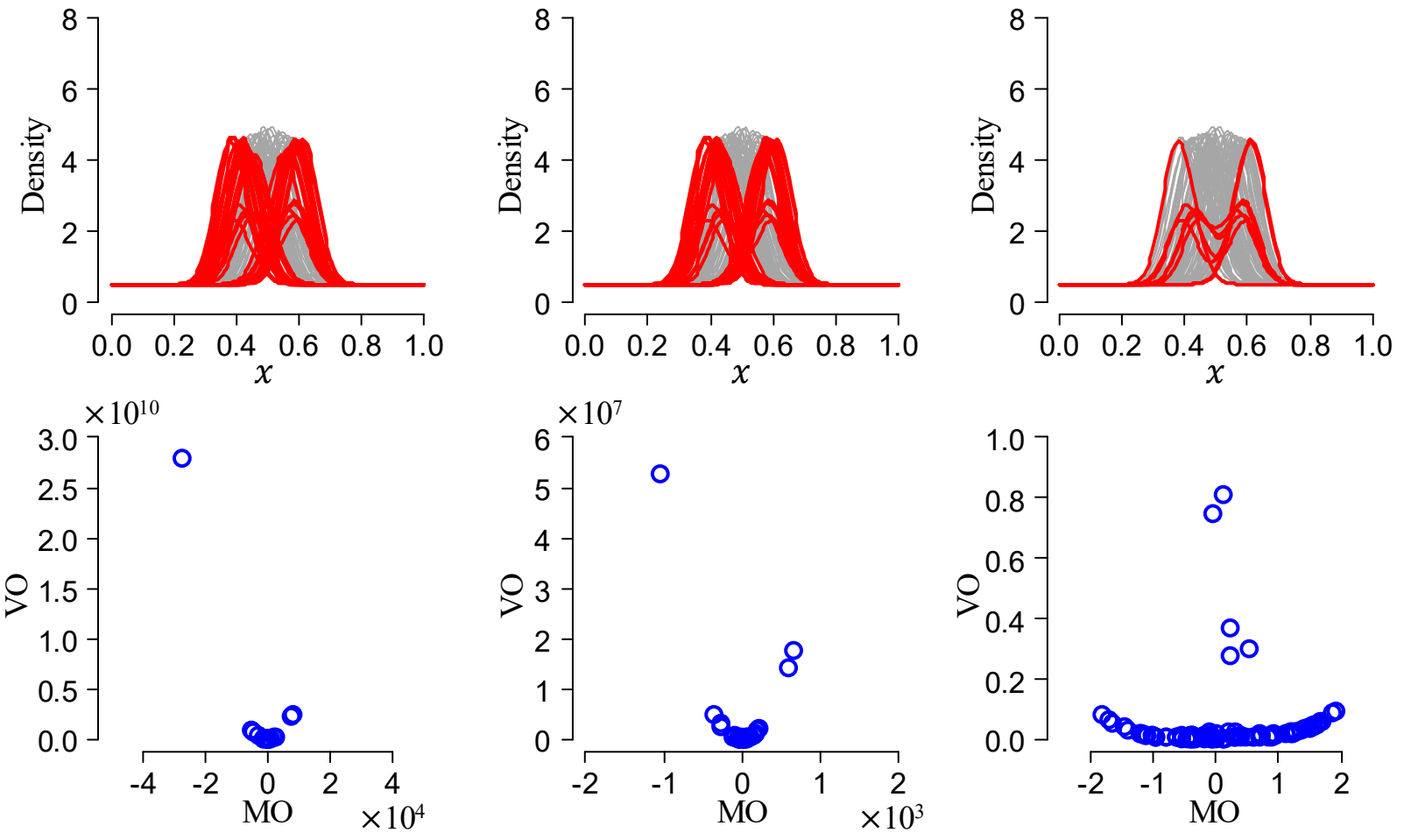

**Fig. A-10.** Same as Fig. A-7 except that the outlier detection is performed to the PDF dataset of Model IV.

In summary, the FDO-based approach can effectively identify the unexposed shape outliers in the bulk of the functional data, which is potentially complementary to the distance-based approach; however, it may also suffer from the risk of instability, an improper detection interval usually leads to poor performance with high false positives. Fortunately, the proposed multiple detection approach described in Subsection 2.3 of the manuscript can provide an elegant strategy to fuse the distance- and FDO-based detectors as well as effectively address the unstable detection.

### *A.3.4. Supplemental materials for the regression outlier detection method*

#### *A.3.4.1 Regularization parameter determination for the LQD-RKHS DtDR model*

This subsection introduces the generalized cross-validation (GCV) for adaptively choosing the regularization parameter in the LQD-RKHS distribution-to-distribution regression (DtDR) model. According to the theory of the GCV (Wahba, 1990; Lian, 2007a), the regularization parameter for the LQD-RKHS DtDR model can be chosen as follows:

$$\lambda_{GCV} = \underset{\lambda_s>0}{\operatorname{argmin}} \frac{\frac{1}{n}\|(\boldsymbol{I}-\boldsymbol{A}(\lambda_s))\operatorname{vec}(\boldsymbol{Y})\|_2^2}{\left[\frac{1}{n}\operatorname{trace}(\boldsymbol{I}-\boldsymbol{A}(\lambda))\right]^2} \quad \text{with } \boldsymbol{A}(\lambda_s) := (\boldsymbol{K}\otimes\boldsymbol{A})[(\boldsymbol{K}\otimes\boldsymbol{A})+\lambda_s\boldsymbol{I}]^{-1} \tag{A-15}$$

where $\boldsymbol{A}$, $\boldsymbol{Y}$ and $\boldsymbol{K}=\boldsymbol{I}_{m\times m}$ are matrixes corresponding to those in Eq. (24) of Chen et al. (2019a), $\boldsymbol{I}$ is an identity matrix of size $mn\times mn$ ($m$ is the FPCA order in the LQD-RKHS DtDR model and $n$ is the number of training PDFs). $\operatorname{GCV}(\lambda_s) = \frac{\frac{1}{n}\|(\boldsymbol{I}-\boldsymbol{A}(\lambda_s))\operatorname{vec}(\boldsymbol{Y})\|_2^2}{\left[\frac{1}{n}\operatorname{trace}(\boldsymbol{I}-\boldsymbol{A}(\lambda_s))\right]^2}$ is also called the GCV statistic. Similar to Lian (2007a), we can compute the GCV statistic at a pre-specified grid of $\lambda_s$ denoted as $\mathcal{C}(\lambda)=\{\lambda_{s,1},\cdots,\lambda_{s,N_\lambda}\}$, then the regularization parameter can be effectively estimated as

$$\lambda_{GCV} = \underset{\lambda_s\in\mathcal{C}(\lambda)}{\operatorname{argmin}}\{\operatorname{GCV}(\lambda_{s,1}),\cdots,\operatorname{GCV}(\lambda_{s,N_\lambda})\} \tag{A-16}$$

*A.3.4.2 Residual calculation using the clr transformation*

The regression error is inevitable, which means the fitted PDF (obtained by the regression model) may deviate from the target PDF. Note that the regression outlier detection is based on residual outlier detection. The horizontal deviation of the predicted PDFs usually will lead to large residual for the PDFs, especially for the “slim” PDF (will be illustrated later). Such a phenomenon can significantly increase the risk of false detection. In this study, we adopt a median alignment (different from the elastic alignment using warping functions described latter in Section A.5) strategy to remedy this issue. Specifically, let $f_i$ denote the target PDF defined on the compact interval $[0,1]$, and let $\mathrm{med}(f_i)$ denote the median defined as

$$\mathrm{med}(f_i) = \inf\left(\left\{x \in [0,1]:\ \int_0^x f_i(t)dt \geq \frac{1}{2}\right\}\right) \tag{A-17}$$

Moreover, let $\hat{f}_i$ denote the fitted result of $f_i$ obtained by the distributional regression model. If $f_i$ and $\hat{f}_i$ are close enough in the horizontal direction (i.e., $|\mathrm{med}(f_i) - \mathrm{med}(\hat{f}_i)| \leq \theta_h$ with $\theta_h$ being a pre-specified threshold), the residual of $f_i$ (with respect to $\hat{f}_i$) will be calculated after horizontally translating $f_i$ to $\hat{f}_i$ to make their median points coincide with each other as illustrated in Fig. A-11, where $f_i^{\#}$ stands for the result of $f_i$ after movement. Based on such a median alignment strategy, the implementation of PDF-residual calculation using the Bayes distance is outlined in Algorithm A. 2.

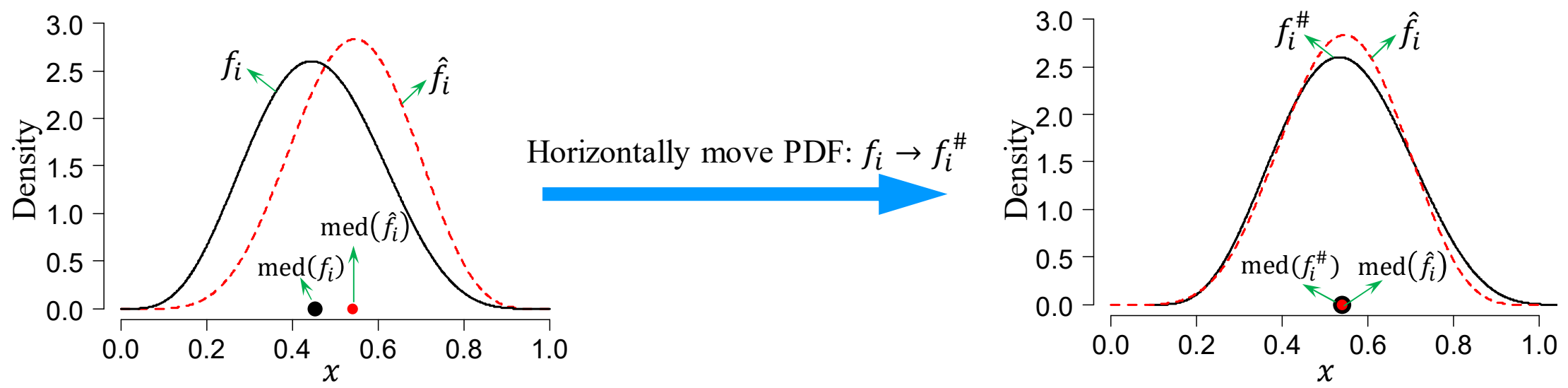

**Fig. A-11.** Illustration of median alignment for two PDFs.

**Algorithm A. 2** Calculation for $\varepsilon_i^{\mathfrak{B}} = d_{\mathfrak{B}}(\hat{f}_i, f_i | \theta_h, \alpha_{mix}^{\mathfrak{B}})$

**Input:** PDF pairs $(\hat{f}_i, f_i)$, threshold $\theta_h$, PDF preprocessing parameter $\alpha_{mix}^{\mathfrak{B}}$

**Output:** Residual $\varepsilon_i^{\mathfrak{B}}$

1: Calculate the medians $\text{med}(f_i)$ and $\text{med}(\hat{f}_i)$

2: **If** $|\text{med}(f_i) - \text{med}(\hat{f}_i)| \leq \theta_h$ **then**

a: Horizontally translate $f_i$ to $\hat{f}_i$ to make their median points coincide with each other, denote the translated PDF as $f_i^{\#}$

b: Find the common support of $f_i^{\#}$ and $\hat{f}_i$, denote the common support as $[c_1, c_2]$

c: Set $f_i^{\#*}(x) = (1 - \alpha_{mix}^{\mathfrak{B}}) f_i^{\#}(x) \chi_{[c_1,c_2]}(x) + \alpha_{mix}^{\mathfrak{B}}$

$\hat{f}_i^*(x) = (1 - \alpha_{mix}^{\mathfrak{B}}) \hat{f}_i(x) \chi_{[c_1,c_2]}(x) + \alpha_{mix}^{\mathfrak{B}}$

where $\chi_{[c_1,c_2]}(\cdot)$ denote the indicator function d: Compute $\varepsilon_i^{\mathfrak{B}} = d_{\mathfrak{B}}(f_i^{\#*}, \hat{f}_i^*) = d_{L_2}(\text{clr}[f_i^{\#*}], \text{clr}[\hat{f}_i^*])$

**else**

a: Set $f_i^*(x) = (1 - \alpha_{mix}^{\mathfrak{B}}) f_i(x) + \alpha_{mix}^{\mathfrak{B}}$, $\hat{f}_i^*(x) = (1 - \alpha_{mix}^{\mathfrak{B}}) \hat{f}_i(x) + \alpha_{mix}^{\mathfrak{B}}$

b: Compute $\varepsilon_i^{\mathfrak{B}} = d_{\mathfrak{B}}(f_i^*, \hat{f}_i^*) = d_{L_2}(\text{clr}[f_i^*], \text{clr}[\hat{f}_i^*])$

**end if**

3: Output $\varepsilon_i^{\mathfrak{B}}$

In the following, we use an example to illustrate the negative effects (in regression outlier detection) caused by the horizontal-shift error, as well as to validate the effectiveness of the recommended remedy. We select 50 PDF-valued samples (corresponding to the response variable) in a regression outlier detection test for demonstration, of which 3 PDFs indexed by $i = 17, 39 \text{ and } 42$ are synthetic abnormal PDFs with their shapes significantly differing from the majority of the data to serve as the regression outliers. After fitting the regression model, we first calculate the residuals for the PDFs without considering the treatment of the median alignment. This can be easily achieved by setting $\theta_h = 0$ in Algorithm A. 2 (the PDF preprocessing parameter $\alpha_{mix}^{\mathfrak{B}}$ is set to 0.1) and the results are shown in Fig. A-12. The comparison of the fitting results for eight selected PDF-valued samples (marked in Fig. A-12 by $i = 6, 17, 18, 22, 27, 39, 42 \text{ and } 49$) are visualized in Fig. A-13. Except the three outlying PDFs (i.e., $f_{17}$, $f_{39}$ and $f_{42}$), two non-outlying PDFs (i.e., $f_6$ and $f_{18}$) also have high residuals which behave like outliers in the residual plot shown in Fig. A-12. It can be seen from Fig. A-13 that the shapes of the fitted PDFs of $f_6$ and $f_{18}$ are highly similar to the target PDFs. Obviously, the high residuals of $f_6$ and $f_{18}$ are attributed to the horizontal shift, which will adversely affect the outlier detection in terms of increasing the risk of false detection. Then, we set $\theta_h = 0.2$ and rerun Algorithm A. 2, the resulting residuals are shown in Fig. A-14. As expected, the negative effects induced by the horizontal-shift error have disappeared, only the three outlying PDFs still hold high residuals.

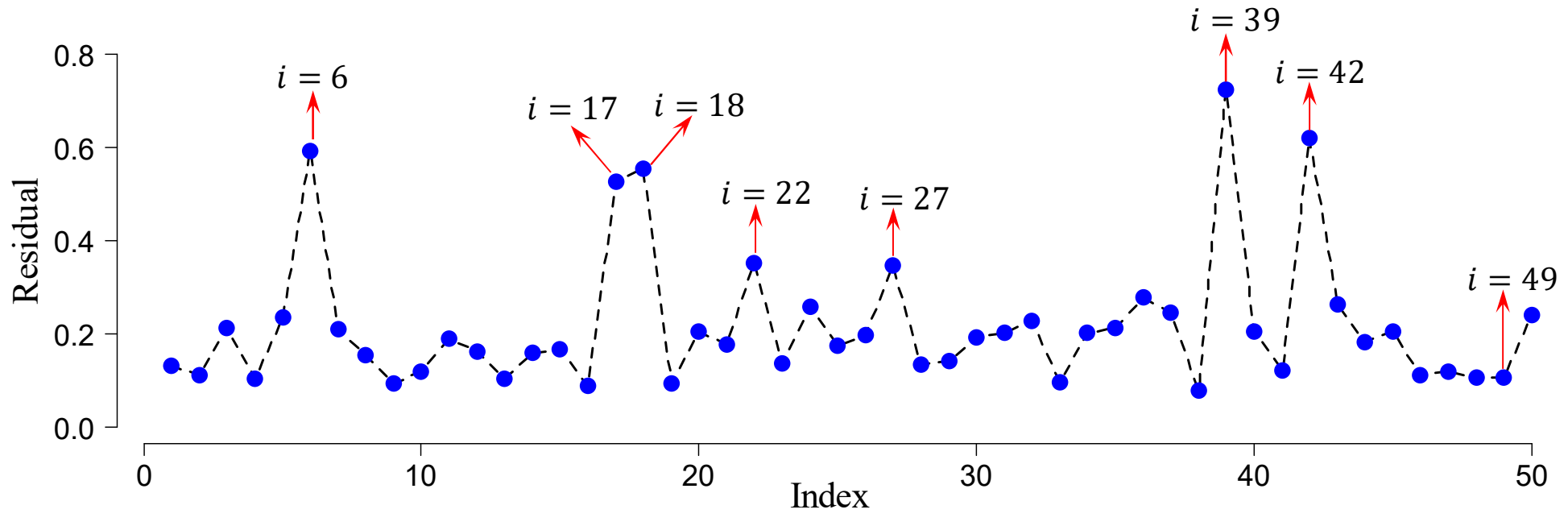

**Fig. A-12.** Residual series $\{\varepsilon_i^{\mathfrak{B}}\}_{i=1}^{30}$ calculated by Algorithm A. 2 with $\theta_h = 0$ and $\alpha_{mix}^{\mathfrak{B}} = 0.1$.

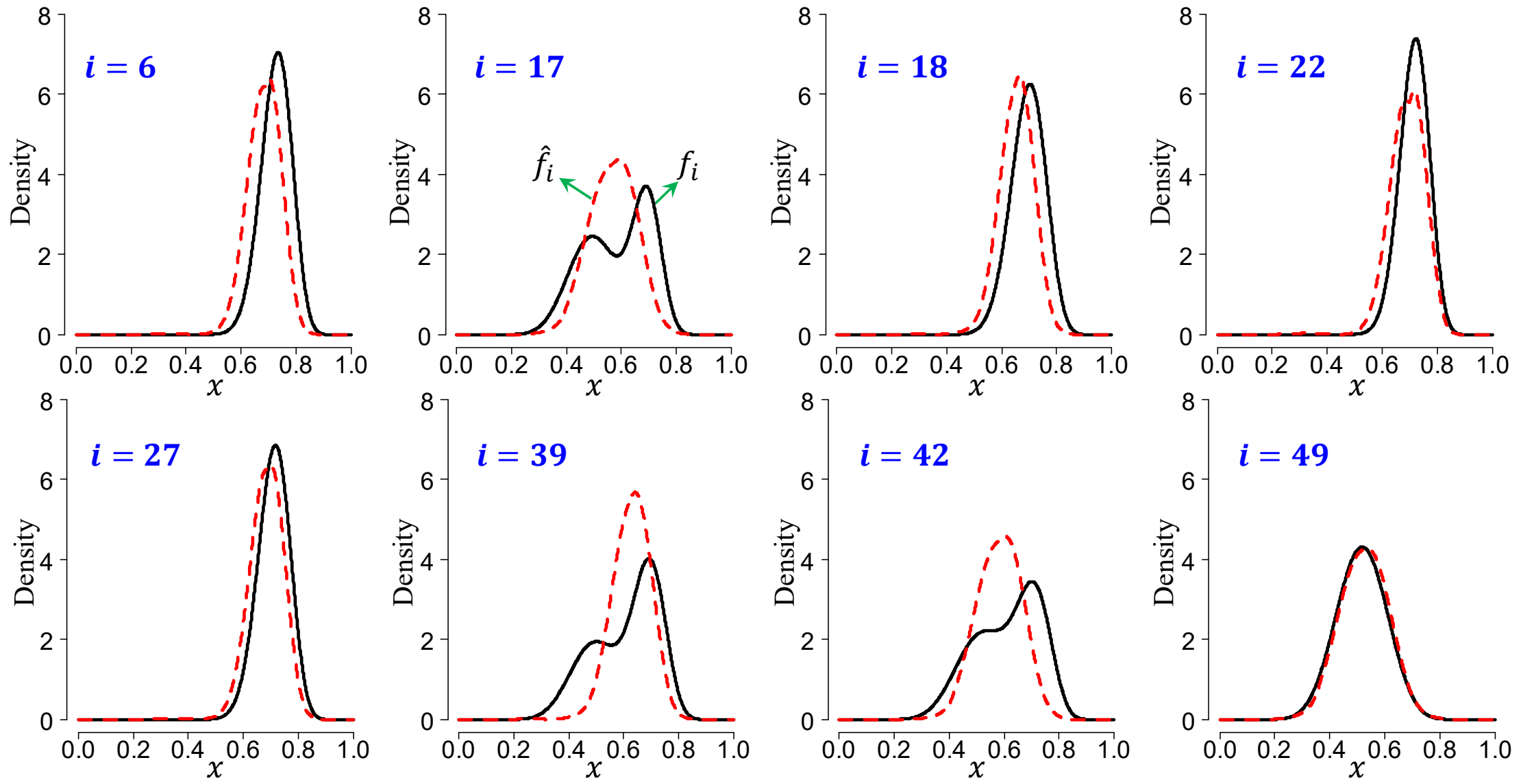

**Fig. A-13.** Comparisons for eight selected PDF samples, the solid line represents the observed PDF to be detected, while the dashed line represents the fitted PDF obtained by the regression model.

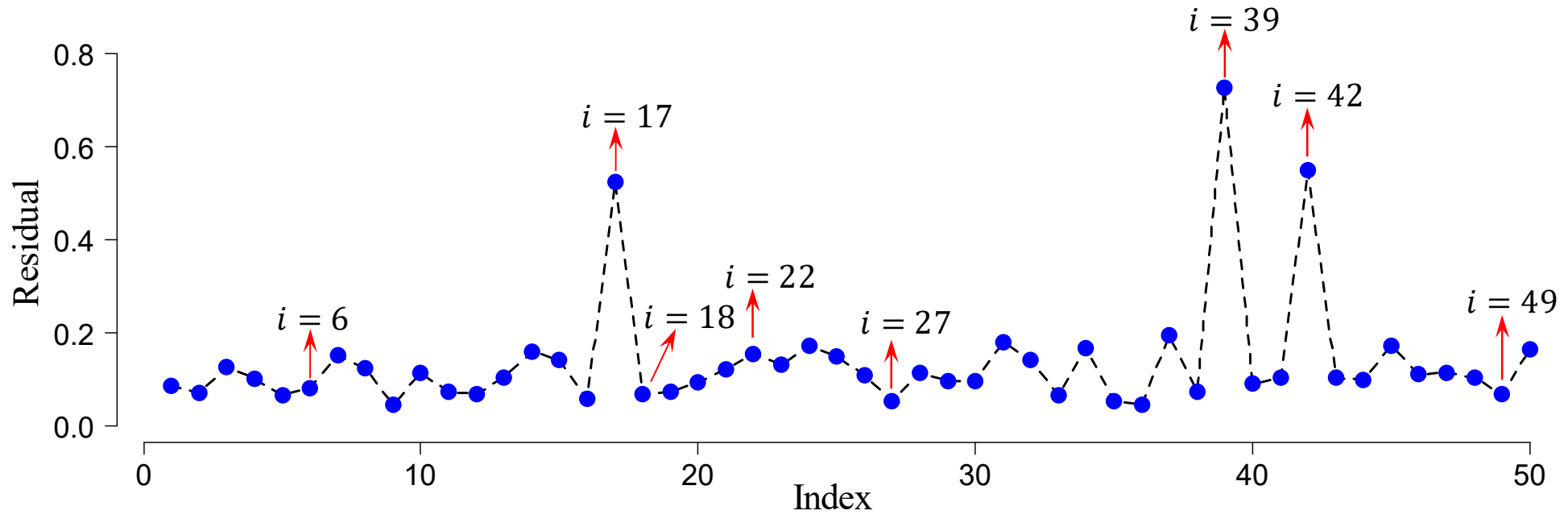

**Fig. A-14.** Residual series $\{\varepsilon_i^{\mathfrak{B}}\}_{i=1}^{30}$ calculated by Algorithm A. 2 with $\theta_h = 0.2$ and $\alpha_{mix}^{\mathfrak{B}} = 0.1$.

*A.3.4.3 Residual calculation using the LQD transformation*

The procedure for calculating $\varepsilon_i^{\mathrm{LQD}}$ is outlined in Algorithm A. 3. As aforementioned, the LQD transformation is insensitive to the horizontal translation of PDFs; thus, we do not consider the aforementioned median alignment in residual calculation when using LQD transformation.

---

**Algorithm A. 3** Calculation for $\varepsilon_i^{\mathrm{LQD}} = d_{L_1}\left(\mathrm{LQD}\left[\hat{f}_i|\alpha_{mix}^{\mathrm{LQD}}\right], \mathrm{LQD}\left[f_i|\alpha_{mix}^{\mathrm{LQD}}\right]\right)$

---

Input: PDF pairs $\left(\hat{f}_i, f_i\right)$, PDF preprocessing parameter $\alpha_{mix}^{\mathrm{LQD}}$

Output: Residual $\varepsilon_i^{\mathrm{LQD}}$

1: Set $f_i^*(x) = \left(1 - \alpha_{mix}^{\mathrm{LQD}}\right)f_i(x) + \alpha_{mix}^{\mathrm{LQD}}$, $\hat{f}_i^*(x) = \left(1 - \alpha_{mix}^{\mathrm{LQD}}\right)\hat{f}_i(x) + \alpha_{mix}^{\mathrm{LQD}}$

2: Compute $\psi_i^*(t) = -\log\{f_i^*(Q_i^*(t))\}$, $\hat{\psi}_i^*(t) = -\log\left\{\hat{f}_i^*\left(\hat{Q}_i^*(t)\right)\right\}$, where $Q_i^*$ and $\hat{Q}_i^*$ are the quantile functions associated with $f_i^*$ and $\hat{f}_i^*$, respectively

3: Compute $\varepsilon_i^{\mathrm{LQD}} = \int_0^1 \left|\psi_i^*(\tau) - \hat{\psi}_i^*(\tau)\right| d\tau$

4: Output $\varepsilon_i^{\mathrm{LQD}}$

---

## A.4. Supplemental materials for robust distributional regression

### *A.4.1. Weight design for robust regression operator estimation*

This subsection discusses how to design the weights used in Eq. (11) (of the manuscript) for dampening the impacts of functional outliers on the distributional regression operator estimation.

Let $\mathcal{G}_{tr} = \{g_1, g_2, \cdots, g_n\}$ and $\mathcal{F}_{tr} = \{f_1, f_2, \cdots, f_n\}$ denote the PDF-valued training samples (corresponding to the predictor and response variables, respectively) used for fitting the distribution-to-distribution regression model. As described in the manuscript, the robustness of the regression model is achieved by downweighting the detected outliers. For this purpose, we perform a two-stage outlier detection to the training samples:

(i) Single dataset outlier detection: outlier detections for the datasets $\mathcal{G}_{tr}$ and $\mathcal{F}_{tr}$ are conducted independently by selecting appropriate detection methods described in Subsection 2.2 (e.g., the Tree-Distance method, the QF-FDO method or a combination of them)

(ii) Regression outlier detection: the outlier detection for the datasets $\mathcal{G}_{tr}$ and $\mathcal{F}_{tr}$ is conducted jointly by using the distributional-regression-based approach described in Subsection 2.4 after the outliers detected in the first stage have been removed.

For convenience, the outliers detected in the first and second stages are called Type I and Type II outliers, respectively. The weights associated with these two types of outliers are designed independently, then combine them to form the final weights, to which we now turn.

*(1) Weight design for Type I outliers*

We consider using the degrees of anomalies computed based on the LQD and clr transformations to design the desired weights for Type I outliers.

We select dataset $\mathcal{G}_{tr} = \{g_1, g_2, \cdots, g_n\}$ to illustrate the weight design procedure, the weights associated with the other dataset (i.e., $\mathcal{F}_{tr} = \{f_1, f_2, \cdots, f_n\}$) can be designed in a similar way.

Let $\Psi_{\mathcal{G}} = \{\psi_1^g, \psi_2^g, \cdots, \psi_n^g\}$ be the functional dataset composed of the LQD

transformations (computed using Eq. (A-8), and the PDF preprocessing parameter $\alpha$ described in Subsection A.2.1 is set to $10^{-10}$) of the elements in $\mathcal{G}_{tr}$. The center function of the functional dataset $\Psi_{\mathcal{G}}$ is estimated by the cross-sectional median function computed by $\psi_c^g(t) = \underset{1\le k\le n}{\text{median}}\{\psi_k^g(t)\}$, $\forall t \in [0,1]$.Then, the dissimilarity of the curve $\psi_i^g$ w.r.t. the center function $\psi_c^g$ is quantified by the $L_1$ distance, and the result is denoted as $\delta_i^g = d_{L_1}(\psi_i^g, \psi_c^g)$. Similarly, let $\mathcal{G}_{tr}^{clr} = \{g_1^{clr}, g_2^{clr}, \cdots, g_n^{clr}\}$ be the functional dataset composed of the clr transformations (computed using Eq.(A-3) with the default PDF preprocessing described in Subsection A.2.2) of the elements in $\mathcal{G}_{tr}$, i.e., $g_i^{clr} = \text{clr}[g_i]$, $i = 1,2,\cdots,n$; then, the dissimilarity of the curve $g_i^{clr}$ w.r.t. the corresponding center function $g_c^{clr}$ (computed in a similar way with $\psi_c^g$) is quantified by the $L_2$ distance, and the result is denoted as $\sigma_i^g = d_{L_2}(g_i^{clr}, g_c^{clr})$.

Let $\mathcal{O}_I(\mathcal{G}) \subset \{1,2,\cdots,n\}$ stand for the index set of the detected Type I outliers contained in $\mathcal{G}_{tr}$. Then, the weight associated with the PDF $g_i$ can be designed as

$$w_I(g_i) = \begin{cases} \left(1 + \dfrac{|\delta_i^g - m(\delta^g)|}{\text{MAD}(\delta^g)}\right)^{-\rho_1} \left(1 + \dfrac{|\sigma_i^g - m(\sigma^g)|}{\text{MAD}(\sigma^g)}\right)^{-\rho_1}, & i \in \mathcal{O}_I(\mathcal{G}) \\ 1, & \text{otherwise} \end{cases} \tag{A-18}$$

where $m(\delta^g) = \underset{1\le i\le n}{\text{median}}\{\delta_i^g\}$ is the median of the scalar dataset $\{\delta_i^g\}_{i=1}^n$, and $\text{MAD}(\delta^g)$ stands for the median absolute deviation (MAD) calculated by $\text{MAD}(\delta^g) = c \cdot \underset{1\le k\le n}{\text{median}}\{|\delta_k^g - m(\delta^g)|\}$ ($c$ is a constant, and we set it to be its default value 1.4826 throughout this study), $\rho_1$ is a user prescribed tuning factor, $m(\sigma^g)$ and $\text{MAD}(\sigma^g)$ have the similar meanings with $m(\delta^g)$ and $\text{MAD}(\delta^g)$, respectively. The tuning parameter $\rho_1$ controls the decay rate of the weight function as illustrated in Fig. A-15.

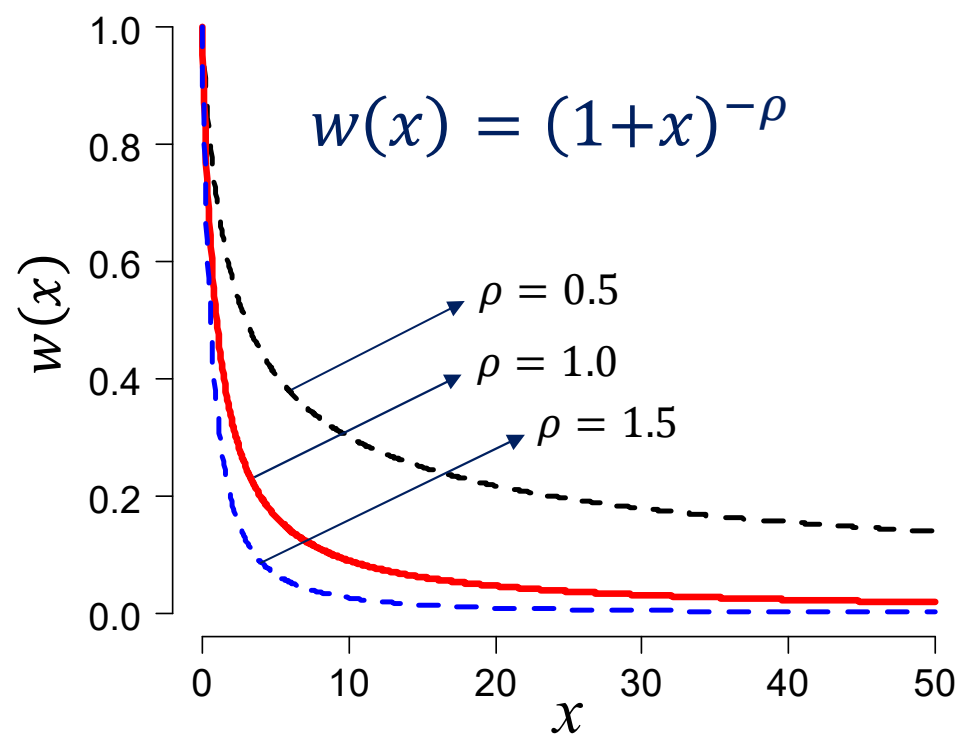

**Fig. A-15.** Schematic illustration of the function $w(x) = (1 + x)^{-\rho}$ with different decay parameters.

Similarly, we can design the weight associated with the PDF $f_i$ based on the Type I outliers detected in $\mathcal{F}_{tr}$, and the result is denoted as $w_I(f_i)$. The final weight associated with the $i$th training sample $\{g_i, f_i\}$ for downweighting the effect of Type I outliers can be obtained by fusing $w_I(g_i)$ and $w_I(f_i)$ as follows (similar to that in Martínez-Hernández et al. (2019)):

$$w_I^i = w_I(g_i) \cdot w_I(f_i), \quad i = 1,2,\cdots,n \tag{A-19}$$

*(2) Weight design for Type II outliers*

We select the dataset $\mathcal{F}_{tr} = \{f_1, f_2, \cdots, f_n\}$ to illustrate the weight design procedure, the weights associated with the other dataset (i.e., $\mathcal{G}_{tr} = \{g_1, g_2, \cdots, g_n\}$) can be designed in a similar way.

Let $\varepsilon_i^{\mathfrak{B}}$ and $\varepsilon_i^{LQD}$ be the calculated residuals associated with the PDF $f_i \in \mathcal{F}_{tr}$ by using the Algorithm A. 2 and Algorithm A. 3, respectively. It is worth noting that the results of $\varepsilon_i^{\mathfrak{B}}$ and $\varepsilon_i^{LQD}$ actually have been calculated in the residual diagnosis of the regression outlier detection stage, see Line 3 of Algorithm 2 in the manuscript. Before using the residuals to design the desired weights, we perform a normalization processing to them as follows:

$$\begin{aligned} \tilde{\varepsilon}_i^{\mathfrak{B}} &= \frac{\varepsilon_i^{\mathfrak{B}} - \min\limits_{1 \le k \le n} \varepsilon_k^{\mathfrak{B}}}{\max\limits_{1 \le k \le n} \varepsilon_k^{\mathfrak{B}} - \min\limits_{1 \le k \le n} \varepsilon_k^{\mathfrak{B}}}, \quad i = 1,2,\cdots,n \\ \tilde{\varepsilon}_i^{LQD} &= \frac{\varepsilon_i^{LQD} - \min\limits_{1 \le k \le n} \varepsilon_k^{LQD}}{\max\limits_{1 \le k \le n} \varepsilon_k^{LQD} - \min\limits_{1 \le k \le n} \varepsilon_k^{LQD}}, \quad i = 1,2,\cdots,n \end{aligned} \tag{A-20}$$

where $\min\limits_{1 \le k \le n} a_k$ and $\max\limits_{1 \le k \le n} a_k$ stand for the minimum and maximum values of the dataset $\{a_1, a_2, \cdots, a_n\}$.

Let $\mathcal{O}_{II}(\mathcal{F}) \subset \{1,2,\cdots,n\}$ stand for the index set of the type II outliers detected in the PDF-valued two-tuples $\{g_i, f_i\}_{i=1}^n, g_i \in \mathcal{G}_{tr}, f_i \in \mathcal{F}_{tr}$. Then, the weight associated with the PDF $f_i$ can be designed as follows:

$$w_{II}(f_i) = \begin{cases} \left(1 + \dfrac{|\tilde{\varepsilon}_i - m(\tilde{\varepsilon})|}{\mathrm{MAD}(\tilde{\varepsilon})}\right)^{-\rho_2}, & i \in \mathcal{O}_{II}(\mathcal{F}) \\ 1, & \text{otherwise} \end{cases} \tag{A-21}$$

where $\tilde{\varepsilon}_i = \left(\tilde{\varepsilon}_i^{\mathfrak{B}} + \tilde{\varepsilon}_i^{LQD}\right)/2$, $m(\tilde{\varepsilon}) = \underset{1 \le i \le n}{\mathrm{median}}\{\tilde{\varepsilon}_i\}$, and $\mathrm{MAD}(\tilde{\varepsilon})$ is the associated median absolute deviation calculated similarly with its counterpart in Eq.(A-18).

Similarly, we can design the weight associated with the PDF $g_i \in \mathcal{G}_{tr}$ based on the Type II outliers, and the result is denoted as $w_{II}(g_i)$. Then, the final weight associated with the $i$th training sample $\{g_i, f_i\}$ for downweighting the effect of Type II outliers can be also obtained by fusing $w_{II}(g_i)$ and $w_{II}(f_i)$ as follows:

$$w_{II}^i = w_{II}(g_i) \cdot w_{II}(f_i), \quad i = 1,2,\cdots,n \tag{A-22}$$

*(3) Final weight*

On the basis of the weights associated with the Type I and Type II outliers designed above, the final weights used in Eq. (11) (of the manuscript) for downweighting the impacts of detected distributional outliers can be obtained as follows:

$$w_i = w_I^i \cdot w_{II}^i, \quad i = 1,2,\cdots,n \tag{A-23}$$

### *A.4.2. Proof of proposition 1*

Proposition 1 can be proofed in a similar way with Theorem 1 in Lian (2007b).

**Proof:** recall that the regression operator $F_{reg}$ is assumed to reside in the RKHS $\mathcal{H}(K_r)$, i.e., $F_{reg} \in \mathcal{H}(K_r)$. According to the property of operator-valued reproducing kernel given in Eq.(A-6), it follows that $K_r\left(\cdot, \psi_j^{g^*}\right) \in \mathcal{H}(K_r), j = 1,2,\cdots,n$. Thus, $\left\{K_r\left(\cdot, \psi_j^{g^*}\right): j = 1,2,\cdots,n\right\}$ can span a subspace of $\mathcal{H}(K_r)$ as follows:

$$\mathcal{H}_0(K_r) = \left\{\sum_{j=1}^{n} K_r\left(\cdot, \psi_j^{g^*}\right)\boldsymbol{\alpha}_j\,,\ \boldsymbol{\alpha}_j \in H\right\} \tag{A-24}$$

where $H$ stands for another Hilbert space. Let $\mathcal{H}_0^{\perp}(K_r) \subset \mathcal{H}(K_r)$ be the orthogonal complement of $\mathcal{H}_0(K_r)$, then $\forall G \in \mathcal{H}_0^{\perp}(K_r)$, we have for any $\boldsymbol{\alpha}_j \in H$ that

$$\langle K_r\left(\cdot, \psi_j^{g^*}\right)\boldsymbol{\alpha}_j,\ G\ \rangle_{\mathcal{H}(K_r)} = 0, \quad j \in \{1,2,\cdots,n\} \tag{A-25}$$

Moreover, the regression operator $F_{reg} \in \mathcal{H}(K_r)$ can be decomposed as

$$F_{reg} = F_0 + G, \quad F_0 \in \mathcal{H}_0(K_r), G \in \mathcal{H}_0^{\perp}(K_r) \tag{A-26}$$

According to the reproducing property given in Eq.(A-7), it follows that

$$\langle G\left(\psi_j^{g^*}\right),\ \boldsymbol{\alpha}_j\ \rangle_H = \langle K_r\left(\cdot, \psi_j^{g^*}\right)\boldsymbol{\alpha}_j,\ G\ \rangle_{\mathcal{H}(K_r)} = 0 \tag{A-27}$$

In view of the arbitrariness of $\boldsymbol{\alpha}_j$, it follows that $G\left(\psi_j^{g^*}\right) = 0$, thus $F_{reg}\left(\psi_j^{g^*}\right) = F_0\left(\psi_j^{g^*}\right) + G\left(\psi_j^{g^*}\right) = F_0\left(\psi_j^{g^*}\right)$. Further note that $G$ is orthogonal to $F_0$, it has $\left\|F_{reg}\right\|_{\mathcal{H}(K_r)} = \|F_0\|_{\mathcal{H}(K_r)} + \|G\|_{\mathcal{H}(K_r)} > \|F_0\|_{\mathcal{H}(K_r)}$ for $G \neq 0$ . Consequently, it holds the following inequality for the objective function in Eq. (11) of the manuscript：

$$J(F_{reg}) = \sum_{i=1}^{n} w_i \left\| \boldsymbol{\xi}_i^{f^*} - F_{reg}\left(\psi_i^{g^*}\right) \right\|_2^2 + \lambda_s \left\| F_{reg} \right\|_{\mathcal{H}(K_r)}^2$$
$$> \sum_{i=1}^{n} w_i \left\| \boldsymbol{\xi}_i^{f^*} - F_0\left(\psi_i^{g^*}\right) \right\|_2^2 + \lambda_s \| F_0 \|_{\mathcal{H}(K_r)}^2 \tag{A-28}$$

which suggests the optimal solution of the regression operator $F_{reg}$ takes the following general form:

$$F_{reg} = \sum_{j=1}^{n} K_r\left(\cdot, \psi_j^{g^*}\right) \boldsymbol{\alpha}_j \tag{A-29}$$

This completes the proof.

## A.5. Computational details for warping functions

This section provides computational details for the warping functions (used in pairwise alignment of cumulative distribution functions (CDFs)) as well as a strategy for dealing with the related numerical issue.

Given two PDFs $f_1(x)$ and $f_2(x)$ defined on the compact interval $[0,1]$, denote their corresponding CDFs as $F_1(x)$ and $F_2(x)$ (i.e., $F_i(x) = \int_{-\infty}^{x} f_i(\tau)d\tau\,, i = 1,2$ ). The problem of interest in this section is determining the warping functions, denoted as $\gamma_{12}(x)$ and $\gamma_{21}(x)$, subject to

$$\begin{aligned} (F_1 \circ \gamma_{12})(x) &= F_2(x),\ x \in [0,1] \\ (F_2 \circ \gamma_{21})(x) &= F_1(x),\ x \in [0,1] \end{aligned} \tag{A-30}$$

respectively. Theoretically, $\gamma_{12}(x)$ is the inverse element w.r.t. $\gamma_{21}(x)$ and vice versa, thus $\gamma_{12} \circ \gamma_{21} = \gamma_{id}$ with $\gamma_{id}$ being the identity warping function defined as $\gamma_{id}(x) = x, x \in [0,1]$. The warping function $\gamma_{12}$ represents the phase information of $F_1(x)$ w.r.t. $F_2(x)$. From the angle of curve alignment (also termed curve registration), $\gamma_{12}$ plays the role of deforming the shape of $F_1(x)$ to reach the shape of $F_2(x)$. In the community of functional data analysis, similar curve alignment is widely used in phase-amplitude separation (Srivastava et al., 2011; Srivastava and Klassen, 2016), and the extracted phase information captured by warping functions can provide a useful feature in shape outlier detection.

It is worth noting that ordinary functional data have both amplitude and phase variabilities (Srivastava and Klassen, 2016), the warping functions used in pairwise alignments usually have to be solved by the dynamic programming (DP) algorithm in the square root slope velocity function (SRVF) framework described in Srivastava et al. (2011). In contrast, the CDFs only have phase variability, the warping functions can also be directly computed as:

$$\begin{aligned} \gamma_{12}(x) &= (F_1^{-1} \circ F_2)(x),\ x \in [0,1] \\ \gamma_{21}(x) &= (F_2^{-1} \circ F_1)(x),\ x \in [0,1] \end{aligned} \tag{A-31}$$

Compared to the time-consuming DP algorithm, this direct computation approach is much more efficient. Denote the computed result of the warping function $\gamma_{12}$ (or $\gamma_{21}$) as $\hat{\gamma}_{12}$ (or $\hat{\gamma}_{21}$). If the PDFs take values near zero, $\hat{\gamma}_{12}$ and $\hat{\gamma}_{21}$ obtained by direct computation may be problematic, that is, they may fail to satisfy $\hat{\gamma}_{12} \circ \hat{\gamma}_{21} = \gamma_{id}$ as illustrated in Fig. A-16(e).

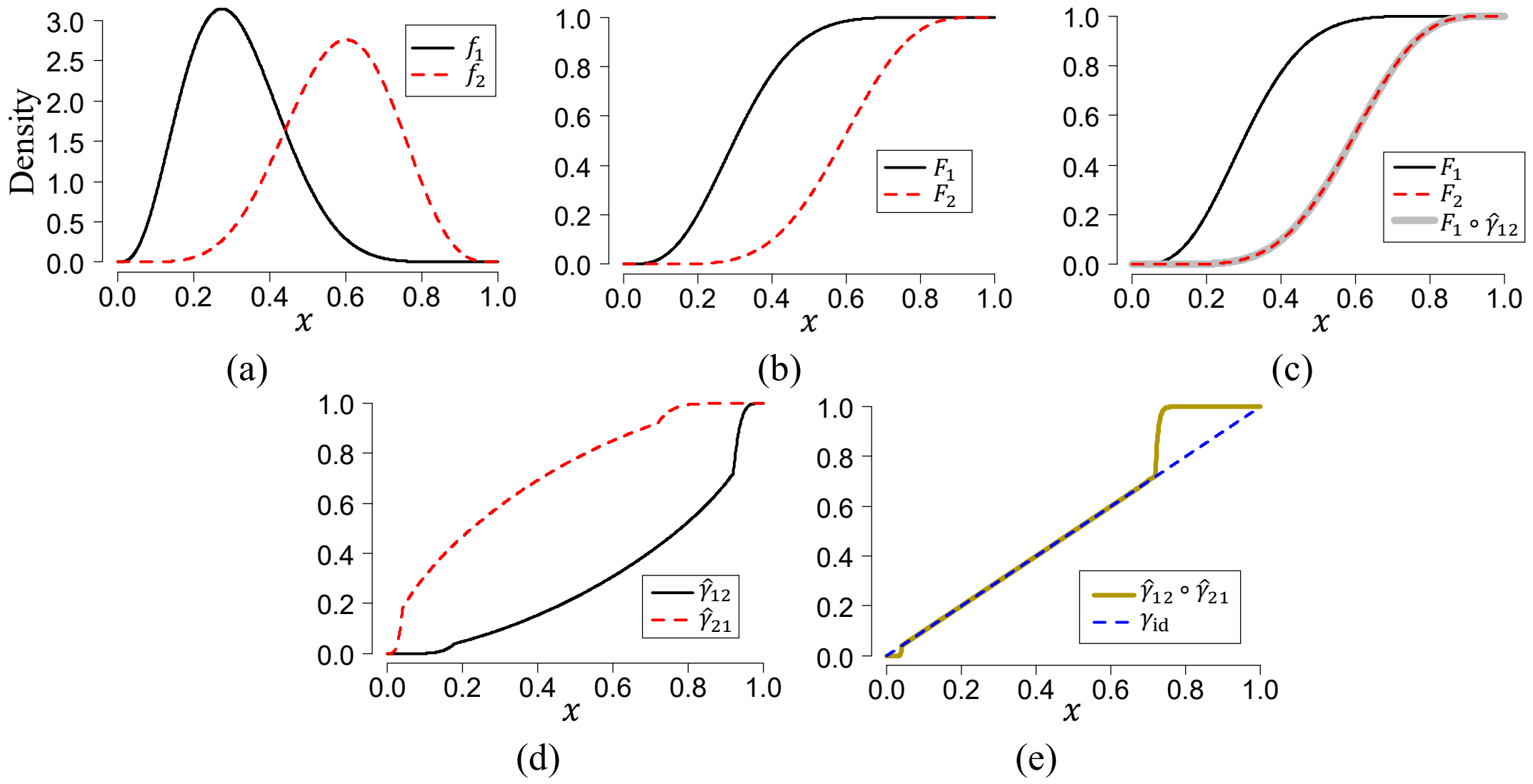

**Fig. A-16.** CDF alignment of the simulated raw distributional data by using direct computed warping function. (a) Original PDFs $f_1$ and $f_2$, (b) corresponding CDFs $F_1$ and $F_2$, (c) comparison of $F_1 \circ \hat{\gamma}_{12}$ and $F_2$, (d) calculated warping functions, and (e) comparison of $\hat{\gamma}_{12} \circ \hat{\gamma}_{21}$ and $\gamma_{id}$.

According to our experience, such a defect can be remedied by adding a small proportion of uniform distribution to the original distributions, that is, the PDFs are proposed to be preprocessed as follows:

$$f_i(x) = (1-\alpha)f_i(x) + \alpha,\ x \in [0,1], i = 1,2 \tag{A-32}$$

where $\alpha$ is a positive constant (termed PDF preprocessing parameter). By setting $\alpha = 0.1$, the re-calculated warping functions by direct computation are shown in Fig. A-17(d). Clearly, the line of $\hat{\gamma}_{12} \circ \hat{\gamma}_{21}$ agrees well with the line of $\gamma_{id}$ (see Fig. A-17(e)), indicating the numerical issue has been effectively overcome.

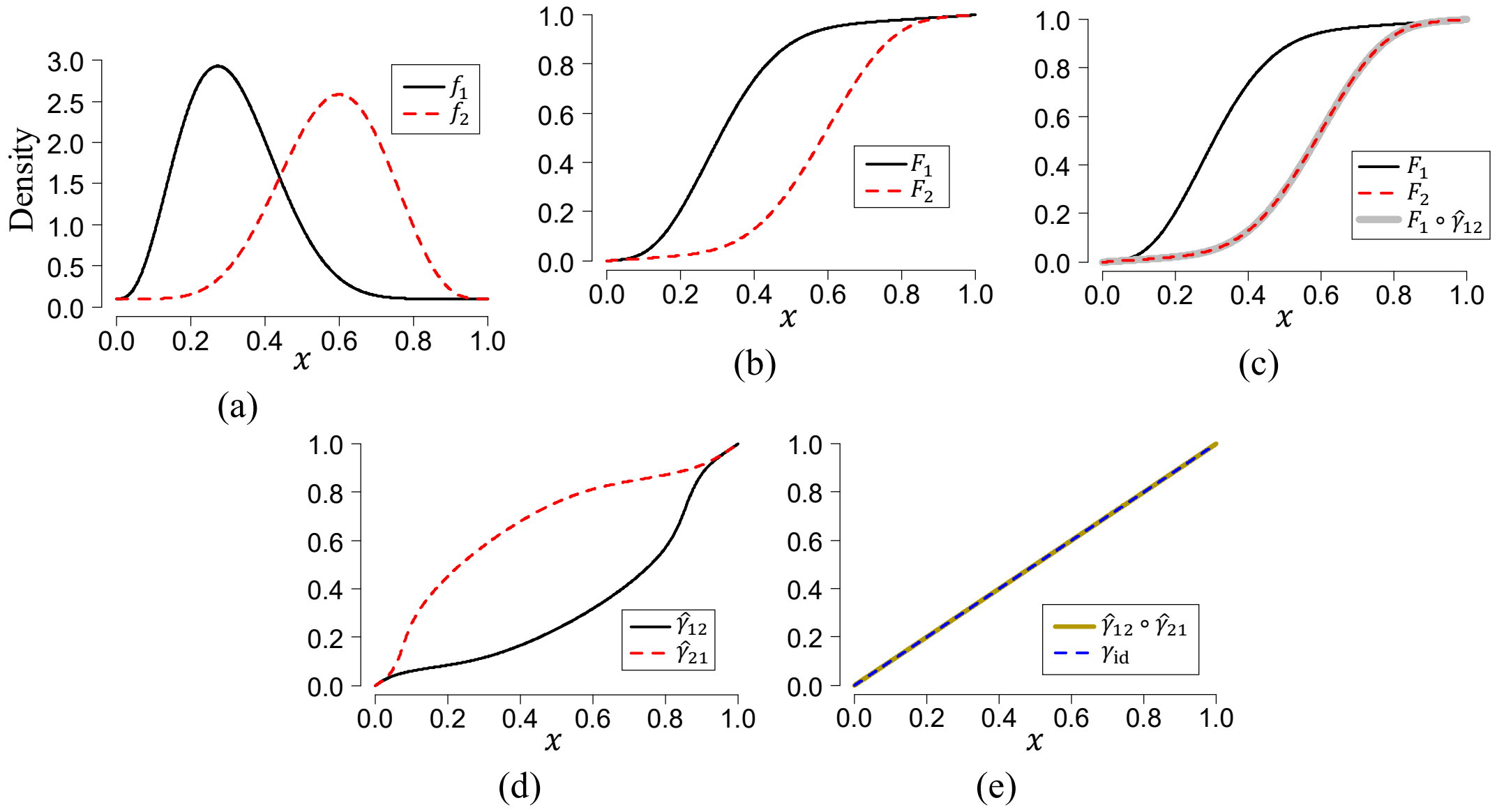

**Fig. A-17.** CDF alignment of the processed distributional data by using direct computed warping function. (a) Processed PDFs $f_1$ and $f_2$ (obtained by using Eq. (A-32) with $\alpha = 0.1$), (b) corresponding CDFs $F_1$ and $F_2$, (c) comparison of $F_1 \circ \hat{\gamma}_{12}$ and $F_2$, (d) calculated warping functions, and (e) comparison of $\hat{\gamma}_{12} \circ \hat{\gamma}_{21}$ and $\gamma_{id}$.

## A.6. Supplemental materials for the simulation and real data studies

### *A.6.1. Basic outlier generation algorithm*

**Algorithm A. 4**: $(S_{PDF}, \text{ IDE})$=PDFoutlier_Insert$(S_{PDF}, N_o, \zeta_{hs}, \varpi)$

**Input:** PDF-valued dataset $\mathrm{S}_{PDF} = \{f_i\}_{i=1}^{n}$, number of outliers $N_o$, and coefficients $\zeta_{hs}$ and $\varpi$

**Output:** The contaminated PDF-valued dataset $\mathrm{S}_{PDF}$, and the index set of outliers denoted as IDE

1: Calculate the modes of the $n$ PDFs $\{f_i\}_{i=1}^{n}$, respectively, and denotes the set of modes as $\Pi = \{\pi_i\}_{i=1}^{n}$ with $\pi_i = \underset{x}{\operatorname{argmax}} f_i(x)$

2: Construct two subsets of PDFs

$$U = \{f_i \in \mathrm{S}_{PDF} | \pi_i \geq q_{1-\varpi}(\Pi)\},\ L = \{f_i \in \mathrm{S}_{PDF} | \pi_i \leq q_{\varpi}(\Pi)\}$$

where $q_{\varpi}(\Pi)$ stands for the $(100\varpi)$th percentile of the dataset $\Pi$

3: Set $\Xi = \{1,2,\cdots,n\}$ and IDE $= \emptyset$

4: **for** $i = 1$ **to** $N_o$ **do**

a: Generate $z \sim \mathrm{U}(0,1)$

b: **if** $z > \zeta_{hs}$ **then** (*generate shape outlier*)

Randomly select one PDF from $U$, and denote it as $h_1 \in U$

Randomly select one PDF from $L$, and denote it as $h_2 \in L$

Compute $h(x) = \varrho h_1(x) + (1-\varrho) h_2(x)$ with $\varrho \sim \mathrm{U}(0.4, 0.6)$

**else** (*generate horizontal-shift outlier*)

Generate $y \sim \mathrm{U}(0,1)$, $a \sim \mathrm{U}(2,5)$, $b \sim \mathrm{U}(13,6)$, $c \sim \mathrm{U}(17,22)$ and $d \sim \mathrm{U}(2,5)$

Compute $h(x) = \mathrm{BetaPdf}(x; a, b) \cdot \chi_{\{y>0.5\}}(y) + \mathrm{BetaPdf}(x; c, d) \cdot \chi_{\{y \leq 0.5\}}(y)$

**end if**

c: Randomly select an element from $\Xi$ denoted as $k_i$

d: Perform PDF replacement $\mathrm{S}_{PDF}[k_i] \leftarrow h$

e: Set $\Xi = \Xi \setminus \{k_i\}$, and put $k_i$ into the index set IDE $\leftarrow k_i$

**end for**

5: Output $\mathrm{S}_{PDF}$ and the index set IDE $= \{k_1, k_2, \cdots, k_{N_o}\}$

### *A.6.2. Initial outlier detection for the real dataset analyzed in the multiple detection*

This subsection conducts an initial outlier detection for the 150 selected PDFs (associated with strain monitoring data collected by the structural health monitoring system of a real bridge) described in Subsection 4.2 of the manuscript. In this initial detection, the Tree-Distance and QF-FDO detection schemes are independently applied to the data, aiming at exploring their complementarity by checking whether they can uncover different types of outliers. The argument settings for the Tree-Distance and QF-FDO detection schemes are listed in Table A-3 and Table A-4, respectively. The other arguments that are not listed in Table A-3 and Table A-4 are the same as the default settings listed in Table A-2. The detection results are visualized in Fig. A-18, the outlying PDFs detected by the two schemes are evidently different, indicating the methods have complementarity with respect to each other. Moreover, the detected outliers using different argument settings are also different, indicating that the detection result contain uncertainty.

**Table A-3**
Argument setting for the Tree-Distance detection scheme in the initial detection.

| | Node | Distance | Whisker | Detection region |
|---|---|---|---|---|
| Scenario I | nLQD | $L_1$ and $L_\infty$ | $L_1$: 2.5IQR<br>$L_\infty$: 3.5IQR | $[0.1, 0.9]$ |
| | CLR | $L_2$ | $L_2$: 2.5IQR | $[u, v]$ |
| | DIFF | $L_1$ and $L_\infty$ | $L_1$: 2.5IQR<br>$L_\infty$: 3.5IQR | $[0, 1]$ |
| | MED | — | 1.5IQR | — |
| Scenario II | Same as Scenario I except the detection region associated with the nLQD node is chosen as $[0.2, 0.8]$ | | | |

**Table A-4**
Argument setting for the QF-FDO detection scheme in the initial detection.

| | Whisker | Detection region |
|---|---|---|
| Scenario I | 1.5IQR (MO direction)<br>2.5IQR (VO direction) | $[0.1, 0.9]$ |
| Scenario II | 1.5IQR (MO direction)<br>2.5IQR (VO direction) | $[0.2, 0.8]$ |

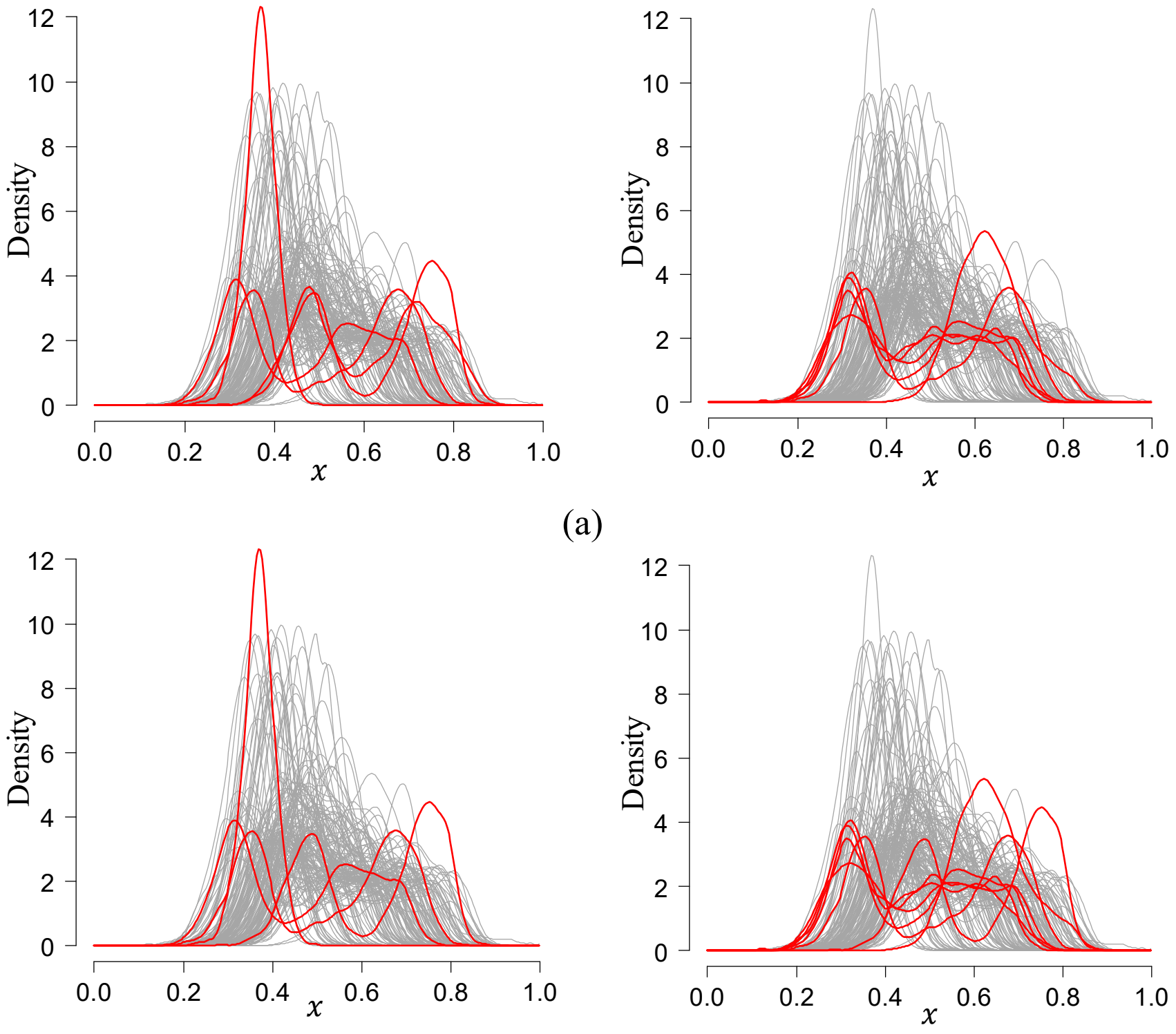

(b)

**Fig. A-18** Comparisons of the detected outlying PDFs (red curves) obtained by the Tree-Distance (left panel) and QF-FDO (right panel) methods in the initial detection with different detection regions. (a) Scenario I and (b) Scenario II.

### *A.6.3. Data-generating process for the regression outlier detection*

We use the mixture beta distribution model to generate $n$ groups of correlated PDF-valued two-tuples denoted as $\{g_i, f_i\}_{i=1}^{n}$. The implementation is summarized in Algorithm A. 5.

---

**Algorithm A. 5**: Generating PDF-valued two-tuples

---

**Input**: Number of two-tuples $n$

**Output**: PDF-valued two-tuples $\{g_i, f_i\}_{i=1}^{n}$

1: **for** $i = 1$ **to** $n$ **do**

a: Generate parameters for $g_i(x)$

$a_i \sim \mathrm{U}(10{,}40)$, $b_i \sim \mathrm{U}(14{,}40)$, $q_i \sim \mathrm{U}(0{,}0.5)$

b: Generate PDF $g_i(x) = (1 - q_i)\mathrm{BetaPdf}(x; a_i, b_i) + q_i \mathrm{BetaPdf}(x; 2a_i, b_i)$

c: Generate parameters for $f_i(x)$

$e_i \sim \mathrm{N}(0, 5^2)$

$c_i = 2.5a_i + \sqrt{a_i} - 15 + e_i$, $d_i = 0.5\sqrt{a_i b_i} + 45 - 0.8a_i + e_i$

$z_i = (c_i + d_i)/2$

d: Generate PDF $f_i(x) = (1 - q_i)\mathrm{BetaPdf}(x; c_i, d_i) + q_i \mathrm{BetaPdf}(x; z_i, z_i)$

**end for**

2: Output $\{g_i, f_i\}_{i=1}^{n}$

---

The PDF-valued two-tuples simulated by Algorithm A. 5 can be equally written as the following structured data:

$$\begin{Bmatrix} g_1 \\ f_1 \end{Bmatrix}, \cdots, \begin{Bmatrix} g_j \\ f_j \end{Bmatrix}, \cdots, \begin{Bmatrix} g_k \\ f_k \end{Bmatrix}, \cdots \begin{Bmatrix} g_n \\ f_n \end{Bmatrix}$$

Clearly, $f_i$ dependents on $g_i$ since their corresponding distributional parameters obey the relationship given in Line c of Algorithm A. 5. If we disorder the matches of the PDF two-tuples via performing an intra-set element exchange as illustrated in Fig. A-19, it can produce two abnormal associations of PDFs. Such a strategy can be used to simulate the abnormal associations for validating the effectiveness of the distributional-regression-based outlier detection method. Algorithm A. 6 presents the implementation of element exchange for single dataset based on the peak information of PDFs, Algorithm A. 7 details the final implementation of the abnormal association generation for contaminating the PDF-valued two-tuples.

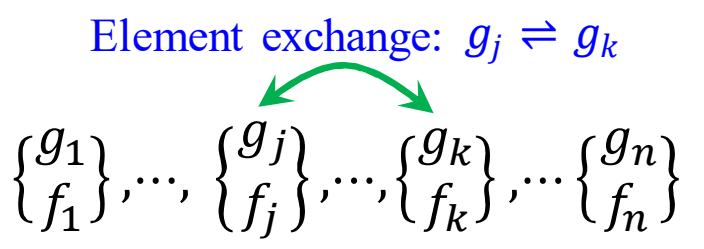

Fig. A-19 Illustration of intra-set element exchange

---

**Algorithm A. 6**: $(S_{PDF}$, IDE)=Element_Exchange$(S_{PDF}, M)$

---

**Input**: PDF-valued dataset $S_{PDF}$ consisting of $n$ PDFs denoted as $\{g_i\}_{i=1}^{n}$, and a parameter $M$ for controlling the number of element exchanges in $S_{PDF}$

**Output**: $S_{PDF}$ after element exchange, and the index set denoted as IDE for locating the places where the element exchanges occur

1: Construct the peak-value set

$$\Pi^g = \{\pi_1^g, \cdots, \pi_n^g\} \text{ with } \pi_i^g = \sup_{x \in [0,1]} g_i(x), i = 1, \cdots, n$$

2: Construct PDF subsets

$$S_1 = \{g_i \in S_{PDF} | \pi_i^g \leq q_{0.2}(\Pi^g)\},\ S_2 = \{g_i \in S_{PDF} | \pi_i^g \geq q_{0.8}(\Pi^g)\}$$

where $q_{0.2}(\Pi^g)$ and $q_{0.8}(\Pi^g)$ are the 20th and 80th percentiles of the dataset $\Pi^g$

3: Randomly select $M$ elements from $S_1$, and denote the selected PDFs as

$$H_L = \{g_{l_1}, g_{l_2}, \cdots, g_{l_M}\} \text{ with } l_1, l_2, \cdots, l_M \text{ being the curve indices in } S_{PDF}$$

4: Randomly select $M$ elements from $S_2$, and denote the selected PDFs as

$$H_U = \{g_{u_1}, g_{u_2}, \cdots, g_{u_M}\} \text{ with } u_1, u_2, \cdots, u_M \text{ being the curve indices in } S_{PDF}$$

5: Perform element exchange in $S_{PDF}$

$$g_{l_1} \rightleftharpoons g_{u_1}, g_{l_2} \rightleftharpoons g_{u_2}, \cdots, g_{l_M} \rightleftharpoons g_{u_M}$$

6: Output the element-exchanged PDF dataset $S_{PDF}$ along with the index set IDE $= \{l_1, \cdots, l_M, u_1, \cdots, u_M\}$

---

**Algorithm A. 7**: Generating abnormal PDF associations by exchanging elements

**Input**: PDF-valued two-tuples $\{g_i, f_i\}_{i=1}^n$, the bivariate parameter $(M_g, M_f)$ for controlling the number of element exchanges in $\{g_i\}_{i=1}^n$ and $\{f_i\}_{i=1}^n$

**Output**: the contaminated PDF-valued two-tuples $\{g_i, f_i\}_{i=1}^n$

1: Denote the PDF-valued datasets $\{g_i\}_{i=1}^n$ and $\{f_i\}_{i=1}^n$ as $S_{PDF}^g$ and $S_{PDF}^f$, respectively

2: **repeat**

a: Perform element exchange on $S_{PDF}^g$ using Algorithm A. 6

$(S_{PDF}^g,\ \mathrm{IDE}_g)$=Element_Exchange$(S_{PDF}^g, M_g)$

b: Perform element exchange on $S_{PDF}^f$ using Algorithm A. 6

$(S_{PDF}^f,\ \mathrm{IDE}_f)$=Element_Exchange$(S_{PDF}^f, M_f)$

**until** $\mathrm{IDE}_g \cap \mathrm{IDE}_f = \emptyset$ ($\emptyset$ denotes the empty set)

3: Output the contaminated PDF-valued two-tuples, i.e.,

$\{g_i, f_i\},\ g_i \in S_{PDF}^g, f_i \in S_{PDF}^g, i = 1,2,\cdots,n$

### *A.6.4. Detailed description for the real data used in the robust distributional regression*

The investigated real data in the robust distributional regression analysis conducted in Subsection 4.4 of the manuscript are strain measurements collected by two strain gauges installed on the bottom plate of the steel box of a long-span bridge. The two sensors are referred to as Sensor A and Sensor B in this study. Sensor A is the one indexed by BII-1-4 in Figure 21 of Chen et al. (2019b), and it is located at the corner of the upstream side of the steel box (of the bridge). Sensor B is installed on the downstream side at the symmetrical position of Sensor A, i.e., the groove next to the U-rib where Sensor BII-1-1 (in Figure 21 of Chen et al. (2019b)) is located. Sensor B has not been selected for investigation in Chen et al. (2019b) because its measurements are seriously contaminated, thus this sensor was not presented in the associated sensor layout diagram.

A total of 8 days measurements collected on August 4,5,9,10,11,12,14 and September 10 of 2012 are selected for investigation. Obviously, the selected days are not continuous, because the data of Sensor B in the other days of the two months are either almost completely contaminated by meaningless outliers or missing. For each sensor, we merge the selected 8-day measurements to form a single time series, and the merged raw data are visualized in Fig. A-20 (a). It can be seen, even in these selected data with relatively "higher quality", there are still many abnormal large values contained in the data of Sensor B. The sample frequency of the strain sensor is 4Hz, thus the 8-day measurements of each sensor include a total of 2764800 data points.

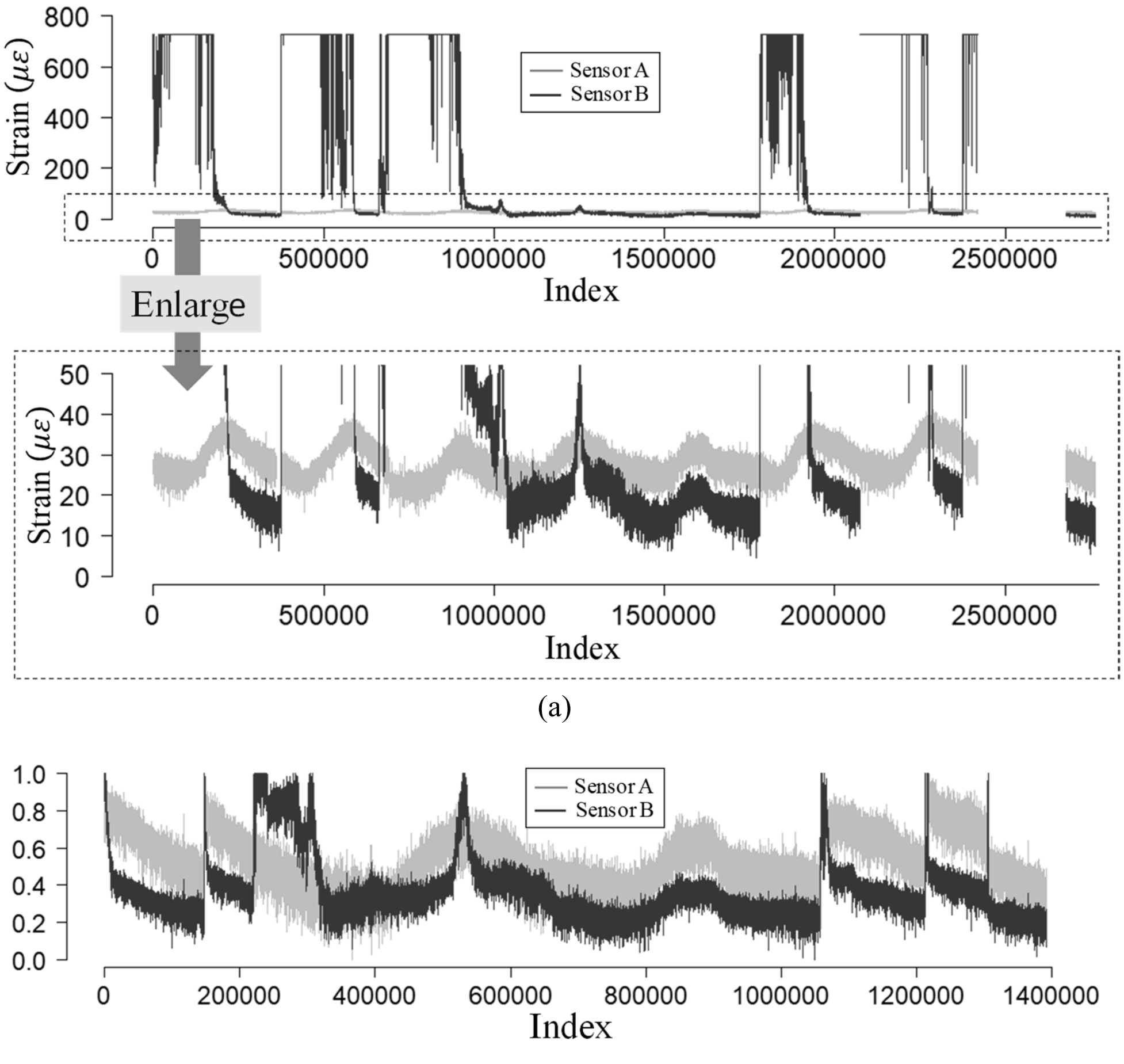

(b)

**Fig. A-20** Investigated strain measurements. (a) Raw data and (b) processed data (eliminating the missing or large values (>50$\mu\varepsilon$), and then scaled to [0,1]).

Before PDF estimation, a rough pretreatment is conducted to the raw data since the data quality of Sensor B is too poor. Specifically, let $\boldsymbol{x}$ and $\boldsymbol{y}$ be the vectors of the selected 8-day time-ordered measurements collected by Sensor A and Sensor B (see Fig. A-20 (a)), respectively, then the two-tuple $(\boldsymbol{x}(k),\ \boldsymbol{y}(k))$ would be removed from $(\boldsymbol{x},\ \boldsymbol{y})$ if the following condition is satisfied:

$$\boldsymbol{x}(k) = \text{NaN} \ \text{ or } \ \boldsymbol{y}(k) = \text{NaN} \ \text{ or } \ \boldsymbol{y}(k) > 50\ (\mu\varepsilon)$$

where NaN stands for the missing data. Finally, the remaining data (consisting of 1392480 data points of each sensor) are merged together in the original time order and then scaled to [0, 1] (similar to that in Chen et al. (2019a)) for each sensor, the resulted data are visualized in Fig. A-20 (b).

For density estimation, the post-processed measurements are divided into 120 segments for each sensor. Each segment consists of 11604 data points (equal to around 48 minutes measurement amount). The PDFs are estimated by kernel density estimator using the segment data as samples (similar to that in Chen et al. (2019a)). The estimated PDFs are denoted as $\{\hat{g}_i\}_{i=1}^{120}$ and $\{\hat{f}_i\}_{i=1}^{120}$ for Sensor A and Sensor B, and visualized in Figs. 10 (a) and (b) of the manuscript, respectively.

### *A.6.5. Results of the two-stage initial outlier detection conducted in Subsection 4.4*

The outlying PDFs detected in the first stage (i.e., the Type I outliers) are visualized in Fig. A-21 as bold colored curves.

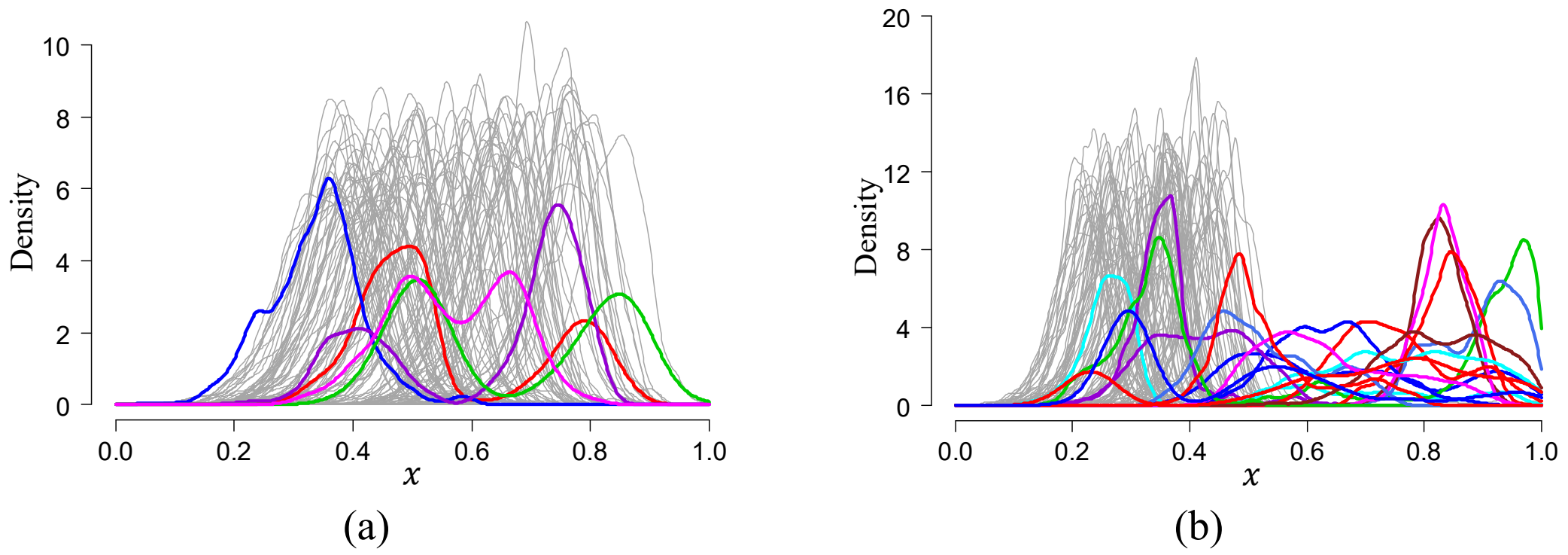

**Fig. A-21** Visualizations of the detected outlying PDFs (represented by bold colored curves) in the first stage. (a) PDFs associated with Sensor A and (b) PDFs associated with Sensor B.

The outlying PDFs detected in the second stage (i.e., the Type II outliers) are visualized in Fig. A-22 as bold colored curves. Recall that the Type II outliers belong to the regression outliers, which correspond to the abnormal associations of the PDF-valued two-tuples. The bold curves in the same color shown in Fig. A-22 belong to the same two-tuple. The anomaly (with respect to the majority of the data) of the detected Type II outlier can only be stand out when viewed in pair. For comparison purposes, Fig. A-23 also displays the curves w.r.t. the bulk of the curves for six selected PDF pairs with normal associations. Comparing the normal PDF pairs in Fig. A-23, one can see that their horizontal positions are correlated with each other. Obviously, the three detected Type II outlying PDFs shown in Fig. A-22 violate the correlation pattern exhibited in Fig. A-23.

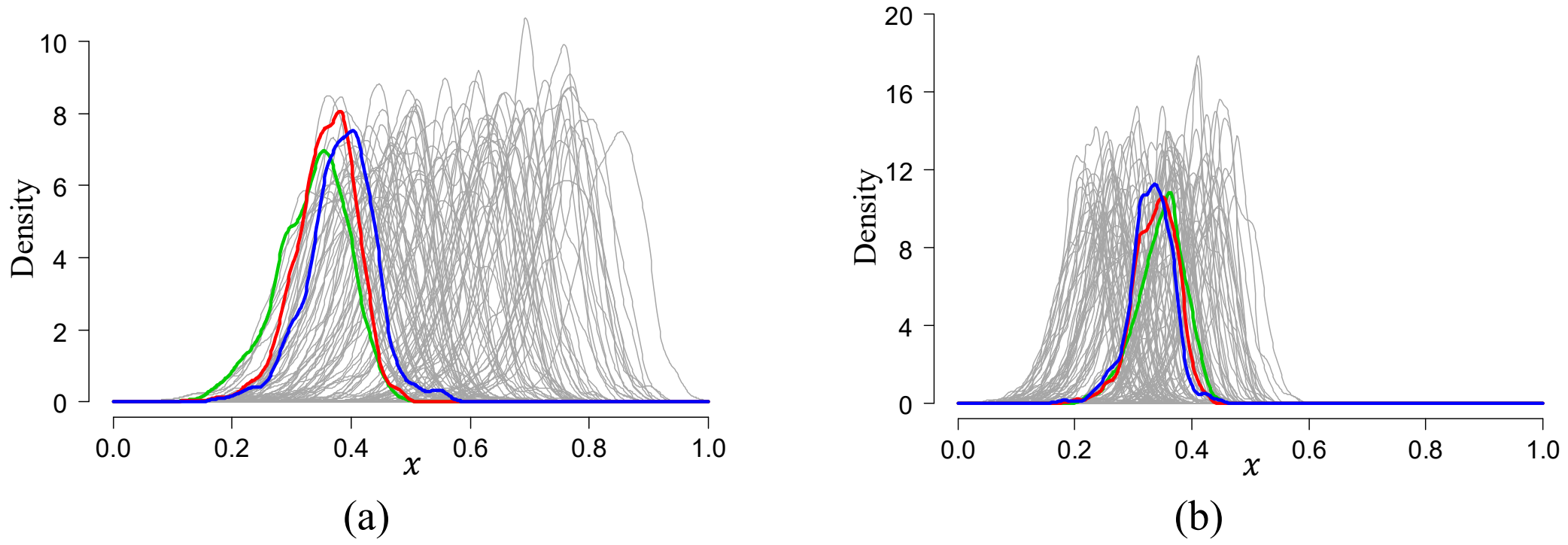

**Fig. A-22** Visualizations of the detected outlying PDFs (represented by bold colored curves) in the second stage (i.e., regression outlier detection). (a) PDFs associated with Sensor A and (b) PDFs associated with Sensor B.

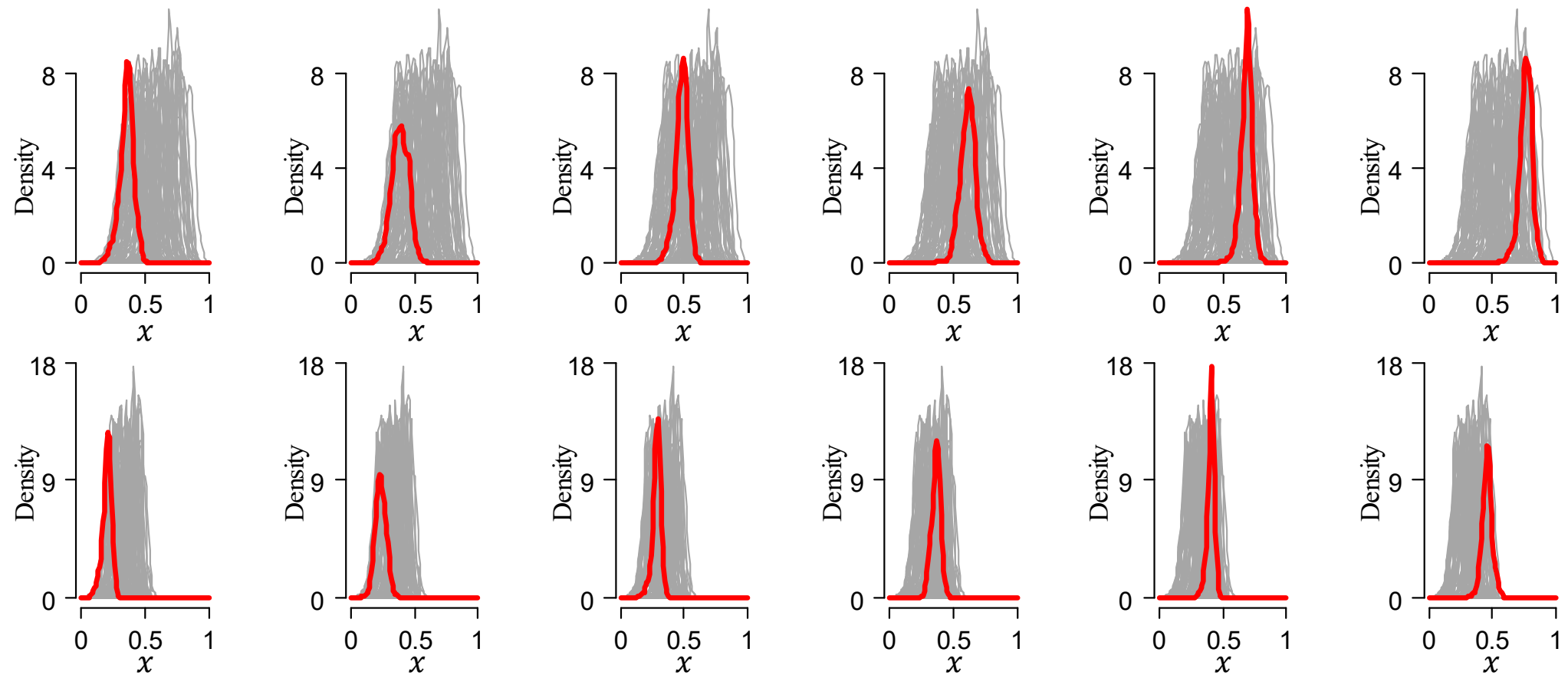

**Fig. A-23** Comparisions for six pairs of PDFs (repesented by bold red curves) with normal associations. The first row corresponds to Sensor A, and the second row corresponds to Sensor B.

## A.7. Additional simulation studies

### *A.7.1. Additional Simulation Study I*

This subsection provides an additional simulation study, and the outlying PDF detection method described in Subsection 4.1 of the manuscript is employed to identify the synthetic outlying curves. Similar to that in Subsection 4.1 of the manuscript, the average detection performance of interest is also based on a series of repeated experiments. In each simulation run, we first independently generate a functional dataset consisting of 100 curves using the following model:

$$v_i(x) = a_i\sin(2\pi x) + b_i\cos(2\pi x), x \in [0,1], i = 1,\cdots,100 \tag{A-33}$$

with $a_i \overset{i.i.d}{\sim} \mathrm{U}(0.012,0.05)$ and $b_i \overset{i.i.d}{\sim} \mathrm{U}(0.012,0.075)$. The simulated functional dataset is referred to as $S_v = \{v_i(x)\}_{i=1}^{100}$. After introducing 10 functional outliers by using Algorithm A. 8, all functions in $S_v$ are converted to PDFs through the following principle:

$$f_i(x) = \frac{v_i(x) - b}{\int_0^1 (v_i(\tau) - b)d\tau}, \quad v_i \in S_v \text{ and } b = \min_{v_k \in S_v} \inf_{x \in [0,1]} \{v_k(x)\} \tag{A-34}$$

The resulting PDF-valued dataset is denoted as $S_f = \{f_i(x)\}_{i=1}^{100}$, representative samples of $S_f$ are displayed in Fig. A-24, where the red lines represent the synthetic outliers.

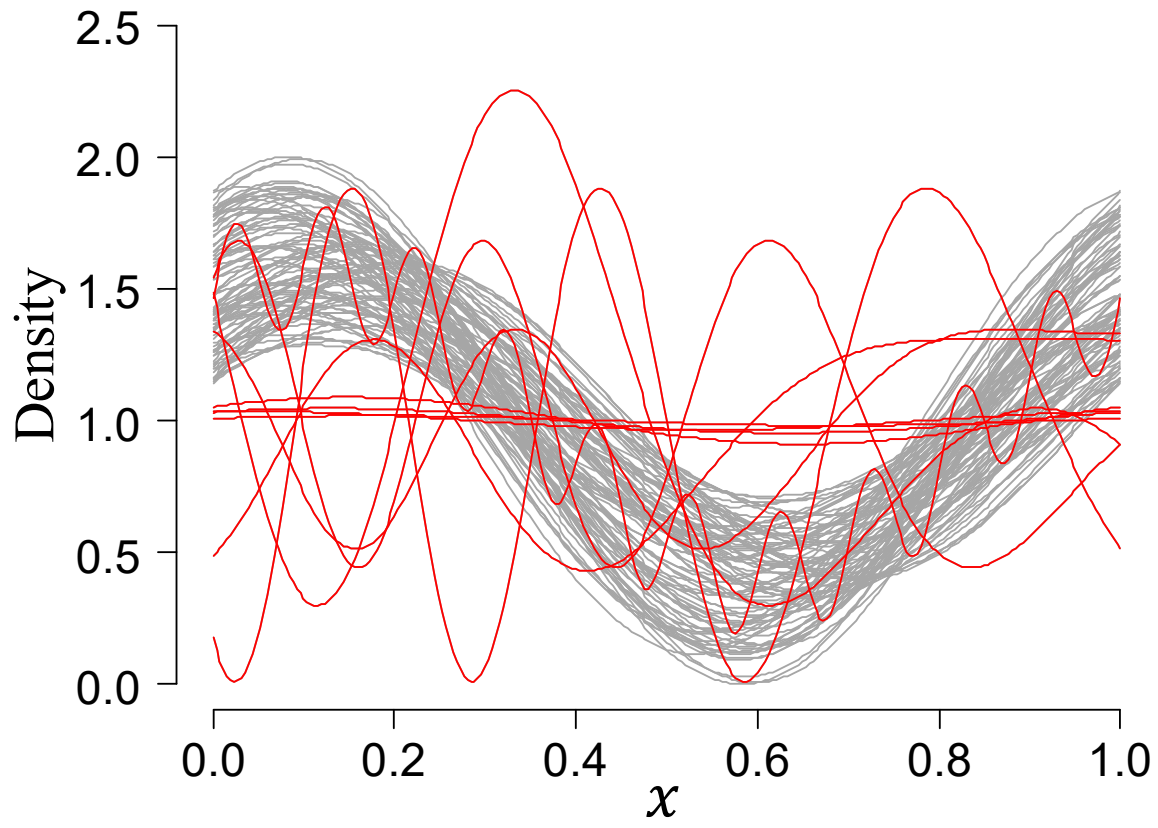

**Fig. A-24.** Representative simulated PDF-valued dataset composed of 90 “good” densities (gray lines) and 10 outlying densities (red lines) using model (A-34) and Algorithm A. 8.

**Algorithm A. 8**: Generate functional outliers

**Input:** Functional dataset $S_v = \{v_i(x)\}_{i=1}^{n}$ with $v_i(x)$ defined on the compact interval $[0,1]$, the number of outliers $N_o$

**Output:** The contaminated functional dataset

1: Initialize the set of outliers $S_{out} = \emptyset$

2: **for** $i = 1$ **to** $N_o - 1$ **do**

  Generate $z \sim \text{U}(0,1)$

  **if** $z < 0.6$ **then**

    Compute $\gamma(x) = c(x-4) + x^3$ with $c \sim \text{U}(-4.5, -2)$, $x \in [0,1]$

    Compute $v_o(x) = a_o \sin(2\pi\gamma(x)) + b_o \cos(2\pi\gamma(x))$

      with $a_o \sim \text{U}(0.02, 0.05)$ and $b_o \sim \text{U}(0, 0.075)$, $x \in [0,1]$

  **else**

    Compute $v_o(x) = a_o \sin(2\pi x) + b_o \cos(2\pi x)$

      with $a_o \sim \text{U}(0, 0.008)$ and $b_o \sim \text{U}(0, 0.008)$, $x \in [0,1]$

  **end if**

  Put $v_o$ into the outlier set $S_{out} \leftarrow v_o$

**end for**

3: Calculate the pointwise median function

$$v_{med}(x) = \underset{1 \le k \le n}{\text{median}}\{v_k(x)\}, \forall x \in [0,1]$$

4: Use the $L_2$ distance to find an element in $S_v$ closest to $v_{med}(x)$, and denote the found element as $v_m$

5: Compute $v_o(x) = v_m(x) + 0.02\sin(20\pi x), x \in [0,1]$, and set $S_{out} \leftarrow v_o$

6: Randomly select $N_o$ elements in $S_v$ to be replaced by the generated outliers stored in $S_{out}$

7: Output the outlier contaminated functional dataset $S_v$

Then, we apply the proposed Tree-Distance detection scheme, with the same argument settings as those in Subsection 4.1 of the manuscript, to the collection of simulated PDFs in $S_f$. We independently repeat such detection experiments for 1000 times, the calculated average values of correct and false detection rates are reported in Table A-5.

**Table A-5**

The calculated average correct detection rates and false detection rates in Additional Simulation Study I. The data in the second, third, fourth and fifth columns correspond to the results detected in the nodes of MED, nLQD, CLR, DIFF, respectively, while the last column corresponds to the merged results detected in the four considered nodes on the tree.

| | MED | nLQD | CLR | DIFF | TREE |
|---|---|---|---|---|---|
| $p_c$(%) | 41.71 | 45.37 | 28.59 | 62.22 | 99.28 |
| $p_f$(%) | 0.00 | 1.18 | 0.02 | 0.00 | 1.20 |

### *A.7.2. Additional Simulation Study II*

Here, we provide an additional comparative study using the same settings with those of Subsection 4.1 in the manuscript, except that the beta densities in E.q. (23) are generated via Algorithm A. 1 by setting $n = 100$, $\delta_1 = 36$ and $\delta_2 = 63$. Using the new data-generating processes, the corresponding results (similar to those of Tables 2~4 in the manuscript) averaging over 1000 repeated experiments are presented in tables from Table A-6 to Table A-8.

**Table A-6**
The calculated average correct detection rates and false detection rates (in brackets) associated with the Tree-Distance detection scheme. The data in the third, fourth, fifth and sixth columns correspond to the results detected in the nodes of MED, nLQD, CLR, DIFF, respectively, while the last column corresponds to the merged results detected in the four considered nodes on the tree. Both $p_c$ and $p_f$ (in brackets) are presented in percentage terms.

| Scenario | Model | MED $p_c$(%) ($p_f$)(%) | nLQD $p_c$(%) ($p_f$)(%) | CLR $p_c$(%) ($p_f$)(%) | DIFF $p_c$(%) ($p_f$)(%) | TREE $p_c$(%) ($p_f$)(%) |
|---|---|---|---|---|---|---|
| Scenario I | Model I | 0.00<br>(0.27) | 99.76<br>(0.25) | 99.99<br>(0.04) | 0.00<br>(0.00) | 100<br>(0.52) |
| | Model II | 0.00<br>(0.30) | 98.16<br>(0.09) | 100<br>(0.02) | 0.00<br>(0.00) | 100<br>(0.39) |
| | Model III | 0.00<br>(0.35) | 96.19<br>(0.06) | 100<br>(0.01) | 0.00<br>(0.00) | 100<br>(0.42) |
| | Model IV | 0.00<br>(0.36) | 92.01<br>(0.00) | 100<br>(0.02) | 0.00<br>(0.00) | 100<br>(0.38) |
| Scenario II | Model I | — | — | — | — | — |
| | Model II | 0.00<br>(0.32) | 96.92<br>(0.00) | 100<br>(0.04) | 0.00<br>(0.00) | 100<br>(0.36) |
| | Model III | 0.00<br>(0.27) | 95.38<br>(0.00) | 100<br>(0.00) | 0.00<br>(0.00) | 100<br>(0.27) |
| | Model IV | 0.00<br>(0.18) | 91.04<br>(0.01) | 99.99<br>(0.00) | 0.00<br>(0.00) | 99.99<br>(0.19) |

**Table A-7**
The calculated average correct detection rates and false detection rates (in brackets) associated with the QF-FDO detection scheme. The data in the third, fourth and fifth columns correspond to the results detected by setting the whisker parameters associated with the VO-outliers to be 1.5, 2.0 and 2.5, respectively, while the whisker parameter associated with the MO-outliers is fixed at 1.5. Both $p_c$ and $p_f$ (in brackets) are presented in percentage terms.

| Scenario | Model | 1.5IQR (VO) $p_c$(%) ($p_f$)(%) | 2.0IQR (VO) $p_c$(%) ($p_f$)(%) | 2.5IQR (VO) $p_c$(%) ($p_f$)(%) |
|---|---|---|---|---|
| Scenario I | Model I | 91.79 (1.02) | 89.83 (0.64) | 87.49 (0.55) |
| | Model II | 92.35 (0.98) | 90.02 (0.63) | 87.64 (0.57) |
| | Model III | 94.74 (0.73) | 93.07 (0.49) | 90.96 (0.44) |
| | Model IV | 93.14 (5.27) | 89.62 (3.78) | 86.16 (2.73) |
| Scenario II | Model I | — | — | — |
| | Model II | 92.75 (0.99) | 90.67 (0.62) | 88.47 (0.53) |
| | Model III | 94.51 (1.59) | 92.61 (1.04) | 90.59 (0.77) |
| | Model IV | 66.52 (16.76) | 64.90 (16.30) | 63.36 (15.93) |

**Table A-8**
The calculated average correct detection rates and false detection rates (in brackets) associated with the warping-function-based detection scheme. Both $p_c$ and $p_f$ (in brackets) are presented in percentage terms.

| Scenario | Model I $p_c$(%) ($p_f$)(%) | Model II $p_c$(%) ($p_f$)(%) | Model III $p_c$(%) ($p_f$)(%) | Model IV $p_c$(%) ($p_f$)(%) |
|---|---|---|---|---|
| Scenario I | 70.46 (14.88) | 70.64 (15.1) | 66.89 (15.40) | 62.34 (15.39) |
| Scenario II | — | 67.99 (15.17) | 65.48 (15.46) | 63.69 (15.39) |

### *A.7.3. Additional Simulation Study III*

This subsection conducts an additional simulation study for regression outlier detection in parallel with the one conducted in Subsection 4.3 of the manuscript.

In each run, 100 groups of correlated PDF-valued two-tuples denoted as $\{g_i, f_i\}_{i=1}^{100}$ are generated by using Algorithm A. 9, and the abnormal associations are generated by using Algorithm A. 10 with $\zeta_{hs} = 0$ and $\varpi = 0.25$. We consider seven different contamination scenarios with $(N_g, N_f)$ (denoting the associated numbers of outliers introduced to $\{g_i\}_{i=1}^n$ and $\{f_i\}_{i=1}^n$) valued at (10, 0), (8, 2), (6, 4), (5, 5), (4, 6), (2, 8) and (0, 10), respectively. In each contamination scenario, the distributional-regression-based detection method with the same argument settings of those in Subsection 4.3 of the manuscript is employed to detect the abnormal associations (i.e., the regression outliers). Based on 500 repeated detection tests, the calculated average correct and false detection rates are listed in Table A-9 for the seven considered contamination scenarios.

---

**Algorithm A. 9**: Generating PDF-valued two-tuples for Additional Simulation Study III

**Input**: Number of two-tuples $n$

**Output**: PDF-valued two-tuples $\{g_i, f_i\}_{i=1}^n$

1: Independently generate parameters for $g_i(x), i = 1,2,\cdots,n$

$\boldsymbol{A} = \{a_1, a_2, \cdots, a_n\}$ with $a_i$ being i.i.d. samples of $\mathrm{U}(14,30)$

$\boldsymbol{B} = \{b_1, b_2, \cdots, b_n\}$ with $b_i$ being i.i.d. samples of $\mathrm{U}(14,20)$

2: Independently generate the errors

$\boldsymbol{E} = \{e_1, e_2, \cdots, e_n\}$ with $e_i$ being i.i.d. samples of $\mathrm{N}(0, 3^2)$

3: **for** $i = 1$ **to** $n$ **do**

a: Generate PDF $g_i(x) = \mathrm{BetaPdf}(x; a_i, b_i)$

b: Generate parameters for $f_i(x)$

$$c_i = 40\frac{a_i - \min(\boldsymbol{A})}{\max(\boldsymbol{A}) - \min(\boldsymbol{A})} + 12 + e_i, \quad d_i = \sqrt{a_i b_i + a_i} + e_i$$

c: Generate PDF $f_i(x) = \mathrm{BetaPdf}(x; c_i, d_i)$

**end for**

4: Output $\{g_i, f_i\}_{i=1}^n$

---

**Algorithm A. 10**: Generating abnormal PDF associations by inserting outliers

**Input**: PDF-valued two-tuples $\{g_i, f_i\}_{i=1}^n$, the bivariate parameter $(N_g, N_f)$ for controlling the number of outliers introduced to $\{g_i\}_{i=1}^n$ and $\{f_i\}_{i=1}^n$, respectively, and coefficients $\zeta_{hs}$ and $\varpi$

**Output**: The contaminated PDF-valued two-tuples $\{g_i, f_i\}_{i=1}^n$

1: Denote the PDF-valued datasets $\{g_i\}_{i=1}^n$ and $\{f_i\}_{i=1}^n$ as $S_{PDF}^g$ and $S_{PDF}^f$, respectively

2: **repeat**

a: Randomly insert $N_g$ outlying PDFs into $S_{PDF}^g$ using Algorithm A. 4

$(S_{PDF}^g, \mathrm{IDE}_g)$= PDFoutlier_Insert$(S_{PDF}^g, N_g, \zeta_{hs}, \varpi)$

b: Randomly insert $N_f$ outlying PDFs into $S_{PDF}^f$ using Algorithm A. 4

$(S_{PDF}^f, \mathrm{IDE}_f)$= PDFoutlier_Insert$(S_{PDF}^f, N_f, \zeta_{hs}, \varpi)$

**until** $\mathrm{IDE}_g \cap \mathrm{IDE}_f = \emptyset$ ($\emptyset$ denotes the empty set)

3: Output the contaminated PDF-valued two-tuples, i.e.,

$\{g_i, f_i\}, g_i \in S_{PDF}^g, f_i \in S_{PDF}^g, i = 1,2,\cdots,n$

---

**Table A-9**
The calculated average correct and false detection rates associated with the regression outlier detection conducted in Additional Simulation Study III for the seven considered contamination scenarios.

| $(N_g, N_f)$ | (10, 0) | (8, 2) | (6, 4) | (5, 5) | (4, 6) | (2, 8) | (0, 10) |
|---|---|---|---|---|---|---|---|
| $p_c$ (%) | 99.82 | 99.22 | 99.18 | 98.54 | 98.90 | 98.98 | 99.70 |
| $p_f$ (%) | 4.08 | 4.79 | 5.22 | 5.48 | 5.30 | 5.00 | 3.97 |

## A.8. Additional figure

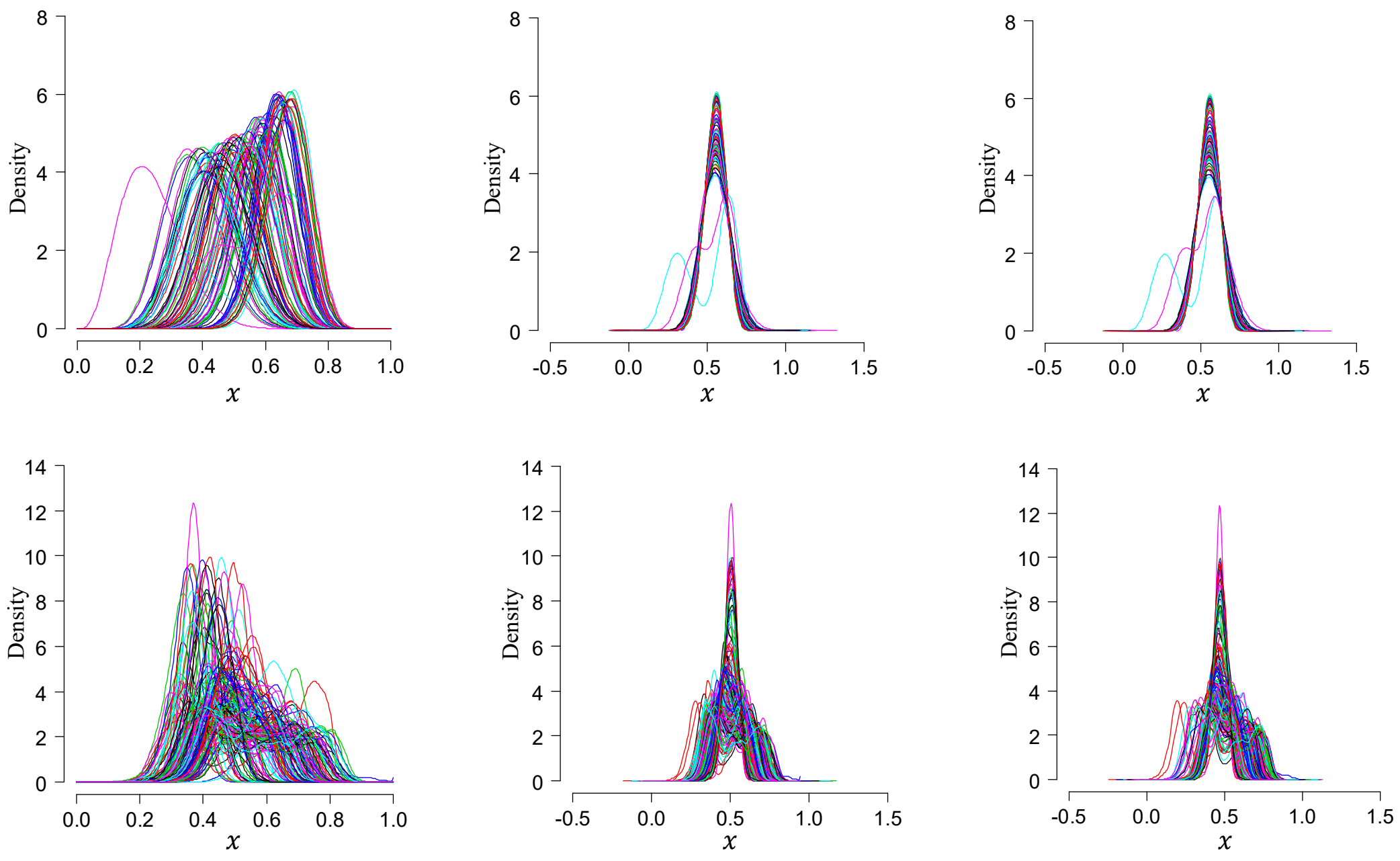

**Fig. A-25.** Illustration of the horizontal centralization processing conducted in the H-CENTR node (on the transformation tree) for two PDF-valued datasets. The first row corresponds to a simulated dataset, while the second row corresponds to a real dataset (same as the one investigated in the multiple detection conducted in Subsection 4.2 of the manuscript). Two centralization scenarios are presented for the raw PDFs (first column): (i) Scenario I (second column): the median is selected as the feature point for horizontal alignment; (ii) Scenario II (third column): the feature point for horizontal alignment is computed by the average of the median and mode.